%% file: ms.tex
\documentclass[10pt,journal,compsoc]{IEEEtran}
\usepackage{amsthm}
\usepackage{amsmath,amssymb,amsfonts}
\usepackage{algorithm}
\usepackage{algpseudocode}
\usepackage{graphicx}
\usepackage{textcomp}
\usepackage{xcolor}
\usepackage{soul}
\usepackage{makecell}
\usepackage{ctable}
\usepackage{booktabs} % For formal tables
\usepackage{multirow}
\usepackage{thm-restate}
\usepackage{sistyle}
\SIthousandsep{,}
\usepackage{calligra}
\usepackage{enumitem}
\usepackage{subfigure}
\usepackage{footmisc} % enabling footnotes.
\usepackage{url}
\usepackage{makecell}
\usepackage{bigstrut}
\usepackage{multicol}
\usepackage{adjustbox}
\usepackage{makecell}
\usepackage[nocompress]{cite}

\ifCLASSINFOpdf
\else
\fi
\begin{document}
\title{A Comprehensive Comparison of Multi-Dimensional Image Denoising Methods}

\author{Zhaoming~Kong,
        Xiaowei~Yang,
        and~Lifang~He% <-this % stops a space
\IEEEcompsocitemizethanks{\IEEEcompsocthanksitem Zhaoming Kong and Lifang He are with with the Computer Science and Engineering Department, Lehigh University,
PA, 18015 USA.\protect\\
E-mail: zhk219@lehigh.edu
\IEEEcompsocthanksitem Xiaowei Yang is with the school of Software Engineering, South China University of Technology, Guangdong Province, China.}% <-this % stops an unwanted space
}

%\markboth{Journal of \LaTeX\ Class Files,~Vol.~18, No.~8, August~2019}%
%{Shell \MakeLowercase{\textit{et al.}}: Bare Demo of IEEEtran.cls for Computer Society Journals}

\IEEEtitleabstractindextext{%
\begin{abstract}
%Filtering multi-dimensional images such as color images, color videos, multispectral images and magnetic resonance images are challenging in terms of both effectiveness and efficiency. Leveraging the nonlocal self-similarity (NLSS) characteristic of images and sparse representation in the transform domain, the block matching 3D (BM3D) methods show powerful denoising performance. Recently, numerous new approaches with different regularization terms, transforms and advanced deep neural network (DNN) architectures are proposed to improve denoising quality. In this paper, we extensively compare 60+ methods for multi-dimensional image denoising tasks. We introduce a new color image and video dataset for benchmarking, and our evaluations are performed from four different perspectives including quantitative metrics, visual effects, human ratings and computational cost. Comprehensive experiments on both synthetic and real-world datasets demonstrate: (i) the effectiveness and efficiency of the BM3D family for various denoising tasks, (ii) a simple matrix-based algorithm could produce similar results compared with tensor-based methods, and (iii) several DNN models trained with synthetic Gaussian noise show state-of-the-art performance on different real-world color image and video datasets. Despite the progress in recent years, we also discuss shortcomings and possible extensions of existing techniques. All our code and datasets are publicly available at https://github.com/ZhaomingKong/Denoising-Comparison.

Filtering multi-dimensional images such as color images, color videos, multispectral images and magnetic resonance images is challenging in terms of both effectiveness and efficiency. Leveraging the nonlocal self-similarity (NLSS) characteristic of images and sparse representation in the transform domain, the block-matching and 3D filtering (BM3D) based methods show powerful denoising performance. Recently, numerous new approaches with different regularization terms, transforms and advanced deep neural network (DNN) architectures are proposed to improve denoising quality. In this paper, we extensively compare over 60 methods on both synthetic and real-world datasets. We also introduce a new color image and video dataset for benchmarking, and our evaluations are performed from four different perspectives including quantitative metrics, visual effects, human ratings and computational cost. Comprehensive experiments demonstrate: (i) the effectiveness and efficiency of the BM3D family for various denoising tasks, (ii) a simple matrix-based algorithm could produce similar results compared with its tensor counterparts, and (iii) several DNN models trained with synthetic Gaussian noise show state-of-the-art performance on real-world color image and video datasets. Despite the progress in recent years, we discuss shortcomings and possible extensions of existing techniques. Datasets and codes for evaluation are made publicly available at https://github.com/ZhaomingKong/Denoising-Comparison.
\end{abstract}

}

\maketitle

\IEEEdisplaynontitleabstractindextext

\IEEEpeerreviewmaketitle

\IEEEraisesectionheading{\section{Introduction}\label{sec:introduction}}

\IEEEPARstart{I}mage denoising plays an important role in modern imaging systems, and it has attracted much attention from both academia and industry. Theoretically, image denoising is a special case of inverse problems \cite{tirer2020back}, which aims to estimate the underlying clean image from the noisy observation. In real-world applications, it can be used as a pre-processing step for subsequent tasks, such as visual tracking \cite{isard1998condensation}, segmentation \cite{pal1993review} and classification \cite{raczkowska2019influence}. Also, it may serve as the ultimate goal of image quality enhancement to produce more visually pleasant images.\\
\indent Image denoising enjoys a rich history and the earliest works may date back to the 1950s \cite{wiener1950extrapolation}. The recent surge of denoising methods is mainly credited to the famous block-matching 3D (BM3D) \cite{dabov2007image} framework, which combines the nonlocal similarity characteristic \cite{buades2005review} of natural images and the sparse representation in the transform domain \cite{yaroslavsky2001transform}. At an early stage, many related works focus on filtering single-channel grayscale images \cite{dabov2009bm3d, katkovnik2010local}. Recently, the advancement of imaging systems and technologies has largely enriched the information preserved and presented by multi-dimensional images, which could deliver more faithful representation for real scenes. For example, the popularity of mobile camera phones facilitates the capturing and recording of colorful and dynamic objects. The multispectral imaging (MSI) and hyperspectral imaging (HSI) techniques provide image information in both the spatial and spectral domain. Magnetic resonance imaging (MRI) utilizes a non-invasive imaging technology that produces three dimensional detailed anatomical images often used for disease detection, diagnosis, and treatment monitoring. \\
\indent The growth of image size and dimension also posts greater demands on denoising. One major challenge of handling higher dimensional images is to effectively exploit the inter-correlated information of multiple channels or spectrums, and also find a balance between noise removal and detail preservation. In the past two decades, the representative BM3D method has been successfully extended to multi-dimensional images in two different ways. The first strategy is to apply a certain decorrelation transform such that in the transformed space each channel or spectral band may be filtered independently by some effective single-channel denoiser. For example, the color-BM3D (CBM3D) method \cite{dabov2007color} shows that the established opponent or YCbCr color transform of natural RGB images could provide near-optimal decorrelation of the color data \cite{danielyan2010denoising}. An alternative solution is to model and utilize the channel- or band-wise correlation by processing the full set of the multi-dimensional image data jointly. To achieve this goal, Maggioni et at. \cite{maggioni2012nonlocal} extend BM3D to BM4D by using 3D cubes of voxels, which are then stacked into a 4D group.\\
\indent Since the birth of BM3D, there is no shortage of denoising methods originating from different disciplines. Interestingly, from the traditional Gaussian denoisers \cite{shao2013heuristic} with matrix and tensor representations to recently developed approaches using different deep neural network (DNN) architectures \cite{tian2019deep}, nearly all the newly proposed methods claim to outperform the BM3D family \cite{dabov2007image, dabov2007color, kostadin2007video, maggioni2012video, maggioni2012nonlocal}. However, some recent studies \cite{plotz2017benchmarking, kong2019color} come to a different conclusion, and it is observed that many methods are normally verified based on a limited number (often less than three) of datasets, and the parameters of BM3D-based methods may not be fine-tuned with certain noise estimation techniques \cite{liu2013single, chen2015efficient}. With a large number of existing methods \cite{Collection_denoising_methods} and outstanding survey papers \cite{yaroslavsky2001transform, buades2005review, milanfar2012tour, thanh2019review, shao2013heuristic, mohan2014survey, schmidhuber2015deep, kong2018brief, tian2019deep}, there still lacks a thorough comparison for the multi-dimensional image denoising tasks. In this paper, we intend to fill this gap by collecting and comparing different methods. The main contributions of this paper are three-fold: \\
%\footnote{https://github.com/wenbihan/reproducible-image-denoising-state-of-the-art}
\indent 1) We construct a new real-world indoor-outdoor color image (IOCI) and color video (IOCV) dataset with multiple different cameras. The dataset consist of images and video sequences captured based on both artificial indoor subjects and natural outdoor scenes under different lighting conditions, which are a good complement to current benchmark datasets with various challenging realistic backgrounds. \\
\indent 2) We compare over 60 multi-dimensional image denoising methods, and perform extensive experiments on both synthetic and real-world datasets. We adopt both subjective and objective metrics, and evaluate the denoising quality of compared methods with quantitative results and human visual ratings. \\
\indent 3) We make three interesting observations based on comprehensive experimental results. First the BM3D family still demonstrate very competitive performance for multiple denoising tasks in terms of both effectiveness and efficiency. Second, for traditional denoisers that learn solely on the noisy observation, we argue that a modified singular value decomposition (M-SVD) method is able to produce similar results with many tensor-based approaches. Furthermore, several DNN models trained with synthetic noise demonstrate impressive generalizability to real-world cases. Specifically, a fusion of collaborative and convolutional filtering (FCCF) model \cite{yue2019high}, and a U-Net \cite{ronneberger2015u} and ResNet \cite{he2016deep} based denoising network (DRUNet) \cite{zhang2020plug} produce the best results for the color image denoising task. The fast deep video denoising network (FastDVDNet) \cite{Tassano_2020_CVPR} and a nonlocal video denoising network (VNLNet) \cite{davy2018non} show significant improvements for color video denoising compared with the benchmark video-BM4D (VBM4D) \cite{maggioni2012video} implementation.\\
\indent The rest of this paper is organized as follows. Section \ref{section_background} introduces background knowledge, and Section \ref{section_related_works} gives a brief review on related multi-dimensional image denoising techniques and datasets. Section \ref{section_experiments} provides experimental results and discussions. Section \ref{section_conclusion} concludes the paper.
%and the U-Net \cite{ronneberger2015u} and ResNet \cite{he2016deep} based denoising network DRUNet \cite{zhang2020plug} produce the best results for the color image denoising task, while FastDVDNet \cite{Tassano_2020_CVPR} and VNLNet \cite{davy2018non} show significant improvements for color video denoising compared with the state-of-the-art video-BM4D (VBM4D) \cite{maggioni2012video} implementation.\\
%\indent The rest of this paper is organized as follows. Section \ref{section_background} introduces background knowledge, and Section \ref{section_related_works} gives a brief review on related multi-dimensional image denoising techniques and datasets. Section \ref{section_experiments} summarizes experimental results and discussions. Section \ref{section_conclusion} concludes the paper.

\section{Background} \label{section_background}

\subsection{Symbols and Notations}
\indent In this paper, to represent multi-dimensional images, we mainly adopt the mathematical notations and preliminaries of tensors from \cite{kolda2009tensor}. Vectors and matrices are first- and second- order tensors which are denoted by boldface lowercase letters $\mathbf{a}$ and capital letters $\mathbf{A}$, respectively. A higher order tensor (the tensor of order three or above) is denoted by calligraphic letters, e.g., $\mathcal{A}$. An $N$th-order tensor is denoted as $\mathcal{A} \in \mathbb{R}^{I_1\times I_2\times\cdots\times I_N}$. The $n$-mode product of a tensor $\mathcal{A}$ by a matrix $\mathbf{U}\in \mathrm{R}^{P_n\times I_n}$, denoted by $\mathcal{A}\times _n\mathbf{U} \in \mathbb{R}^{I_1 \cdots I_{n-1} P_n I_{n+1} \cdots I_N}$ is also a tensor. The mode-$n$ matricization or unfolding of $\mathcal{A}$, denoted by $\mathbf{A}_{(n)}$, maps tensor elements $(i_1,i_2,\ldots,i_N)$ to matrix element $(i_n,j)$ where $j=1+\sum_{k=1,k\neq n}^{N}(i_k-1)J_k$, with $J_k = \prod_{m=1,m\neq n}^{k-1}I_m$. The Frobenius norm of a tensor $\mathcal{A} \in \mathbb{R}^{I_1\times I_2\times\cdots\times I_N}$ is defined as $\|\mathcal{A}\|_F = \sqrt{\sum_{i_1=1}...\sum_{i_N=1}\mathcal{A}_{i_1...i_N}^2}$. The nuclear norm of a matrix is denoted by $\|\mathbf{A}\|_\ast$, which is the sum of all singular values of $\mathbf{A}$. The symbol of $\otimes $ denotes the Kronecker product of matrices.

\subsection{Noise modeling}
Let us consider a noisy observation $\mathcal{Y}$ and its underlying clean image $\mathcal{X}$, a general assumption of noise distribution is additive white Gaussian noise (AGWN) with variance $\sigma^2$ represented by $\mathcal{N}(0, \sigma^2)$, and the degradation process is then given as
\begin{equation}\label{awgn}
  \mathcal{Y} = \mathcal{X} + \mathcal{N}
\end{equation}
According to \cite{ramanath2005color}, noise modeling can be more complex and challenging since noise in real-world images may be multiplicative and signal dependent. Therefore, there are some non i.i.d Gaussian models designed for multi-dimensional images, such as the mixed Gaussian impulse noise of color images \cite{yan2013restoration, huang2017mixed}, sparse random corruptions of video data \cite{zhang2014novel}, strip noise removal of MSI \cite{chen2017denoising}, and Rician noise reconstruction of MRI \cite{awate2007feature}. In this paper, our synthetic experiments and analysis are mainly based on the AGWN model because (i) a majority of compared methods could handle Gaussian noise, (ii) certain types of noise could be transformed to Gaussian distribution, and (iii) Romano et al. \cite{romano2017little} recently point out that the removal of AGWN from an image is largely a solved problem, which may help explain the effectiveness of the simplified noise modeling in Eq. (\ref{awgn}).
\subsection{Nonlocal self-similarity}
\input{Fig_NSS}
\noindent The nonlocal self-similarity (NLSS) feature of natural images is probably the most important prior adopted by many different denoising methods. Briefly, NLSS refers to the fact that a local image patch often has many nonlocal similar patches to it across the image \cite{xu2015patch}. Usually, the similarity between two patches $\mathcal{P}_A$ and $\mathcal{P}_B$ is measured by their Euclidean distance $d_{AB} = \|\mathcal{P}_A - \mathcal{P}_B\|$. In practice, to save some time, the search for similar patches is restricted to a local window $\Omega_{SR}$ with pre-defined size. As illustrated in Fig. \ref{Fig_NSS}, the patch representation and the rule of similar patch search (SPS) may vary for different types of multi-dimensional images. For example, SPS of color images can be conducted only on the luminance channel to reduce complexity; for video sequences, SPS is performed along both temporal and spatial dimensions; and for MRI and MSI data, a patch could be represented by a 3D cube or a square tube with multiple spectral bands.
\section{Related Works} \label{section_related_works}
In this section, we briefly introduce related multi-dimensional denoising methods and datasets, which are summarized in Table \ref{Table_traditional_method}, Table \ref{Table_DNN_method} and Table \ref{Table_dataset_description}, respectively. The timeline of representative denoising methods and datasets is illustrated in Fig. \ref{Fig_timeline}. In our paper, the denoising methods are roughly divided into two categories, namely 'traditional denoisers' and 'DNN methods', depending on whether neural network architectures are used. More detailed description of related works are available in \cite{yaroslavsky2001transform, buades2005review, milanfar2012tour, thanh2019review, shao2013heuristic, mohan2014survey, schmidhuber2015deep, kong2018brief, tian2019deep}.
\input{Table_traditional_method}

\subsection{Traditional denoisers}
\vspace{-0.1cm}
\input{Fig_traditional_framework}
\vspace{-0.1cm}
\noindent For traditional denoisers, training and denoising are usually accomplished only with the noisy image. To utilize the NLSS prior, the most popular and successful framework attributed to BM3D \cite{dabov2007image} mainly follows three consecutive stages: grouping, collaborative filtering and aggregation. The flowchart of this effective three-stage paradigm is illustrated in Fig. \ref{Fig_traditional_framework}.
\subsubsection{Grouping}
For every $d$-dimensional noisy image patch $\mathcal{P}_{n}$, based on certain patch matching criteria \cite{foi2007pointwise, rubel2014metric, buades2016patch, foi2016foveated, ehret2017global, Foi2020}, the grouping step stacks $K$ similar (overlapping) patches located within a local window $\Omega_{SR}$ into a $d+1$-dimensional group. For example, consider a 3D patch $\mathcal{P}_{n} \in \mathbb{R}^{H \times W \times N}$, where $H$, $W$ and $N$ represents height, width and the number of channels or spectral bands, respectively, the 4D group of $K$ patches can be directly represented by a fourth-order tensor $\mathcal{G}_n \in \mathbb{R} ^{H \times W \times N \times K}$, or a 2D matrix $\mathbf{G}_n \in \mathbb{R}^{HWN \times K}$ if every patch $\mathcal{P}_n$ is reshaped into a long vector $\mathbf{p}_n \in \mathbb{R}^{HWN}$.
\subsubsection{Collaborative filtering}
Collaborative filters operate on the noisy patch group $\mathcal{G}_n$ to estimate the corresponding underlying clean group $\mathcal{G}_c$ via
\begin{equation}\label{tensor_collaborative_filtering}
  \hat{\mathcal{G}}_c = \mathop{\arg\min_{\mathcal{G}_c}} \| \mathcal{G}_n - \mathcal{G}_c \|_{F}^2 + \rho\cdot\Psi(\mathcal{G}_c)
\end{equation}
or in the matrix form
\begin{equation}\label{matrix_collaborative_filtering}
  \hat{\mathbf{G}}_c = \mathop{\arg\min_{\mathbf{G}_c}} \| \mathbf{G}_n - \mathbf{G}_c \|_{F}^2 + \rho\cdot\Psi(\mathbf{G}_c)
\end{equation}
where $\| \mathcal{G}_n - \mathcal{G}_c \|_{F}^2$ or $\| \mathbf{G}_n - \mathbf{G}_c \|_{F}^2$ indicates the conformity between the clean and noisy groups, and $\Psi(\cdot)$ is a regularization term for certain priors. For example, to model the nonlocal redundancies, the low-rank prior is adopted in \cite{xu2017multi, dong2012nonlocal, dong2015low, chang2017hyper} with $\Psi(\mathbf{G}_c) = \|\mathbf{G}_c\|_\ast$ for matrix and $\Psi(\mathcal{G}_c) = \sum_{n = 1}^4 a_{(n)}\|\mathbf{G}_{c_{(n)}}\|_\ast$ for tensor \cite{liu2012tensor}. In addition, the dictionary learning model with over-complete representations \cite{elad2006image,mairal2007sparse,xu2018trilateral} is utilized to reconstruct $\mathbf{G}_c$ with a dictionary $\mathbf{D}$ and sparse coding coefficients $\mathbf{C}$ via
\begin{equation}\label{matrix_dictionary}
  \hat{\mathbf{C}} = \mathop{\arg\min_{\mathbf{C}}}\|\mathbf{G}_n - \mathbf{D} \mathbf{C}\|_F^2 + \lambda\|\mathbf{C}\|_1
\end{equation}
where $\lambda \| \cdot \|$ is the regularization term that enforces sparsity constraint on $\mathbf{C}$. Once $\hat{\mathbf{C}}$ is computed, the latent clean patch group $\hat{\mathbf{G}}_c$ can be estimated as $\hat{\mathbf{G}}_c = \mathbf{D}\hat{\mathbf{C}}$. In \cite{peng2014decomposable} and \cite{Zhang2015KTSVD}, Eq. (\ref{matrix_dictionary}) is extended to tensor for MSI denoising with higher-order SVD (HOSVD) \cite{tucker1966some, de2000multilinear} and tensor-SVD (t-SVD) \cite{kilmer2011factorization, kilmer2013third} transforms, respectively. A simple and effective method is to model the sparsity with certain thresholding techniques \cite{donoho1994ideal, donoho1995noising} to attenuate the noise. For example, the hard-thresholding technique is adopted by the BM3D family and some state-of-the-art tensor-based methods \cite{rajwade2012image, kong2019color}, which attempts to shrink the coefficients $\mathcal{T}(\mathcal{G}_n)$ in the transform-domain \cite{yaroslavsky2001transform} under a threshold $\tau$ via
 \begin{equation}\label{hard_thresholding}
  \mathcal{G}_{t}=\left\{
    \begin{aligned}
    \mathcal{T}(\mathcal{G}_n), \quad |\mathcal{T}(\mathcal{G}_n)| \geq \tau \\
    0, \quad |\mathcal{T}(\mathcal{G}_n)| < \tau
    \end{aligned}
    \right.
 \end{equation}
where $\mathcal{T}$ represents a pre-defined or learned transform. The estimated clean group $\hat{\mathcal{G}}_c$ is obtained by inverting the transform via
\begin{equation}\label{clean_group_estimate}
  \hat{\mathcal{G}}_c = \mathcal{T}^{-1} (\mathcal{G}_{t})
\end{equation}
The popularity of SVD-based transforms in traditional denoisers largely results from its invertible orthogonal bases.
\subsubsection{Aggregation}
To further smooth out noise, the estimated clean patches of $\hat{\mathcal{G}}_c$ are averagely written back to their original location. More specifically, at the pixel level, every pixel $\hat{p}_i$ of the denoised image is the (weighted) average of all pixels at the same position of filtered group $\hat{\mathcal{G}}_c$, which can be formulated as
 \begin{equation}\label{aggregation}
   \hat{p}_i = \sum_{\hat{p}_{i_k} \in \hat{\mathcal{G}}_c} w_{i_k} \hat{p}_{i_k}
 \end{equation}
 where $w_{i_k}$ and $\hat{p}_{i_k}$ denote weight and local pixel of $\hat{\mathcal{G}}_c$, respectively. \\
 \indent The major difference among various traditional denoisers mainly derives from the collaborative filtering scheme, which is often decided by the choice of transforms and algebraic representation. Intuitively, reshaping the 4D group $\mathcal{G}$ of (\ref{tensor_collaborative_filtering}) into the 2D matrix $\mathbf{G}$ of (\ref{matrix_collaborative_filtering}) may break the internal structure of multi-dimensional images. Therefore, a typical assumption \cite{muti2008lower, rajwade2012image, peng2014decomposable, zhang2015denoising, Zhang2015KTSVD, chang2017hyper, kong2019color, gong2020low} is that tensor representation and decomposition techniques could help preserve more structure information, based on the fact that multi-dimensional images could be naturally represented by multi-dimensional array. However, the conventional and widely-used tensor model may fall into the unbalance trap \cite{oseledets2011tensor, bengua2017efficient}, leading to unsatisfactory image restoration performance. In this paper, we further show that with slight modifications, a simple modified SVD (M-SVD) implementation could produce competitive results for color image denosing compared with tensor-based methods. More details are given in Appendices.
\subsection{DNN methods}
\input{Table_DNN_method}

The most recent development of image denoising is mainly brought by the applications of deep neural networks (DNNs), which demonstrate excellent performance in the latest denoising competition \cite{abdelhamed2019ntire} based on the smartphone image denoising dataset (SIDD) \cite{abdelhamed2018high}. Different from traditional denoisers that use only internal information of the noisy observation, DNN methods usually adopt the supervised training strategy guided by external priors and datasets to minimize the distance $\mathcal{L}$ between predicted and clean images via
\begin{equation}\label{DNN_denoising_equation}
  %\min \| \mathcal{P}_{c} - f(\mathcal{P}_{n})\|_2 + \rho\cdot\Psi(f(\mathcal{P}_{n}))
  \underset{\theta}{\min} \sum_{i} \mathcal{L} (\mathcal{F}_{\theta}(\mathcal{Y}_i), \mathcal{X}_i) + \rho\cdot\Psi(\mathcal{F}_{\theta}(\mathcal{Y}_i))
\end{equation}
where $\mathcal{X}_{i}$ and $\mathcal{Y}_{i}$ are clean/noisy image (patch) pairs, $\mathcal{F}_{\theta}$ with parameter $\theta$ is a nonlinear function that maps noisy patches onto predicted clean patches, and $\Psi(\cdot)$ represents certain regularizers \cite{lefkimmiatis2017non, ulyanov2018deep}. Early methods \cite{zhou1987novel, chiang1989multi} work with the known shift blur function and weighting factors, and Burger et al. \cite{burger2012image} show that a plain multi-layer perceptron (MLP) network is able to compete with BM3D at certain noise levels. To extract latent features and exploit the NLSS prior, a more commonly used model is the convolutional neural network (CNN) \cite{lecun1998gradient}, which is suitable for multi-dimensional data processing\cite{ji20123d, moeskops2016automatic, shi2016real, windrim2018pretraining} with flexible size of convolution filters and local receptive fields. Convolution operations are first applied to image denoising in \cite{jain2009natural}, and Fig. \ref{Fig_CNN} illustrates a simple CNN denoising framework with three convolution layers.
\vspace{-0.36cm}
\input{Fig_CNN}\\
\indent Due to the simplicity and effectiveness of CNN, it is widely adopted by different DNN denoising algorithms, and the variations of CNN-based networks are quite extensive. For example, for color images, the denoising CNN (DnCNN) \cite{zhang2017beyond} incorporates batch normalization (BN) \cite{ioffe2015batch}, rectified linear unit (ReLU) \cite{nair2010rectified} and residual learning \cite{he2016deep} into the CNN model. A generative adversarial network (GAN) blind denoiser (GCBD) \cite{chen2018image} introduces GAN\cite{radford2015unsupervised} to resolve the problem of unpaired noisy images. Recently, a graph convolution denoising network (GCDN) is presented in \cite{valsesia2020deep} to capture self-similar information.
%Imaging signal pipeline (ISP) is modeled by CycleISP \cite{zamir2020cycleisp} in forward and reverse directions with CNN and a special dual attention mechanism.
For color videos, VNLNet \cite{davy2018non} proposes a new approach to combine video self-similarities with CNN. FastDVDNet \cite{Tassano_2020_CVPR} achieves real-time video denoising without a costly motion compensation stage. For MSI, HSID-CNN \cite{yuan2018hyperspectral} and HSI-sDeCNN \cite{maffei2019single} assign both spatial and spectral information of MRI data simultaneously to CNN. A 3D Quasi-Recurrent Neural Network (QRNN3D) \cite{wei20203} utilizes 3D convolutions to extract structural spatio-spectral correlation of MSI data, and employ a quasi-recurrent pooling function to capture the global correlation along spectrum. For MRI denoising, a multi-channel DnCNN (MCDnCNN) \cite{jiang2018denoising} network extends DnCNN to handle 3D volumetric data, and the prefiltered rotational invariant patch-based CNN (PRI-PB-CNN) \cite{manjon2018mri} trains the CNN model with MRI data pre-filtered by PRI nonlocal principal component analysis (PRI-NLPCA) \cite{manjon2015mri}.\\
\indent Despite the effectiveness of DNN methods, they could be two-folded weapons, which enjoy three major advantages and also face the same challenges. First, DNN methods are able to utilize external information to guide the training process, and thus may not be restricted to the theoretical and practical bound of traditional denoisers \cite{chatterjee2009denoising}. However, DNN methods rely heavily on the quality of the training datasets, and also certain prior information such as ISO, shutter speed and camera brands, which are not always available in practice. Second, DNN methods could take advantage of the advanced GPU devices for acceleration and achieve real-time denoising \cite{zhang2016fast, Tassano_2020_CVPR} for certain tasks. But the expensive computational resources may not be accessible to ordinary users and researchers. Third, the deep, flexible and sophisticated structure of DNN methods is capable of extracting the latent features underlying the noisy images, but compared with the implementation of CBM3D that only needs to store four small pre-defined transform matrices, the complex networks with millions of parameters may drastically increase the storage cost.
\vspace{-0.2cm}
\input{Fig_timeline}
\vspace{-0.1cm}
\subsection{Datasets}
\input{Fig_dataset_illus}
\noindent In this section, we briefly introduce popular multi-dimensional image datasets used for synthetic and real-world experiments. The information is summarized in Table \ref{Table_dataset_description}, with more details given in \cite{kong2018brief, nam2016holistic, anaya2018renoir, xu2018real, abdelhamed2018high, yue2019high, Perazzi2016, yue2020supervised, CAVE_0293, arad_and_ben_shahar_2016_ECCV, chakrabarti2011statistics, cocosco1997brainweb, marcus2007open}. Some sample images of the datasets are illustrated in Fig. \ref{Fig_dataset_illus}. Usually, datasets used for synthetic experiments consist of noise-free images acquired under ideal conditions with sufficient light and careful camera settings, whereas real-world images are inevitably contaminated by noise to various degrees, decided by the environments and equipments.
\input{Table_dataset_description}

\vspace{-0.2cm}
\subsubsection{Color image and video datasets}
Some recent works \cite{wei2020physics} target raw image denosing based on the characteristics of photosensors, but the raw data are not always accessible, so we focus on the standard RGB (sRGB) color space. Compared to other types of images, the abundance of real-world color image datasets largely results from the relatively low cost of generating clean reference images from noisy observations. Briefly, a real low-light image noise reduction dataset (RENOIR) \cite{anaya2018renoir} is the first attempt to obtain low-noise reference images with low light sensitivity (ISO = 100) and long exposure time. The Darmstadt noise dataset (DND) \cite{plotz2017benchmarking} utilizes a more careful post-processing technique to produce higher quality reference images based on low-ISO images. A more common strategy adopted by a variety of datasets\cite{kong2018brief, xu2018real,abdelhamed2018high, yue2019high, yue2020supervised} is called 'image averaging', which captures the same and unchanged scene for many times and computes their mean value to obtain the corresponding noise-free image. The rationale of this simple strategy is that for each pixel, the random noise is assumed to be larger or smaller than 0, and thus can be greatly suppressed by sampling the same pixel for many times \cite{xu2018real}.\\
\indent Compared to the passion for collecting real-world color image datasets, fewer efforts have been made to produce realistic clean color videos. The image averaging strategy can not be directly applied to videos, and the difficulty lies mainly in continuously capturing noisy-clean video pairs for dynamic scenes. Besides, the strategy of generating clean dynamic videos using low ISO and high exposure time may cause motion blur. Recently, the captured raw video denoising (CRVD) dataset \cite{yue2020supervised} is introduced based on a wise frame-by-frame strategy. The authors propose to capture raw video frames by moving controllable static objects to manually create motions for them. For each motion, $M$ raw noisy images are captured and their mean image is regarded as the noise-free video frame. Finally, all the noise-free frames are grouped together according to their temporal order to generate the clean video.
\subsubsection{MSI and MRI datasets}
In this subsection, we briefly introduce the MSI and MRI datasets listed in Table \ref{Table_dataset_description}. Briefly, the emergence of MSI and MRI reflects the development of new imaging techniques, but compared to color images, collecting MSI and MRI data are of greater difficulty and also more expensive. To the best of our knowledge, there is no publicly available realistic MSI and MRI dataset with noisy/clean pairs.\\
\indent MSI cameras are mostly spectroscopic devices, which need to be carefully calibrated in order to obtain a reliable spectral information \cite{amigo2020configuration}. The representative CAVE dataset \cite{CAVE_0293} is constructed with a generalized assorted pixel (GAP) camera, it includes 32 scenes of a wide variety of real-world materials and objects. Each image has the size of $512 \times 512$ in space and includes full spectral resolution reflectance data ranging from 400 nm to 700 nm at 10 nm steps, which leads to 31 bands. The ICVL dataset \cite{arad_and_ben_shahar_2016_ECCV} is acquired using a Specim PS Kappa DX4 camera, and it includes more than 200 different images collected at $1392\times1300$ spatial resolution, and the data are downsampled to 31 spectral channels from 400nm to 700nm at 10nm increments. The real-world harvard hyperspectral dataset (HHD) dataset \cite{chakrabarti2011statistics} includes 75 images of size $1040 \times 1392 \times 31$ collected under daylight, artificial and mixed illumination. The images were captured using a commercial hyperspectral camera (Nuance FX, CRI Inc) with 31 wavelength bands ranging from 420nm to 720nm at 10nm steps. \\
\indent The MRI technique uses a strong magnetic field and radio waves to create detailed images of the organs and tissues within the body, and it could be regarded as an extension of white light imaging to incorporate better spectral resolution \cite{elson2020interventional}. The acquisition of MRI data is often expensive and time-consuming. For synthetic experiments, the Brainweb dataset contains a set of MRI data volumes produced by an MRI simulator, which are often used as 'ground-truth' MR images for different validation purpose. Full 3D data volumes have been simulated using three sequences (T1-, T2-, and proton-density- (PD-) weighted) and a variety of slice thicknesses, noise levels, and levels of intensity non-uniformity. For realistic OASIS brain image datasets, the selected 3D T1-weighted (T1w) data of size 256$\times$256$\times$128 are acquired by an MP-RAGE volumetric sequence on a Siemens 1.5T Vision scanner, with repetition time (TR) 9.7 msec, echo time (TE) 4.0 msec, flip angle 10 degrees, inversion time (TI) 10 msec, duration time of 200 ms, and voxel resolution of 1$\times$1$\times$1.25 mm$^3$.
%MRI could be regarded as an extension of white light imaging to incorporate better spectral resolution, resulting in improved ability to distinguish different chromophores in the tissue \cite{elson2020interventional}, and MSI cameras are mostly spectroscopic devices, which need to be carefully calibrated in order to obtain a reliable spectral information \cite{amigo2020configuration}. The representative CAVE dataset consist of scenes including full spectral resolution reflectance data from 400nm to 700nm at 10nm steps.\\
%\indent MRI uses a strong magnetic field and radio waves to create detailed images
%of the organs and tissues within the body\footnote{https://www.medicalnewstoday.com/articles/146309}. The acquisition of MR images is expensive and time-consuming. To provide 'ground-truth' MR images for validation purpose, the Brainweb dataset contains a set of realistic MRI data volumes produced by an MRI simulator. For real OASIS brain datasets, the selected T1-weighted images are acquired by an MP-RAGE volumetric sequence on a Siemens 1.5T Vision scanner, with repetition time (TR) 9.7 msec, echo time (TE) 4.0 msec, flip angle 10 degrees, inversion time (TI) 10 msec, and trigger delay (TD) 200 msec.
\subsubsection{The proposed color image and video datasets}
\input{Fig_Kong_dataset}
\vspace{-0.2cm}
\noindent In this subsection, we briefly introduce the motivation and details regarding the setup and protocol followed by our indoor-outdoor color image (IOCI) and video (IOCV) datasets, some examples are shown in Fig. \ref{Fig_Kong_dataset}.\\
\indent \textbf{IOCI}. Many existing datasets are mainly restricted to static indoor scenes where the lighting conditions can be manually controlled to simulate different illumination. But in a lot of real-world cases, photos are taken in outdoor environments where the objects and source of light may be more complicated. In our IOCI dataset, nine different cameras\footnote{Six cameras are used in previous experiments \cite{kong2019color}} are used to collect images of both indoor and outdoor scenes. Unlike previous works \cite{abdelhamed2018high, xu2018real, yue2019high} that use pre-defined camera settings such as ISO, shutter speed and aperture, we mostly adopt the settings of the cameras' 'auto mode', which minimizes human interference, and is also suitable for the uncontrollable outdoor environments. Captured images that display obvious misalignment and illumination differences are discarded. Each indoor and outdoor 'ground-truth' image is obtained by averaging at least 100 and 30 images of the same scene, respectively. \\
\indent \textbf{IOCV}. The carefully-designed frame-by-frame strategy of the CRVD dataset is able to produce high-quality clean reference videos, but it also requires a lot of human effort, and the manually-created motions may not be continuous. Therefore, we adopt a video-by-video strategy, and instead of manually moving controllable objects, we propose to move cameras automatically. The procedure of generating mean videos as ground-truth is illustrated in Fig. \ref{Fig_Video_slider}. Specifically, we choose light-weight cameras to avoid clear shaking and misalignment, and fix the cameras to a rotatable tripod ball-head placed on top of a professional motorized slider. The slider and the tripod ball-head could be set to move and rotate at different speeds, which simulate the movement of observed objects in more than one directions. Both the slider and cameras are controlled remotely to avoid human interference. Noisy sequences are captured repeatedly for at least 30 times to generate the corresponding 'ground-truth' reference video by their mean value.
\input{Fig_Video_slider}
\section{Experiments} \label{section_experiments}
In this section, we compare the performance of different methods for the denoising task of color images, color videos, MSI and MRI data. All implementations are provided by the authors or downloaded from the authors' personal websites. The BM3D family normally consist of two stages, where the first stage utilizes hard-thresholding to provide an initial estimate, which is then smoothed with a nonlocal empirical Wiener filter to produce a final filtered image. In our experiments, we refer to these two output stages as 'BMXD1' and 'BMXD2', respectively. For DNN methods that require GPU devices, we use Google Colab's free GPU. All other experiments are performed on a personal computer equipped with Core(TM) i5-7500 @ 3.4 GHz and 16GB RAM.
 %All results including Tables, denoised images and video sequences are available at XXX.
\subsection{Results for Real-World Color Image Datasets}  \label{color_image_experiments}
\subsubsection{Experimental Settings} \label{color_image_settings}
We compare the performance of over 25 effective methods on four real-world color image datasets, namely CC15, CC60, PolyU and our IOCI. Typically, the decisive parameters for traditional denoisers are input noise level, patch size and the number of local similar patches, etc, whereas the number of layers, learning rate and weight size are essential for DNN methods. It is noticed that many of the compared methods are designed for synthetic datasets, while ideally, in real-world cases, the parameters of all compared methods should be fine tuned to produce the best possible performance, but this could be computationally expensive and also the clean reference images for training and validation are always not available. Therefore, a more practical way is to selectively apply effective pre-trained models. Since most of the DNN implementations offer two to six pre-trained models with different parameter settings for testing, for fair comparison, we tune the input noise level for each Gaussian denoiser from four different values, which could be regarded as four pre-trained models corresponding to 'low', 'medium-low', 'medium-high' and 'high' denoising modes. We report the best average values of all compared methods on each dataset. Also, similar to \cite{kong2019color}, the best result of CBM3D on every image is included to further investigate its effectiveness, and this implementation is termed 'CBM3D\_best'. Peak signal-to-noise ratio (PSNR) and structural similarity (SSIM) \cite{wang2004image} are employed for objective evaluations. Normally, the higher the PSNR and SSIM values, the better the quality of denoised images.

\input{Table_color_all}

\subsubsection{Objective metric}
Detailed average PSNR and SSIM results are given in Table \ref{Table_color_all}. Overall, CBM3D is able to produce very competitive performance on almost every dataset, while DNN methods do not always demonstrate advantages over traditional denoisers, largely due to the lack of training data. \\
\indent More specifically, for traditional denoisers, three different transform-domain approaches namely CBM3D, the nonlocal haar (NLH) transform, and color multispectral t-SVD (CMSt-SVD) show similarly good denoising performance on all datasets. Also, the effectiveness of CMSt-SVD and CBM3D1 indicates that a one-step implementation with a small number of local patches and the hard-thresholding technique is able to produce very competitive results. It is noticed that the matrix-based methods such as multi-channel weighted nuclear norm (MCWNNM) and M-SVD present almost the same denoising capability as tensor-based methods such as hyper-Laplacian regularized unidirectional low-rank tensor (LLRT) and 4DHOSVD, which indicates that the use of tensor representation may not boost structural information retrieval in realistic cases. \\
\indent For DNN methods, the dual adversarial network (DANet) utilizes the PolyU dataset for training, on which it shows outstanding denoising results, but it struggles on other datasets when the training data are unavailable. Another important and interesting observation is that DNN models trained with AWGN noise are more competitive and robust than those trained with specific real-world datasets, as can be seen from the excellent performance of FFDNet, FCCF and DRUNet on all datasets, which strongly supports the effectiveness of the Gaussian noise modeling in Eq. (\ref{awgn}). \\
%For example, FCCF benefits from the high quality estimate of CBM3D as input for further enhancement, and on CC15 and CC60 datasets, FCCF produces the best performance and significantly improves CBM3D by over 1.3dB, however, on IPHONE 5S and SONY A6500 datasets, it shows slight degradation and on all other datasets, it only exihibits marginal improvements.\\
\indent From the perspective of denoising speed reported in Table \ref{Table_color_all}, Self2Self requires thousands of iterations to train its network with the noisy observation, resulting in the highest cost even run on advanced GPU devices. Other more efficient DNN models are able to handle color images of size $512 \times 512 \times 3$ within 0.2 seconds. For traditional denoisers, the computational cost lies mainly in the iterative local patch search and training. For example, M-SVD spends 26 and 42 seconds on grouping and learning local SVD transform, respectively, but it is slightly faster than 4DHOSVD, because it avoids folding and unfolding operations of tensor data along different modes. Among all the traditional denoisers, the state-of-the-art CBM3D is the most efficient because it does not have to train local transforms, and its grouping step is performed only on the luminance channel.

%From the perspective of denoising speed, it is noticed that M-SVD is slightly faster than 4DHOSVD, because it avoids folding and unfolding of tensor data along different modes. Taking the advantage of modern GPU architecture, deep learning models are much faster than most of the traditional Gaussian denoisers which need to iteratively perform local SVD based on a (large) number of similar patches. But less computational time is not equivalent to less complexity. For CBM3D, the computational cost lies mainly in local grouping and matrix multiplication, which are easily parallelized, and our GPU implementation of grouping could reduce 60$\%$ of the computational time. Also some intermediate results could be preserved for parameter-tuning to further reduce complexity. Furthermore, it is noticed that CBM3D only needs to store four small-size predefined local transform matrices, thus its storage cost is much lower than that of pre-trained deep learning models.
\subsubsection{Visual evaluation}
Visual evaluations of compared methods on CC15, PolyU and the proposed IOCI datasets are given in Fig. \ref{Fig_CC15}, Fig. \ref{Fig_Xu_100} and Fig. \ref{Fig_my_own}, respectively. Two commercial softwares Neat Image (NI) (https://ni.neatvideo.com/) and Topaz DeNoiseAI (https://topazlabs.com/) are included in our comparison, and their optimized denoising speed enables users to fine-tune parameters for satisfactory visual effect. From Fig. \ref{Fig_CC15}, it can be seen that except for DRUNet and DBF, all other methods produce color artifacts, especially for patch-based approaches such as CBM3D, CMSt-SVD and NLH. Comparison in Fig. \ref{Fig_Xu_100} indicates that an improvement of objective metric like PSNR does not always guarantee enhancement of visual effect. Except for NI, DeNoiseAI, and NLH, all other methods suffer over-smoothness to certain extents due to improper choice of parameters. From Fig. \ref{Fig_CC15} and Fig. \ref{Fig_Xu_100}, it seems that DeNoiseAI improve the visual quality in terms of noise removal and detail recovery because of its unique sharpening technique. Nevertheless, it also shows some color artifacts, and an interesting example of our IOCI datatset in Fig. \ref{Fig_my_own} shows that the advanced DeNoiseAI software tends to produce unwanted artifacts along the edges, which are barely noticable in mean image, noisy observation and results of other compared methods.
\input{Fig_CC15}
\input{Fig_Xu_100}
\input{Fig_my_own}
\vspace{-0.069cm}
\subsection{Results for Real-World Color Video Dataset}
\subsubsection{Experimental settings}
We evaluate the performance of VBM4D \cite{maggioni2012video}, CVMSt-SVD \cite{kong2019color}, VNLNet \cite{davy2018non} and FastDVDNet \cite{Tassano_2020_CVPR} with our IOCV dataset. For VBM4D, to save some time, we only use its first stage implementation named 'VBM4D1'. The parameters of compared methods are chosen in the same way as section \ref{color_image_settings}, and similar to \cite{Tassano_2020_CVPR}, the PSNR and SSIM of a sequence are computed as their average values of each frame.
\subsubsection{Objective metric}
Table \ref{Table_video_psnr_ssim} lists the average PSNR and SSIM results of compared methods. From Table \ref{Table_video_psnr_ssim}, it can be seen that two traditional denoisers CVMSt-SVD and VBM4D show similar performance in terms of both efficiency and effectiveness. For DNN methods, VNLNet produces the best results by integrating the NLSS prior into CNN models, but its computationally expensive patch search process also results in high computational complexity. By contrast, FastDVDNet achieves almost real-time video denoising by getting rid of the time-consuming flow estimation step. Specifically, it takes FastDVDNet less than 0.1 seconds to process a single video frame, which is
8 times faster than VNLNet, and at least 13 times faster than the benchmark traditional denoisers VBM4D and CVMSt-SVD. The denoising speed of FastDVDNet is remarkable considering its competitive denoising performance in all cases. Interestingly, comparing the results of color image denoising in Table \ref{Table_color_all} and video denoising in Table \ref{Table_video_psnr_ssim}, it is noticed that the CNN-based models present more dominating performance in the video denoising task, which may be explained by the fact that videos contain more correlated information among different frames that could be exploited by the CNN architecture.
\input{Table_video_psnr_ssim}

\subsubsection{Visual evaluation}
Visual comparison is given in Fig. \ref{Fig_color_video}. It can be clearly observed that the noisy video is severely corrupted. The patch-based traditional denoisers CVMSt-SVD and VBM4D1 leave behind medium-to-low-frequency noise, which would lead to video sequences with noticeable flickering, but VBM4D1 looks more visually pleasant than CVMSt-SVD because of its built-in de-flickering function. The DNN methods shows better results in this case. Specifically, VNLNet gives the best and most impressive visual effects by suppressing noise while recovering true colors and details. FastDVDNet is also able to effectively remove noise but also shows slight over-smooth effects.
\input{Fig_color_video}
\vspace{-0.3cm}
\subsubsection{Human ratings}
Due to the limitations of hardware equipments and environment, the mean videos generated by the motorized slider with our video averaging strategy also inevitably exhibit some noise, flickering, staircase effects, motion blur and misalignment that would undermine the accuracy of objective evaluations. Also, the quality of videos may not be judged frame by frame. Therefore, we further conduct subjective evaluations by collecting human opinions \cite{fang2019perceptual}. More specifically, we randomly select 10 videos from our IOCV dataset, and invite 10 volunteers to rate the mean, noisy and denoised sequences. The invited volunteers have very little background knowledge of image or video denoising, and they are not aware of how the presented video sequences are processed. For each of the 10 videos, the volunteers are asked to choose at least 2 best sequences, which then earn 1 point for the corresponding methods. The detailed human rating results are reported in Table \ref{Table_color_video_rating}. First, it is noticed that our mean-video strategy has the highest score on 9 of 10 videos, which justifies the effectiveness of the proposed IOCV dataset for objective evaluation. Furthermore, the human rating results are consistent with our evaluations in Table \ref{Table_video_psnr_ssim} and Fig. \ref{Fig_color_video}, which suggests that DNN methods output higher quality videos than the benchmark traditional denoisers on our IOCV dataset. Also, human eyes prefer VBM4D over CVMSt-SVD because CVMSt-SVD presents more temporally decorrelated low-frequency noise in flat areas, which will appear as particularly bothersome for the viewer.\\
\vspace{-0.25cm}
\input{Table_color_video_rating}

\indent To conclude, the CNN-based models FastDVDNet and VNLNet show clear advantages over the representative traditional denoisers. VNLNet demonstrates state-of-the-art video denoising and enhancing performance by objective and subjective measures, and FastDVDNet performs consistently well with impressively low computational time.
 %Two networks, FastDVDNet and VNLNet outperform the benchmark VBM4D in terms of computational cost and denoising capability, respectively.
\vspace{-0.02cm}
\subsection{Results for MSI Datasets}
\subsubsection{Experimental Settings} \label{MSI_settings}
In this section, we evaluate the performance of various MSI denoising methods on CAVE, ICVL and HHD datasets. For synthetic experiments, due to the high computational cost of some implementations, we mainly use the CAVE dataset for comparison. Similarly, considering the large size of the ICVL dataset, we select the first 20 MSI data and evaluate two representative methods MSt-SVD and QRNN3D. Apart from the classical spatial-based quality indices PSNR and SSIM, we adopt two widely used spectral-based quality indicators for MSIs, namely spectral angle mapper (SAM) \cite{yuhas1990determination} and dimensionless global relative error of synthesis (ERGAS) \cite{wald2002data}. Different from PSNR and SSIM, recovered MSI data with lower SAM and ERGAS are considered of better quality.

\subsubsection{Synthetic MSI dataset}
\input{Table_msi_psnr_ssim}

We assume that entries in all slices of MSI data are corrupted by zero-mean i.i.d Gaussian noise with different noise levels. Detailed denoising results on CAVE and ICVL dataset are listed in Table \ref{Table_msi_psnr_ssim} and \ref{Table_ICVL}, respectively.
\input{Table_ICVL}
Benefiting from the advanced GPU devices, QRNN3D is 60 times faster than the benchmark traditional denoisers MSt-SVD and BM4D. However, like many DNN methods, QRNN3D is also subject to the influence of training datasets. Its models trained with the ICVL dataset can not handle noisy images from the CAVE dataset, as can be seen from the difference of Table \ref{Table_msi_psnr_ssim} and Table \ref{Table_ICVL}. Similar to QRNN3D, HSI-SDeCNN also fails to produce competitive performance, because its network models are trained only with certain noise levels and bandwidths. For traditional denoisers, results in Table \ref{Table_msi_psnr_ssim} shows the advantage of tensor-based methods by exploiting spatial and spectral correlation. Specifically, NGMeet, LLRT and LTDL consistently outperform other compared approaches at all noise levels by more than 1db. However, it takes LLRT and LTDL more than 40 minutes to process a single image, because their iterative strategy with a large number of local similar patches significantly increases the computational burden. By comparison, MSt-SVD is more efficient, since it is a one-step approach that utilizes global patch representation.
\input{Fig_CAVE}\\
\indent Fig. \ref{Fig_CAVE} compares competitive tensor-based denoisers at high noise level $\sigma = 100$ on the CAVE dataset. In this extreme case, the pre-defined transform bases of BM4D may not take advantage of the correlation among all the spectral bands, while all other compared methods can effectively remove noise and retain details to certain extents. Interestingly, although from Table \ref{Table_msi_psnr_ssim}, MSt-SVD is outperformed by NGMeet, LLRT and LTDL, it shows the best detail recovery ability, as can be seen from the white text in Fig. \ref{Fig_CAVE}. This observation indicates increasing the number of iterations and local similar patches may not help preserve fine details and structure of MSI data at high noise levels.
\subsubsection{Real-world HHD Dataset}
\noindent Inevitably, images of HHD dataset are contaminated by noise to varying degrees under different illumination. Considering the large size of noisy images, we examine the effectiveness of five efficient benchmarks on handling real-world MSI data. Since there is no clean reference image, we conduct visual evaluations, and for fair comparison, the parameters of compared methods are carefully tuned. The results are visualized in Fig. \ref{Fig_HHD}. Without realistic training data, the CNN model of QRNN3D produces obviously blurry image. Interestingly, different from our observations of synthetic AWGN comparison in Fig. \ref{Fig_CAVE}, no compared methods could consistently outperform BM4D in this realistic case. As can be seen in Fig. \ref{Fig_HHD}, all state-of-the-art algorithms struggle to adapt to different local image contents to balance smoothness and sharp details. For patch-based methods, applying a same set of parameters or transforms to multiple spectral bands may result in similar denoise patterns across the whole image. Therefore, apart from the high computational burden, another challenge of filtering MSI data lies in dealing with the spatial and spectral variation of noise levels and distributions.
\input{Fig_HHD}
\vspace{-0.1cm}
\subsection{Results for MRI datasets}  \label{MR_image_experiments}
\subsubsection{Experimental Settings} \label{MRI_settings}
Different from the AWGN noise modeling of MSI, MRI data are often corrupted by Rician noise \cite{awate2007feature}. Specifically, Let $\mathcal{X}$ be the original noise-free signal, the noise Rician MRI data $\mathcal{Y}$ can be obtained via
\begin{equation}\label{rician_noise}
  \mathcal{Y} = \sqrt{(\mathcal{X} + \sigma n_1)^2 + ( \sigma n_2)^2}
\end{equation}
where $n_1$,$n_2$ $\sim$ $N(0,1)$ are i.i.d. random vectors following the standard normal distribution, $\sigma$ is the standard deviation in both real and imaginary channels of the noisy 3D MR images. To handle Rician noise, the technique of forward and inverse variance stabilizing transform (VST) \cite{foi2011noise} is often adopted by Gaussian denoisers via
\begin{equation}\label{equ_vst_gaussian}
    \hat{\mathcal{X}} = \text{VST}^{-1}(denoise(\text{VST}(\mathcal{Y}, \sigma), \sigma_{\text{VST}}), \sigma)
\end{equation}
where $\text{VST}^{-1}$ denotes the inverse of VST, $\sigma_{\text{VST}}$ is the stabilized standard deviation after VST transform, and $\sigma$ is the standard deviation of the noise in Eq. (\ref{rician_noise}). According to Eq. (\ref{equ_vst_gaussian}), the noisy Rician data $\mathcal{Y}$ is first stabilized by the VST and then filtered by certain Gaussian denoisers using a constant noise level $\sigma_{\text{VST}}$, and then the final estimate is obtained by applying the inverse VST to the output of the denoising result \cite{maggioni2012nonlocal}. \\
 \indent In our experiments, we use PSNR and SSIM as the objective metrics, and similar to \cite{maggioni2012nonlocal}, \cite{manjon2012new}, the PSNR value is only computed on the foreground defined by $\mathcal{X}_f = \{x \in \mathcal{X}: x > 10 \cdot D/255\}$, where $D$ is the peak of clean data $\mathcal{X}$.
\subsubsection{Synthetic Brainweb dataset}
The volume data are added with varying levels of stationary Rician noise from 1$\%$ to 19$\%$ of the maximum intensity with an increase of 2 $\%$. In real world cases, the noise level is usually much lower than 19$\%$, but we are also interested in the denoising capability of compared methods under extreme conditions. Table \ref{Table_MRI} lists detailed quantitative results, and Fig. \ref{Fig_illus_PSNR_SSIM_MRI} compares the denoising performance at high noise levels when $\sigma \geq 11\%$. At lower noise levels, PRI-NLPCA is able to take advantage of the high-quality initial estimate of NLPCA, and achieves the best performance when $\sigma \leq 9\%$. As noise level increases, tensor-based methods demonstrate advantages of extracting latent features, and the iterative low-rank HOSVD (ILR-HOSVD) produces the best results when $\%11 \leq \sigma \leq 15\%$. However, the iterative learning strategy is subject to the presence of extremely high noise, while BM4D benefits from its pre-defined transforms and outperforms other methods when $\sigma \geq 17\%$. \\
\input{Table_MRI}

\input{Fig_illus_PSNR_SSIM_MRI}
%For computational speed, PRINLM is at least 6 times faster than state-of-the-art BM4D methods.\\
\indent Visual evaluation on T1w data when $\sigma = 19\%$ is illustrated in Fig. \ref{Fig_Brainweb}, and it shows that BM4D2 is successful in both noise suppression and detail preservation. Compared to NLPCA and ILR-HOSVD, MSt-SVD is less affected by severe noise, as it may benefit from global patch representation with fast Fourier transform (FFT). The state-of-the-art NLM methods, namely PRINLM and PRI-NLPCA exhibit pleasant visual effects in homogeneous areas at the cost of slight over-smoothness along the edges. Furthermore, PRI-NLPCA shows some artifacts on the background due to the influence of noise on learning local PCA transform.
\input{Fig_Brainweb}
\subsubsection{Real-world MRI dataset}
To evaluate the performance of compared methods on real-world 3D MR data, we carry out experiments on T1w MR images of the OASIS dataset \cite{marcus2007open}. The Rician noise levels of two selected T1w data, namely OAS1\_0112 and OAS1\_0092 are estimated to be 3$\%$ and 4.5$\%$ of the maximum intensity, respectively \cite{zhang2015denoising}. The filtered images of different methods are compared in Fig. \ref{Fig_real_MRI_0112} and Fig. \ref{Fig_real_MRI}, respectively. Based on our synthetic experiments in Table \ref{Table_MRI}, we learn that in most cases all methods could achieve competitive denoising performance at low noise levels, as can also be seen from the excellent visual effects on realistic OAS1\_0112 data in Fig. \ref{Fig_real_MRI_0112}. Thus, from the perspective of real-world MRI denoising, BM4D1 and PRINLM should be preferred due to their low computational cost reported in Table \ref{Table_MRI}. Despite the similar denoising results of compared methods, we also observe that on certain slices of the 3D MR data, NLPCA produces the best results in terms of fine detail preservation, while other methods exhibit over-smoothness to varying degrees, as is illustrated in Fig. \ref{Fig_real_MRI}. This interesting observation is another vivid example to show that the use of tensor representation does not always help preserve more structural information of the underlying clean images.
\input{Fig_real_MRI_0112}
\vspace{-0.1cm}
\input{Fig_real_MRI}
\vspace{-0.18cm}
\subsection{Discussion}
Our discussion in this section will mainly focus on color image denoising due to the abundance of advanced compared methods and real-world datasets.

\subsubsection{The robustness of CBM3D}
\noindent Recently, there is some criticism \cite{chen2019real} of CBM3D that to achieve competitive denoising performance, it may need to traverse different Gaussian noise levels $\sigma$, which could be potentially time-consuming and impractical without ground-truth. Such criticism may be unfair for CBM3D for two reasons. First, compared to nearly all methods listed in Table \ref{Table_color_all}, CBM3D does not involve any training process which could be influenced by the noise distributions of specific datasets. Second, in our experiments, we discover that the pre-defined transforms of CBM3D are not very sensitive to the choice of the input noise level parameter $\sigma$ within a reasonable range. Fig. \ref{Fig_CBM3D_noise_level_influence} compares the PSNR and SSIM values of CBM3D1 with $\sigma \in [10, 30]$ on five datasets. It is noticed that choosing a noise level $\sigma$ between 10 and 20 could yield similarly good performance for CBM3D, making it an exceptionally robust method. Furthermore, from the perspective of blind denoising \cite{rabie2005robust, lebrun2015multiscale, majumdar2018blind, guo2019toward}, CBM3D could be viewed as a competitive blind denoiser with input noise level fixed to be $\sigma = 15$.
\input{Fig_CBM3D_noise_level_influence}
\vspace{-0.3cm}
\subsubsection{Understanding the denoising performance}
The theoretical bound of compared image denoising methods is hard to obtain, but it is interesting to investigate the denoising capability of compared methods under the challenging practical cases when prior information, such as training datasets and camera settings are unavailable. We use our IOCI's FUJIFILM X100T camera dataset, and for each scene, we generate three new mean images by averaging 3, 5 and 10 noisy images, and they are named 'Mean\_3', 'Mean\_5' and 'Mean\_10', respectively. They could be regarded as images captured via different continuous shooting modes with 'high', 'medium' and 'low' noise levels. Fig. \ref{Fig_PSNR_diff_Fujifilm_test} illustrates the average PSNR differences of six implementations compared with 'Mean\_3' on FUJI dataset. Interestingly, Fig. \ref{Fig_PSNR_diff_Fujifilm_test} shows that state-of-the-art denoising methods only produce marginal improvements compared with Mean\_3, which indicates that their denoising performance is similar to obtaining the mean image by averaging 3 consecutive noisy observations.
\input{Fig_PSNR_diff_Fujifilm_test}

\subsubsection{Denoising with resizing}
To boost the denoising performance of traditional patch-based denoisers, many methods \cite{dabov2007color, rajwade2012image, manjon2015mri, zhang2015denoising, xu2017multi, xu2018trilateral, chang2017hyper, wu2018weighted} propose to iteratively combine the filtered image with the corresponding noisy observation in different ways \cite{chen2006new, luo2012generalized, romano2017little, tirer2018image}. A simple approach is to add back the noisy observation $\mathcal{Y}$ to the denoised image $\hat{\mathcal{X}}$ via
\begin{equation}\label{simple_enhancement_approach}
  \hat{\mathcal{Y}} = \lambda \mathcal{Y} + (1-\lambda)\hat{\mathcal{X}}
\end{equation}
where $\lambda \in [0,1]$ is a relaxation parameter reflecting the importance of $\mathcal{Y}$. The new noisy image $\hat{\mathcal{Y}}$ is then fed into the denoiser for final estimate. The iterative denoising strategies may significantly increase the computational burden, and the results reported in Table \ref{Table_msi_psnr_ssim} and Table \ref{Table_MRI} show that a simple one-stage implementation such as MSt-SVD and BMXD1 could produce competitive performance at moderate noise levels. It is therefore interesting to investigate how to efficiently handle severe noise with traditional denoisers. Recently, Zontak et al. \cite{zontak2013separating} show that in the down-sampled noisy image, patches tend to be noiseless and share similar patterns with underlying clean patches. Therefore, instead of directly filtering the large-size noisy observation, an alternative is to first handle downsized image, and then upscale the denoised image back to its original size with some effective image super-resolution algorithms \cite{dong2015image, wang2020deep}. This idea is similar to the encoder-decoder scheme of DNN methods. For simplicity we use MATLAB's built-in 'imresize' function and applies it to the efficient CMSt-SVD. This special implementation is termed as 'CMSt-SVD\_R'. The visual effects of CMSt-SVD with and without the image resizing strategy are compared in Fig. \ref{Fig_discussion_visual_enhancement}. As can be seen, the resizing strategy could effectively reduce color artifacts at the cost of slight over-smooth effects, thus it may be more effective and suitable for very noisy images with fewer fine details and textures.
\input{Fig_discussion_visual_enhancement}
\vspace{-0.56cm}
\section{Conclusion}  \label{section_conclusion}
Recently, multi-dimensional image denoising has become an attractive research topic with potentially important applications. The great success of the BM3D-based methods contributes significantly to the emergence of numerous related denoising methods, varying from traditional Gaussian denoisers to advanced DNN methods. In this paper, new benchmark datasets for color image and video are introduced, and comprehensive comparison of representative denoising methods are conducted with both synthetic and real-world experimental settings. Throughout extensive objective and subjective evaluations, we show: (i) the overall powerful denoising ability of the BM3D family, (ii) the use of tensor representation does not guarantee better results compared to matrix implementations, and (iii) the outstanding generalizabiity and denoising performance of some DNN methods such as FCCF, DRUNet, FastDVDNet and VNLNet to real-world color image/video datasets.
\\
\indent Despite the achievements in recent years, the state-of-the-art approaches still suffer from limitations that restrict their application to key real-world scenarios. For example, from Fig. \ref{Fig_Xu_100} and Fig. \ref{Fig_my_own}, it is noticed that many image denoising methods show only limited enhancement and may even degrade the image quality with over-smoothness and annoying aftifacts. As image size grows rapidly with the development of new imaging technologies, a challenging problem is to decide if it is worth applying filtering to an image at hand \cite{abramov2020prediction}, and if denoising is desired, which of the filters listed in Table \ref{Table_traditional_method} and Table \ref{Table_DNN_method} should be used. Recently, a new trend focuses on how to jointly handle image denoising and other computer vision tasks such as dehazing \cite{wu2019learning}, demoisacking \cite{liu2020joint} and classification \cite{liu2020connecting}. It is interesting to collect realistic bencthmark datasets to further explore their mutual influence.

\vspace{-0.08cm}
\begin{appendices}
\section{The modified SVD method}\label{Appendix_m_svd}
In this section, we introduce the motivation and detials of the modified SVD (M-SVD) method briefed in Algorithm \ref{m-svd}. For simplicity, our analysis is based on the color image denoisng task. Given the local group $\mathcal{G} \in \mathbb{R}^{ps \times ps \times 3 \times K}$ of $K$ patches, the original SVD algorithm operates directly on the matrix representation $\mathbf{G}_{(4)}$ of $\mathcal{G}$ via
\begin{equation}\label{traditional_SVD}
  \mathbf{G}_{(4)} = \mathbf{U}\mathbf{S} \mathbf{V}^T
\end{equation}
where $\mathbf{S}$ is singular value matrix, $\mathbf{U}$ and $\mathbf{V}$ are group and patch level transform matrices, respectively. A more effective way to obtain $\mathbf{U}$ is to perform SVD on the opponent color space \cite{dabov2007color, kong2019color} represented by $\mathbf{G}_{opp_{(3)}}$, with $\mathcal{G}_{opp} = \mathcal{G}(:,:,1,:) + \mathcal{G}(:,:,2,:) + \mathcal{G}(:,:,3,:)$. Then the core matrix $\mathbf{C}$ of $\mathbf{G}_{(4)}$ can be calculated by
%where $\mathbf{S}$ is singular value matrix, $\mathbf{U}$ and $\mathbf{V}$ are group and patch level transform matrices, respectively. Obtaining group-level transform $\mathbf{U}$ in (\ref{traditional_SVD}) may be subject to the influence of noise. Inspired by \cite{dabov2007color, kong2019color}, an alternative is to train $\mathbf{U}$ based on the new color space component $\mathbf{G}_{new_{(3)}} \in \mathbb{R}^{K \times ps^2}$ via SVD, with $\mathcal{G}_{new} = \mathcal{G}(:,:,1,:) + \mathcal{G}(:,:,2,:) + \mathcal{G}(:,:,3,:)$. The core matrix $\mathbf{C}$ of $\mathbf{G}_{(4)}$ can be calculated by
\begin{equation}\label{core_matrix}
  \mathbf{C} = \mathbf{U}^T \mathbf{G}_{(4)} \mathbf{V}
\end{equation}
After the truncation of $\mathbf{C}$ and the inverse transform of (\ref{core_matrix}), the estimated clean group $\hat{\mathbf{G}}_c = \mathbf{U} \mathbf{C}_{trun} \mathbf{V}^T$ is obtained.
\input{Algorithm_M_SVD}

\vspace{-0.26cm}
\section{A closer look at the effectiveness of HOSVD for denoising}\label{Appendix_proof_relationship}
Our real-world experiments of color image and MRI denoising in section \ref{color_image_experiments} and \ref{MR_image_experiments} show that the SVD-based methods could produce very competitive performance compared with HOSVD-based approaches, which are also prone to over-smooth textures. With such observation, we further investigate and discuss the effectiveness of HOSVD for denoising. Our analysis in this section is developed mainly based on the results of \cite{kamm1998kronecker, de2000multilinear, de2000best, kolda2009tensor, cichocki2015tensor, bengua2017efficient}, with more details available in the supplementary file.
 %Briefly, we first discuss the relationship between HOSVD and SVD, and then show that the use of HOSVD may not guarantee better performance for image denoising.

\subsection{The relationship between HOSVD and SVD}
The relationship between HOSVD and SVD is established mainly based on the following Theorems.
\newtheorem{myThm}{Theorem}[]

%\begin{myThm} Given an $N$-th order tensor $\mathcal{G} \in \mathbb{R}^{I_1 \times I_2 \times \cdots \times I_N}$, if $\mathbf{U}_i$, $\mathbf{V}_i$ and $\mathbf{S}_i$ are the singular vectors and singular value matrix of $\mathbf{G}_{(i)}$ obtained by $\mathbf{G}_{(i)} = \mathbf{U}_i \mathbf{S}_i \mathbf{V}_i^T$, then $\mathbf{U}_1 \otimes \cdots \otimes \mathbf{U}_N$, $\mathbf{V}_1 \otimes \cdots \otimes \mathbf{V}_N$ and $\mathbf{S}_1 \otimes \cdots \otimes \mathbf{S}_N$ are singular vectors and singular value matrices of $\mathbf{G}_{(1)} \otimes \cdots \otimes \mathbf{G}_{(N)}$. More specifically
%\begin{equation}\label{HOSVD_SVD_kronecter_product}
%\begin{split}
%  &\mathbf{G}_{(1)} \otimes \cdots \otimes \mathbf{G}_{(N)} \\
%   & = (\mathbf{U}_{1} \otimes \cdots \otimes \mathbf{U}_{N}) (\mathbf{S}_1 \otimes \cdots \otimes \mathbf{S}_N) (\mathbf{V}_1 \otimes \cdots \otimes \mathbf{V}_N)^T \\
%   & = (\mathbf{U}_1 \mathbf{S}_1 \mathbf{V}_1^T) \otimes \cdots \otimes (\mathbf{U}_N \mathbf{S}_N \mathbf{V}_N^T)
%\end{split}
%\end{equation}
%
%\end{myThm}

\begin{myThm}
Given an $N$-th order tensor $\mathcal{G} \in \mathbb{R}^{I_1 \times I_2 \times \cdots \times I_N}$, its full HOSVD is
\begin{equation}\label{n_th_order_HOSVD}
  \mathcal{G} = \mathcal{C} \times _1\mathbf{U}_1 \times _2\mathbf{U}_2 \times \cdots \times _N\mathbf{U}_N
\end{equation}
where $\mathcal{C} \in \mathbb{R}^{I_1 \times I_2 \times \cdots \times I_N} $ represents core tensor. If $\mathbf{S}_i$ is the singular value matrix of $\mathbf{G}_{(i)}$, then
\begin{equation}\label{equal_C_S_general}
  \mathbf{C}_{(i)}\mathbf{C}_{(i)}^T = \mathbf{S}_i\mathbf{S}_i^T
\end{equation}
\end{myThm}

%\begin{myThm}
%Given an $N$-th order tensor $\mathcal{G} \in \mathbb{R}^{I_1 \times I_2 \times \cdots \times I_N}$ and its core tensor $\mathcal{C}$ of the same size, with $\mathbf{G}_{(i)} =  \mathbf{U}_i \mathbf{S}_i \mathbf{V}_i^T$ via SVD, then the $j$-th column of $\mathbf{V}_i$ can be represented as
%\begin{equation}\label{eq_relation_V_U_N_order_tensor}
%  \mathbf{V}_i(:,j) = \frac{\sum_{k = 1}^{I_1\cdots I_{i-1}I_{i+1}\cdots I_N} \mathbf{C}_{(i)} (j,k) \bigcup \underset{t \neq i}{\otimes}
% \mathbf{U}_t (:,k) }{\|\mathbf{C_{(i)}}(j,:)\|}
%\end{equation}
%\end{myThm}
\begin{myThm}
Given an $N$-th order tensor $\mathcal{G} \in \mathbb{R}^{I_1 \times I_2 \times \cdots \times I_N}$ and its core tensor $\mathcal{C}$ of the same size, with $\mathbf{G}_{(i)} =  \mathbf{U}_i \mathbf{S}_i \mathbf{V}_i^T$ via SVD, then the $j$-th column of $\mathbf{V}_i$ can be represented as
\begin{equation}\label{eq_relation_V_U_N_order_tensor}
 % \mathbf{V}_i(:,j) = \frac{(\mathbf{U}_N \otimes \cdots \otimes \mathbf{U}_{i+1} \otimes \mathbf{U}_{i-1} \otimes \cdots \otimes \mathbf{U}_1) \mathbf{C}_{(i)}(j,:)^T}{\|\mathbf{C_{(i)}}(j,:)\|}
 \mathbf{V}_i(:,j) = \frac{\hat{\mathbf{U}}_i\mathbf{C}_{(i)}(j,:)^T}{\|\mathbf{C}_{(i)}(j,:)\|}
\end{equation}
\end{myThm}
\noindent where $\hat{\mathbf{U}}_i = \mathbf{U}_N \otimes \cdots \otimes \mathbf{U}_{i+1} \otimes \mathbf{U}_{i-1} \otimes \cdots \otimes \mathbf{U}_1$.\\
\indent For simplicity, consider the case of third-order tensor group $\mathcal{G} \in \mathbb{R}^{I_1 \times I_2 \times K}$ with $K$ patches of size $I_1 \times I_2$, its full HOSVD is given by
 \begin{equation}\label{third_order_HOSVD}
   \mathcal{G} = \mathcal{C} \times _1 \mathbf{U}_1 \times _2\mathbf{U}_2 \times _3\mathbf{U}_3
 \end{equation}
The full SVD of the matrix representation $\mathbf{G}_{(3)}$ is
\begin{equation}\label{SVD_for_HOSVD}
  \mathbf{G}_{(3)} = \mathbf{U}_3 \mathbf{S}_3 \mathbf{V}_3^T
\end{equation}
Theorem 1 indicates that truncating the columns of the group-level transform $\mathbf{U}_3$ is equivalent to shrinking the rows of $\mathbf{C}_{(3)}$ and also the singular value matrix $\mathbf{S}_3$ of $\mathbf{G}_{(3)}$. Theorem 2 demonstrates how the patch-level transform $\mathbf{V}_3$ of $\mathbf{G}_{(3)}$ can be derived from the first and second mode projection matrices $\mathbf{U}_1$ and $\mathbf{U}_2$ of $\mathcal{G}$.
\subsection{The risk of truncated HOSVD for denoising}
%\indent For simplicity, we use the case of third order tensor. Based on the above theorems, it is noticed that truncating the columns of the group-level transform $\mathbf{U}_3$ is equivalent to truncating the rows of $\mathbf{C}_{(3)}$ and the singular values of $\mathbf{S}$, which assumes that a small subset of patches could represent the whole group $\mathbf{G}$. Similarly, the column-wise truncation of the patch-level transforms $\mathbf{U}_1$ and $\mathbf{U}_2$ will also result in the column-wise truncation of $\mathbf{C}_{(3)}$, which performs the patch-based feature selection. Furthermore, the hard-thresholding technique is equivalent to the element-wise shrinkage of $\mathbf{C}_{(3)}$. More specifically, consider a noise-free group $\mathcal{G}\in \mathbb{R}^{3\times 3 \times 3}$ consisting of three same patches of size $3\times 3$, with $\mathcal{G}(:,:,1) = \mathcal{G}(:,:,2) = \mathcal{G}(:,:,3)$, and the corresponding third mode unfolding of $\mathcal{G}$ is
HOSVD-based methods remove noise by truncating the columns of projection matrices via low-rank approximation, or by shrinking the core tensor via the hard-thresholding technique. To demonstrate the over-smooth effects of truncated HOSVD (T-HOSVD), we use a noise-free group $\mathcal{G} \in \mathbb{R}^{3 \times 3 \times 3}$ consisting of three same patches with $\mathcal{G}(:,:,1) = \mathcal{G}(:,:,2) = \mathcal{G}(:,:,3)$, and the third-mode unfolding of $\mathcal{G}$ is
\begin{equation}\label{original_unfolding}
\mathbf{G}_{(3)} =
\begin{pmatrix}
3 & 1 & 5 & 6 & 4 & 8 & 9 & 2 & 6\\
3 & 1 & 5 & 6 & 4 & 8 & 9 & 2 & 6\\
3 & 1 & 5 & 6 & 4 & 8 & 9 & 2 & 6
\end{pmatrix}
\end{equation}
where every row of $\mathbf{G}_{(3)}$ corresponds to a vectorized image patch. According to Eq. (\ref{third_order_HOSVD}), the core tensor $\mathcal{C}$ is
\begin{equation}\label{core_tensor_third_order}
\mathbf{C}_{(3)} =
\begingroup % keep the change local
\setlength\arraycolsep{2.19pt}
\begin{pmatrix}
-27.98 & 0 & 0 & 0 & -5.38 & 0 & 0 & 0 & 2.07\\
0 & 0 & 0 & 0 & 0 & 0 & 0 & 0 & 0\\
0 & 0 & 0 & 0 & 0 & 0 & 0 & 0 & 0
\end{pmatrix}
\endgroup
\end{equation}
Assume the pre-defined multi-rank $r_1 = r_2 = 2$ is applied to truncate $\mathbf{U}_1$ and $\mathbf{U}_2$ with $\mathbf{U}$ = $\mathbf{U}(:,1:2)$, then after the inverse transform of Eq. (\ref{third_order_HOSVD}), the corresponding denoised group $\hat{\mathcal{G}}_c$ is
\begin{equation}\label{denoised_unfolding}
%\hat{\mathbf{G}}_{{(3)}} =
\begingroup % keep the change local
\setlength\arraycolsep{1.69pt}
\begin{pmatrix}
3.09 & 1.92 & 4.54 & 5.95 & 3.46 & 8.26 & 9.00 & 2.03 & 5.98\\
3.09 & 1.92 & 4.54 & 5.95 & 3.46 & 8.26 & 9.00 & 2.03 & 5.98\\
3.09 & 1.92 & 4.54 & 5.95 & 3.46 & 8.26 & 9.00 & 2.03 & 5.98
\end{pmatrix}
\endgroup
\end{equation}
\noindent Comparing Eq. (\ref{original_unfolding}) and Eq. (\ref{denoised_unfolding}), the loss of information is $ \|\mathbf{G}_{(3)} - \hat{\mathbf{G}}_{c_{(3)}}\|_F^2 = 4.29$.
%\begin{equation}
%    loss = \|\mathbf{G}_{(3)} - \hat{\mathbf{G}}_{{(3)}}\|_F^2 = 4.29
%\end{equation}
Also, from (\ref{core_tensor_third_order}) it is noticed that the hard-thresholding technique in Eq. (\ref{hard_thresholding}) will yield the same loss if the threshold parameter $\tau$ is set to $ 2.07 <\tau < 5.38$. Fig. \ref{Fig_oversmooth} illustrates the over-smooth effects of T-HOSVD when applied to a noise-free image with all the patches in a group forced to be the same.
%if the hard-threshold parameter $\tau$ is set to $ 2.07 <\tau < 5.38$, then it will yield the same loss.
%From the perspective of image denoising, the loss of information will result in over-smooth effects as illustrated in Fig. \ref{Fig_oversmooth}, when truncated HOSVD (T-HOSVD) is applied to a noise-free image with all the patches in a group are forced to be the same.
\input{Fig_oversmooth}\\
\indent Our analysis in this section is based on an extreme and ideal case that all patches in a noise-free group are the same. In real-world applications, the difference among nonlocal similar patches is not negligible, and the use of tensor decomposition techniques may help filter out more redundant information and extract latent features, as is observed in the MRI denoising experiment illustrated in Fig. \ref{Fig_Brainweb}. Also, it is noticed from Fig. \ref{Fig_oversmooth} that certain strutural shapes and textures are preserved. However, despite recent development of the tensor theory \cite{zhang2018tensor}, the choice of the best multi-rank and threshold parameter $\tau$ for the tensor truncation strategy still remains a challenge.
%the major challenge of the tensor truncation strategy for denoising lies in the choice of proper multi-rank and threshold parameter $\tau$.
\end{appendices}

\section*{Acknowledgment}

The authors would like to thank those researchers who make their code, software packages and datasets publicly available. We also appreciate the genorosity of our friends and volunteers for sharing their camera devices and participating in the video quality assessment task.

\ifCLASSOPTIONcaptionsoff
  \newpage
\fi

\bibliographystyle{IEEEtran}
\bibliography{IEEEabrv,mybib_abrv_new3}

%\end{IEEEbiography}
%% insert where needed to balance the two columns on the last page with
%%\newpage

\end{document}

%% file: Fig_NSS.tex
\begin{figure}[htbp]
\graphicspath{{Figs/Fig_frameworks/}}
\centering
\subfigure[Color image]{
\label{Fig4}
\includegraphics[width=0.81in]{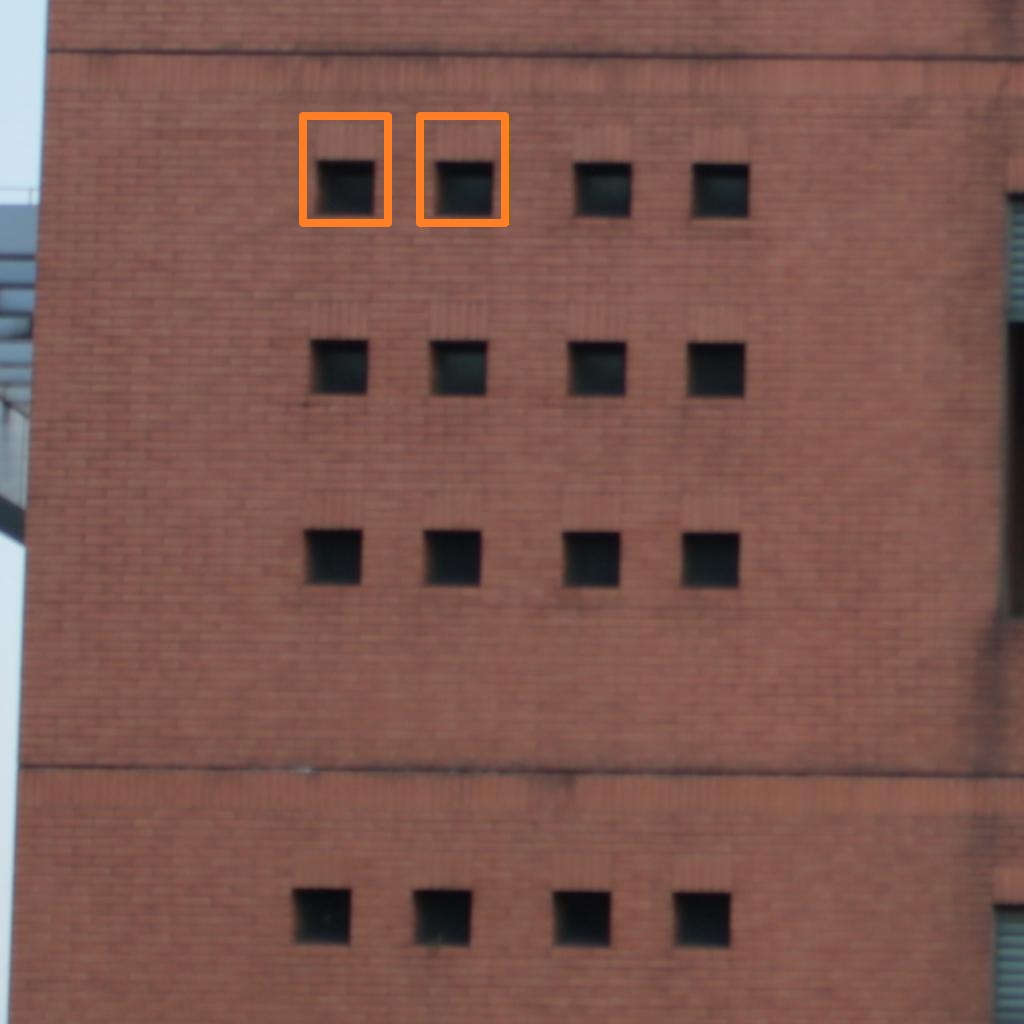}}
\subfigure[Color video]{
\label{Fig4}
\includegraphics[width=0.81in]{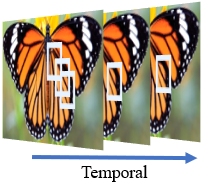}}
\subfigure[MRI]{
\label{Fig4}
\includegraphics[width=0.81in]{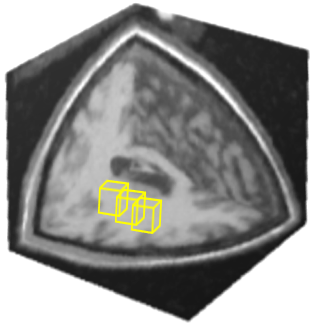}}
\subfigure[MSI]{
\label{Fig4}
\includegraphics[width=0.81in]{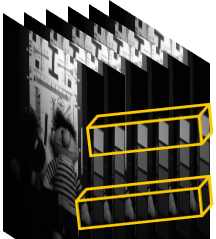}}

\caption{The nonlocal self-similarity (NLSS) prior and patch representation of different multi-dimensional images.}
\label{Fig_NSS}
\end{figure}

%% file: Table_traditional_method.tex
\begin{table*}[htbp]
\scriptsize
\centering
\caption{Related traditional denoisers for multi-dimensional image data with different noise modeling techniques and applications. 'CI': color image, 'CV': color video, 'MSI': multispectral imaging, 'HSI': hyperspectral imaging, 'MRI': magnetic resonance imaging.}
\scalebox{0.849}{
\begin{tabular}{|c|c|c|c|c|c|}
\hline
\textbf{Category}                      & \textbf{Representation}           & \textbf{Methods}       & \textbf{Noise Modeling}  & \textbf{Applications}      & \textbf{Key Words}                                                     \\
\hline
\multirow{41}{*}{\shortstack[l]{Traditional \\ denoisers}} & \multirow{26}{*}{Matrix} & AWT \cite{donoho1995noising, huang2005color,othman2006noise}           & AWGN                        & CI, MSI, HSI      & Adaptive wavelet thresholding methods    \\
\cline{3-6}
                              &                          & DCT \cite{yaroslavsky1996local, foi2007pointwise}           & AWGN                        & CI                & Image denoising with discrete cosine transform (DCT)   \\
\cline{3-6}
                              &                          & NLM \cite{buades2005review, dai2013multichannel}           & AWGN                        & CI                & nonlocal means (NLM) algorithm with Gaussian kernels                           \\
\cline{3-6}
                              &                          & K-SVD \cite{elad2006image, mairal2007sparse, fu2015adaptive}           & AWGN                        & CI, MSI               & Over-complete dictionary learning and sparse coding                \\
\cline{3-6}
                              &                          & TCVD \cite{liu2010high}         & AWGN                        & CV                & Combination of robust optical flow with NLM                   \\
\cline{3-6}
                              &                          & NLPCA \cite{zhang2010two, dong2012nonlocal, phophalia20173d}        & AWGN                        & CI, MRI           & Applications of SVD transform                             \\
\cline{3-6}
                              &                          & EPLL \cite{zoran2011learning, hurault2018epll}           & AWGN                        & CI                & A Bayesian method for whole image restoration                          \\
\cline{3-6}
                              &                          & LRMR \cite{zhang2013hyperspectral}         & AWGN                        & MRI               & A low-rank matrix decomposition model                         \\
\cline{3-6}
                              &                          & GID \cite{xu2018external}          & AWGN                        & CI                & External data guided and internal prior learning              \\
\cline{3-6}
                              &                          & SPTWO \cite{buades2016patch}         & AWGN                        & CV                & NLPCA with optical flow estimation                            \\
\cline{3-6}
                              &                          & LSCD \cite{rizkinia2016local}         & AWGN                        & CI, MSI, HSI      & Spectral component decomposition with line feature        \\
%\cline{3-6}
%                              &                          & Global-Search \cite{ehret2017global} & AWGN                        & CV                & Efficient approximate global patch search                     \\
\cline{3-6}
                              &                          & MCWNNM \cite{gu2014weighted, xu2017multi}        & AWGN                        & CI                & Extension of WNNM to color images                             \\
\cline{3-6}
                              &                          & TWSC \cite{xu2018trilateral}         & AWGN                        & CI              & A trilateral weighted sparse coding scheme                   \\
\cline{3-6}
                              &                          & VNLB \cite{arias2018video}         & AWGN                        & CV                & A patch-based Bayesian model for video denoising                   \\
\cline{3-6}
                              &                          & NLH \cite{hou2020nlh}      & AWGN                        & CI                & Nonlocal pixel similarity with Haar transform                 \\
\cline{3-6}
                              &                          & AMF \cite{chan2005salt}          & Impulsive                   & CI                & Impulsive noise removal by adaptive median filter             \\
%\cline{3-6}
%                              &                          & IAFF \cite{ahmed2013removal}         & Impulsive                   & CI                & Impulsive noise removal by iterative adaptive fuzzy filter    \\
\cline{3-6}
                              &                          & NoiseClinic \cite{lebrun2015multiscale}         & AWGN and Realistic                   & CI                & A multiscale blind Bayes denoising algorithm    \\
\cline{3-6}
                              &                          & LSM-NLR \cite{huang2017mixed}      & AWGN, Poisson and Impulsive & CV                & Low-rank with laplacian scale mixture modeling                \\
\cline{3-6}
                              &                          & FastHyde \cite{zhuang2018fast}     & AWGN and Poisson            & MSI, HSI          & Low-rank and sparse representation                            \\
\cline{3-6}
                              &                          & NMoG \cite{chen2017denoising}         & AWGN, Stripe and Impulsive  & MSI, HSI          & Low-rank matrix model with non i.i.d Gaussian noise           \\
\cline{3-6}
                              &                          & ODCT \cite{manjon2012new}          & AWGN and Rician             & MRI               & A 3D DCT implementation                                       \\
\cline{3-6}
                              &                          & ONLM \cite{coupe2008optimized}         & AWGN and Rician             & MRI               & An optimized blockwise NLM Filter                             \\
\cline{3-6}
                              &                          & AONLM \cite{manjon2010adaptive}       & AWGN and Rician             & MRI               & Adaptive NLM with spatially varying noise levels              \\
\cline{3-6}
                              &                          & RLMMSE \cite{aja2008noise}       & Rician                      & MRI               & A novel linear minimum mean square error estimator    \\
\cline{3-6}
                              &                          & PRI-NLM \cite{manjon2012new}      & AWGN and Rician             & MRI               & A rotationally invariant version of the NLM filter            \\
\cline{3-6}
                              &                          & PRI-NLPCA \cite{manjon2015mri}    & AWGN and Rician             & MRI               & A two stage filter based on NLPCA and PRI-NLM                 \\
\cline{2-6}
                              & \multirow{17}{*}{Tensor} & CBM3D \cite{dabov2007color}        & AWGN                        & CI, CV                & BM3D with oppenent and YUV color mode transforms              \\
%\cline{3-6}
%                              &                          & CVBM3D \cite{dabov2007video}        & AWGN                        & CV                & Extention of CBM3D to color videos                            \\
\cline{3-6}
                              &                          & LRTA \cite{renard2008denoising}         & AWGN                        & MSI, HSI          & A low Tucker rank tensor decomposition  model                 \\
\cline{3-6}
                              &                          & PARAFAC \cite{liu2012denoising}       & AWGN                        & MSI               & A rank-one candecomp/parafac (CP) model                             \\
\cline{3-6}
                              &                          & BM4D \cite{maggioni2012nonlocal}         & AWGN                        & MSI, HSI, MRI     & Extention of BM3D to MSI and MRI using 3D patch               \\
\cline{3-6}
                              &                          & VBM4D \cite{maggioni2012video}        & AWGN                        & CV                & Extention of BM3D to color videos using 3D patch              \\
\cline{3-6}
                              &                          & 4DHOSVD \cite{rajwade2012image, zhang2015denoising, zhang2017denoise}      & AWGN                        & CI, MSI, HSI, MRI & Applications of 4DHOSVD transform                             \\
\cline{3-6}
                              &                          & TDL \cite{peng2014decomposable}           & AWGN                        & MSI, HSI          & A Tucker based tensor dictionary learning algorithm           \\
\cline{3-6}
                              &                          & KTSVD \cite{Zhang2015KTSVD}         & AWGN                        & MSI, HSI          & A t-SVD based tensor dictionary learning algorithm            \\
\cline{3-6}
                              &                          & LRTV \cite{he2015total}         & AWGN                        & MSI, HSI          & Low-rank model with total variation (TV) regularization            \\
\cline{3-6}
                              &                          & ITSReg \cite{xie2016multispectral}        & AWGN                        & MSI, HSI          & Intrinsic Tensor Sparsity Regularization                      \\
\cline{3-6}
                              &                          & LLRT \cite{chang2017hyper}        & AWGN                        & CI, MSI          & low-rank tensor with hyper-laplacian regularization                      \\
\cline{3-6}
                              &                          & LLRGTV \cite{He2018LLRGTV}       & AWGN                        & MSI, HSI          & Low Tucker rank decomposition with total variation            \\
\cline{3-6}
                              &                          & WTR1 \cite{wu2018weighted}         & AWGN                        & CI                & A weighted rank-one CP decomposition model                    \\
\cline{3-6}
                              &                          & ILR-HOSVD \cite{lv2019denoising}    & AWGN                        & MRI               & A recursive low Tucker rank model with rank estimation        \\
\cline{3-6}
                              &                          & MSt-SVD \cite{kong2017new, kong2019color}       & AWGN                        & CI, MRI, MSI, HSI & An efficient one-step t-SVD implementation  \\
\cline{3-6}
                              &                          & NGMeet \cite{he2019non}       & AWGN                        & MSI, HSI          & Low-rank tensor model with iterative regularization           \\
\cline{3-6}
                              &                          & LTDL \cite{gong2020low}         & AWGN                        & MSI, HSI          & A low-rank tensor dictionary learning method                  \\
\cline{3-6}
                              &                          & NLTA-LSM \cite{dong2015low}      & AWGN and Poisson            & MSI, HSI, MRI     & Low Tucker rank with laplacian scale mixture modeling         \\
\hline
\end{tabular}
}
\label{Table_traditional_method}
\end{table*}

%% file: Fig_traditional_framework.tex
\begin{figure}[htbp]
  \centering
  \graphicspath{{Figs/Fig_frameworks/}}
  \includegraphics[width=3.46in]{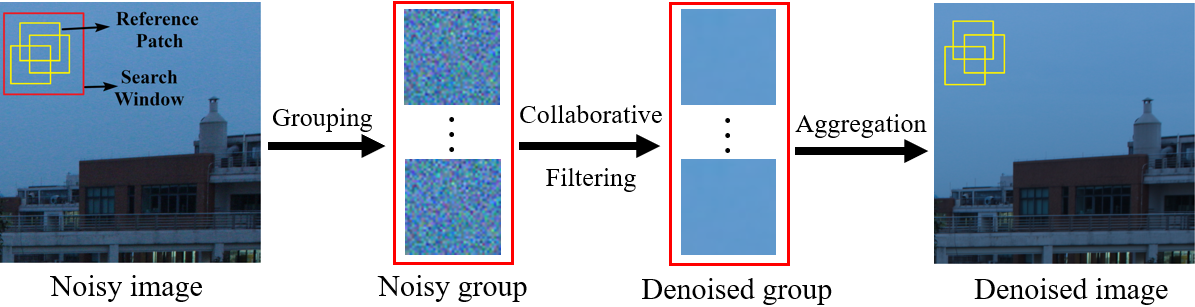}\\
  \caption{Illustration of the grouping-collaborative filtering-aggregation framework for traditional denoisers.}
  \label{Fig_traditional_framework}
\end{figure}

%% file: Table_DNN_method.tex
\begin{table*}[htbp]
\scriptsize
\centering
\caption{Related DNN denoising methods for multi-dimensional image data with different noise modeling techniques and applications. 'CI': color image, 'CV': color video, 'MSI': multispectral imaging, 'HSI': hyperspectral imaging, 'MRI': magnetic resonance imaging.}
\scalebox{0.869}{
\begin{tabular}{|c|c|c|c|c|c|}
\hline
\textbf{Category}                      & \textbf{Network  Architectures}   & \textbf{Methods}       & \textbf{Noise Modeling}  & \textbf{Applications}      & \textbf{Key Words}                                                  \\
\hline
\multirow{38}{*}{\shortstack[l]{DNN \\ methods}}         & \multirow{36}{*}{CNN/NN}   & MLP \cite{burger2012image}          & AWGN                        & CI                & A multilayer perceptron model                                 \\

\cline{3-6}
                              &                          & TNRD \cite{chen2016trainable}         & AWGN                        & CI                & A trainable nonlinear difussion model                         \\
\cline{3-6}
                              &                          & DnCNN \cite{zhang2017beyond, jiang2018denoising}          & AWGN                        & CI, MRI                 & CNN with Residual learning and batch normalization                       \\
\cline{3-6}
                              &                          & NLNet \cite{lefkimmiatis2017non}         & AWGN                        & CI                & A nonlocal CNN model                                          \\
\cline{3-6}
                              &                          & UDNet \cite{Lefkimmiatis_2018_CVPR}        & AWGN                        & CI                & A robust and flexible CNN model                               \\
\cline{3-6}
                              &                          & FFDNet \cite{zhang2018ffdnet}       & AWGN                        & CI                & A flexible model with tunable input noise levels              \\
\cline{3-6}
                              &                          & HSID-CNN \cite{yuan2018hyperspectral}      & AWGN                        & MSI, HSI          & A 2D and 3D combined CNN model                                \\
\cline{3-6}
                              &                          & HSI-SDeCNN \cite{maffei2019single}    & AWGN                        & MSI, HSI          & A single CNN model with spatial and spectral information      \\
\cline{3-6}
                              &                          & VNLNet \cite{davy2018non}       & AWGN                        & CV                & The first nonlocal CNN model for video                        \\
\cline{3-6}
                              &                          & ViDeNN \cite{claus2019videnn}       & AWGN                        & CV                & Combination of spatial and temporal filtering with CNN        \\
\cline{3-6}
                              &                          & SADNet \cite{chang2020spatial}        & AWGN                        & CI                & A novel spatial-adaptive CNN model                            \\
\cline{3-6}
                              &                          & FastDVDNet \cite{Tassano_2020_CVPR}   & AWGN                        & CV                & Real-time video denoising without flow estimation                    \\
\cline{3-6}
                              &                          & LIDIA \cite{Vaksman_2020_CVPR_Workshops}        & AWGN                        & CI                & A lightweight denoiser with instance adaptation               \\
\cline{3-6}
                              &                          & DRUNet \cite{zhang2020plug}     & AWGN                   & CI                & A plug-and-play method with deep denoiser prior     \\
\cline{3-6}
                              &                          & ADRN \cite{zhao2020adrn}          & AWGN                   & MSI, HSI                & A deep residual network model with channel attention scheme            \\
\cline{3-6}
                              &                          & HSI-DeNet \cite{chang2018hsi}    & AWGN and Strip              & MSI, HSI          & A CNN method with mixed noise modeling                        \\
\cline{3-6}
                              &                          & Noise2noise \cite{lehtinen2018noise2noise}       & AWGN and Poisson                     & CI              & A CNN model trained only with noisy data.                         \\
\cline{3-6}
                              &                          & VDNet \cite{yue2019variational}        & AWGN and Realistic          & CI                & A variational inference method with noise estimation          \\
\cline{3-6}
                              &                          & CBDNet \cite{guo2019toward}       & Realistic                   & CI                & a convolutional blind denoising network                       \\
\cline{3-6}
                              &                          & PRI-PB-CNN \cite{manjon2018mri}   & AWGN an Rician              & MRI               & Combination of sliding window scheme and 3D CNN               \\
\cline{3-6}
                              &                          & MIFCN \cite{abbasi2019three}        & Realistic                   & MRI               & A fully convolutional network model                           \\
\cline{3-6}
                              &                          & DIDN \cite{yu2019deep}         & AWGN and Realistic          & CI                & A deep iterative down-up CNN model                            \\
\cline{3-6}
                              &                          & DRDN \cite{song2019dynamic}         & Realistic                   & CI                & A dynamic residual dense network model                        \\
\cline{3-6}
                              &                          & SGN \cite{gu2019self}          & Realistic                   & CI                & A self-guided network with top-down architecture              \\
\cline{3-6}
                              &                          & FCCF \cite{yue2019high}         & AWGN and Realistic          & CI                & Fusion of collaborative filtering and CNN                     \\
\cline{3-6}
                              &                          & DBF \cite{chen2019real}          & AWGN and Realistic          & CI                & Integration of CNN into a boosting algorithm                    \\
\cline{3-6}
                              &                          & RIDNet \cite{anwar2019real}       & AWGN and Realistic          & CI                & A single stage model with feature attention                   \\
\cline{3-6}
                              &                          & AINDNet \cite{Kim_2020_CVPR}      & AWGN and Realistic          & CI                & Transfer learning with adaptive instance normalization        \\
\cline{3-6}
                              &                          & Self2Self \cite{quan2020self2self}      & AWGN and Realistic          & CI                & A self-supervised model trained only with noisy images        \\
\cline{3-6}
                              &                          & QRNN3D \cite{wei20203}          & AWGN and Stripe                   & MSI, HSI                & A deep recurrent neural network model with 3D convolution           \\
\cline{3-6}
                              &                          & CycleISP \cite{zamir2020cycleisp}     & Realistic                   & CI                & A camera pipeline model in forward and reverse directions     \\
\cline{3-6}
                              &                          & DANet \cite{yue2020dual}        & Realistic                   & CI                & A Bayesian framework for noise removal and generation.                    \\
\cline{3-6}
                              &                          & MIRNet \cite{Zamir2020MIRNet}        & Realistic                   & CI                & A multi-scale model with parallel convolution streams                   \\
\cline{3-6}
                              &                          & GCDN \cite{valsesia2020deep}        & AWGN and Realistic                   & CI                & A graph convolution denoising network                  \\
\cline{2-6}
                              & \multirow{5}{*}{CNN + GAN} & GCBD \cite{chen2018image}         & AWGN and Realistic          & CI                & A GAN-based blind denoiser                               \\
\cline{3-6}
                              &                          & SCGAN \cite{yan2019unsupervised}        & AWGN                        & CI                & Unsupervised modeling with self-consistent GAN                \\
\cline{3-6}
                              &                          & DnGAN \cite{yeh2018image}        & AWGN and Blur               & CI                & GAN with  maximum a posteriori (MAP) framework                \\
%\cline{3-6}
%                              &                          & RED-WGAN \cite{ran2019denoising}     & AWGN and Rician             & MRI               & A residual encoder decoder Wasserstein GAN model              \\
\cline{3-6}
                              &                          & ADGAN \cite{Lin_2019_CVPR_Workshops}        & Realistic                   & CI                & Attentive GAN with noise domain adaptation                    \\
\hline
\end{tabular}
}
\label{Table_DNN_method}
\end{table*}

%% file: Fig_CNN.tex
\begin{figure}[htbp]
  \centering
  \graphicspath{{Figs/Fig_frameworks/}}
  \includegraphics[width=2.8in, height=1.416in]{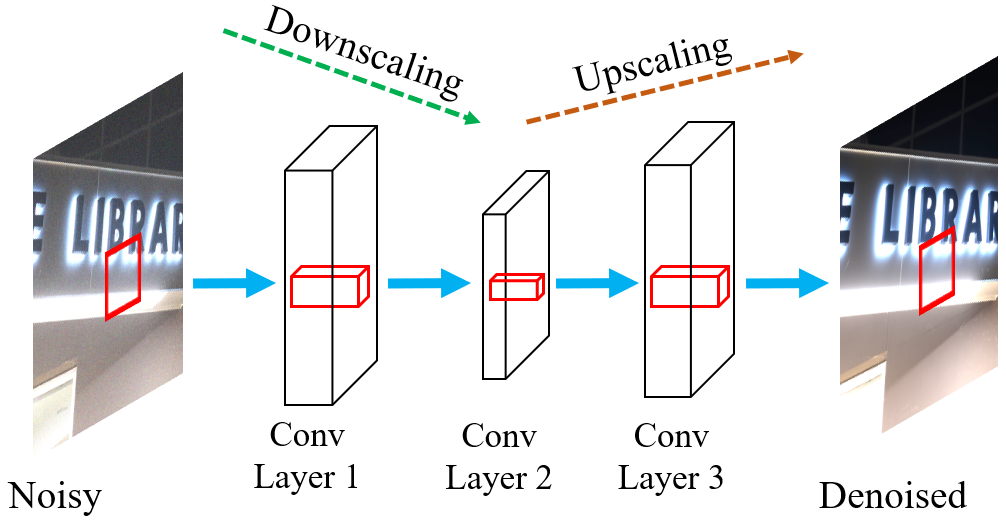}\\
  \caption{Illustration of a simple CNN denoising framework with three convolution layers.}
  \label{Fig_CNN}
\end{figure}

%% file: Fig_timeline.tex
\begin{figure*}[htbp]

\graphicspath{{Figs/Fig_frameworks/}}

\centering
\includegraphics[width=6.99in]{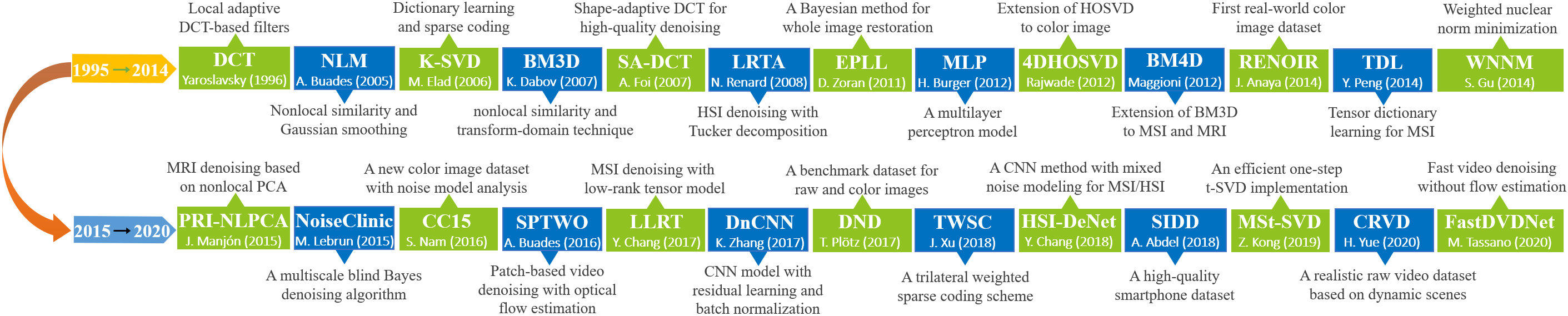}

\caption{Timeline of representative multi-dimensional image denoising methods and datasets.}

\label{Fig_timeline}
\end{figure*}

%% file: Fig_dataset_illus.tex
\begin{figure}[htbp]

\graphicspath{{Figs/Fig_dataset_illus/}}

\centering
\includegraphics[width=3.39in]{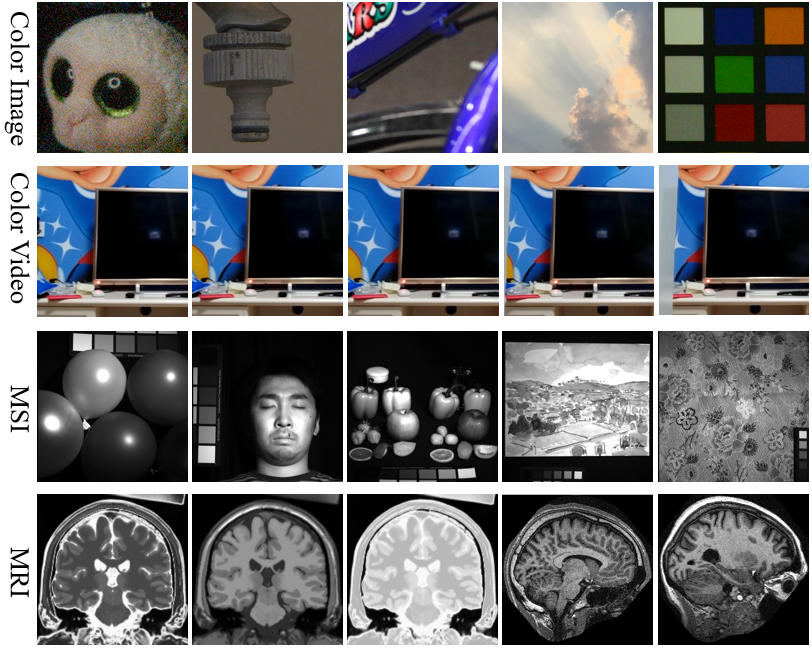}

\caption{Illustration of multi-dimensional image datasets. From the first row to the fourth row: color image, color video, MSI and MRI data.}

\label{Fig_dataset_illus}
\end{figure}

%% file: Table_dataset_description.tex
\begin{table}[htbp]
\tiny
  \centering
  \caption{Information of popular multi-dimensional image denoising datasets for synthetic and real-world experiments. 'GT': ground-truth, 'A': Available, 'N/A': Not-Available, 'F': number of frames.}
  \scalebox{0.88}{
   \renewcommand{\arraystretch}{0.69}
    \begin{tabular}{ccccccc}
    \toprule
    Type  & Name  & Experiments & GT    & \# Cameras & Image size & \# Images \\
    \midrule
    \multirow{8}[26]{*}{Color image} & Kodak \cite{Kodak} & Synthetic & A     & /     & 768 $\times$ 512 $\times$ 3 & 24 \\
\cmidrule{2-7}          & RENOIR \cite{anaya2018renoir} & Real-world & A     & 3     & 3684 $\times$  2760 $\times$ 3 & 120 \\
\cmidrule{2-7}          & Nam-CC15 \cite{nam2016holistic} & Real-world & A     & 3     & 512 $\times$  512 $\times$ 3 & 15 \\
\cmidrule{2-7}          & Nam-CC60 \cite{nam2016holistic} & Real-world & A     & 3     & 500 $\times$  500 $\times$  3 & 60 \\
\cmidrule{2-7}          & PolyU \cite{xu2018real}& Real-world & A     & 5     & 512 $\times$  512 $\times$  3 & 100 \\
\cmidrule{2-7}          & DND   \cite{plotz2017benchmarking} & Real-world & N/A    & 4     & 512 $\times$  512 $\times$  3 & 1000 \\
\cmidrule{2-7}          & SIDD  \cite{abdelhamed2018high}& Real-world & A    & 5     & 256 $\times$  256 $\times$  3 & 1280 \\
\cmidrule{2-7}          & High-ISO \cite{yue2019high}  & Real-world & A    & 2     & 512 $\times$  512 $\times$  3 & 90 \\
\cmidrule{2-7}          & Our IOCI & Real-world & A     & 9     & 1024 $\times$  1024 $\times$  3 & 409 \\
    \midrule
    \multirow{4}[12]{*}{Color video} & Set8 \cite{Set8} & Synthetic & A     & /     & 960 $\times$ 540 $\times$ 3 $\times$ F & 4 \\
\cmidrule{2-7}          & DAVIS \cite{Perazzi2016} & Synthetic & A     & /     & 854 $\times$ 480 $\times$ 3 $\times$ F & 30 \\
\cmidrule{2-7}          & CRVD \cite{yue2020supervised} & Real-world & A     & 1     & 1920 $\times$ 1080 $\times$ 3 $\times$ F & 61 \\
\cmidrule{2-7}          & Our IOCV & Real-world & A     & 4     & 512 $\times$  512 $\times$  3 $\times$  F & 39 \\
    \midrule
    \multirow{3}[8]{*}{MSI} & CAVE \cite{CAVE_0293}  & Synthetic & A     & 1     & 512 $\times$ 512 $\times$ 31 & 32 \\
\cmidrule{2-7}          & ICVL \cite{arad_and_ben_shahar_2016_ECCV} & Synthetic & A    & 1     & 1392 $\times$ 1300 $\times$  31 & 201 \\
\cmidrule{2-7}          & HHD \cite{chakrabarti2011statistics}  & Real-world & N/A    & 1     & 1392 $\times$ 1040 $\times$ 31 & 77 \\
    \midrule
    \multirow{2}[4]{*}{MRI} & BrainWeb \cite{cocosco1997brainweb} & Synthetic & A     & /     & 181 $\times$ 217 $\times$ 181 & 3 \\
\cmidrule{2-7}          & OASIS \cite{marcus2007open}  & Real-world & N/A    & /     & 256 $\times$ 256 $\times$ 128 & 2 \\
    \bottomrule
    \end{tabular}
    }%
  \label{Table_dataset_description}%
\end{table}%

%\vspace*{-\baselineskip}

%% file: Fig_Kong_dataset.tex
\begin{figure}[htbp]
\graphicspath{{Figs/Fig_dataset_illus/}}
\centering
\subfigure[The proposed IOCI dataset]{
\label{Fig4}
\includegraphics[width=3.38in]{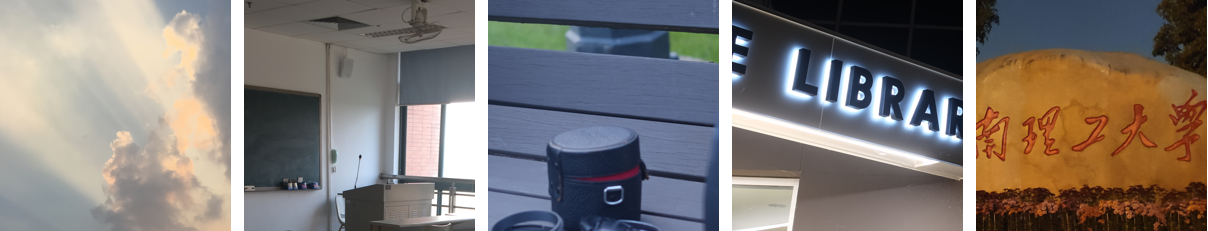}}
\subfigure[The proposed IOCV dataset]{
\label{Fig4}
\includegraphics[width=3.38in]{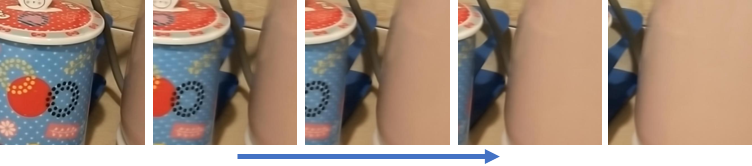}}

\caption{Illustration of the proposed IOCI and IOCV dataset.}
\label{Fig_Kong_dataset}
\end{figure}

%% file: Fig_Video_slider.tex
\begin{figure}[htbp]

\graphicspath{{Figs/Fig_frameworks/}}

\centering
\includegraphics[width=3.58in]{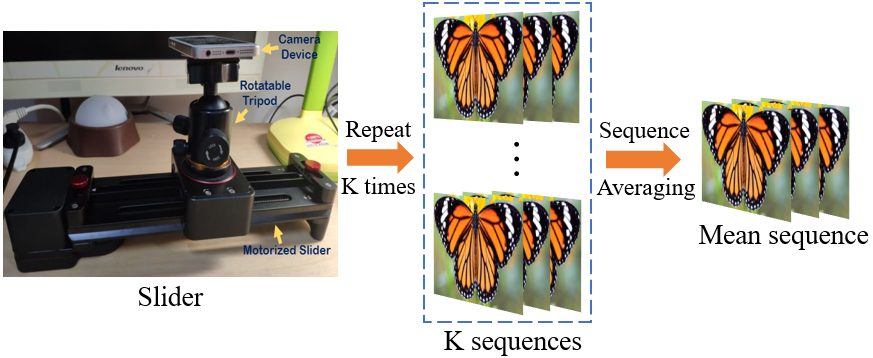}

\caption{The procedure of generating mean video sequences with a motorized slider. The camera is fixed to a rotatable tripod ball-head placed on
top of the slider.}

\label{Fig_Video_slider}
\end{figure}

%% file: Table_color_all.tex
\begin{table*}[htbp]
\tiny
  \centering
  \caption{Average PSNR, SSIM values and computational time (second) on four realistic color image datasets. The average time is calculated based on the PolyU dataset, and 'N/A' means the models can not handle certain size of images. The best results (excluding CBM3D\_best) are black bolded.}
    \scalebox{0.806}{
    \renewcommand{\arraystretch}{0.399}
    \begin{tabular}{cccccccccccccccc}
    \toprule
    \multirow{3}[6]{*}{Method} & \multirow{2}[4]{*}{Dataset} & \multirow{2}[4]{*}{CC15} & \multirow{2}[4]{*}{CC60} & \multirow{2}[4]{*}{PolyU} & \multicolumn{9}{c}{IOCI}                                        & \multirow{3}[6]{*}{Implementation} & \multirow{3}[6]{*}{Time (s)} \\
\cmidrule{6-14}          &       &       &       &       & CANON 100D & CANON 600D & HUAWEI Honor6X & IPHONE 5S & IPHONE 6S & SONY A6500 & XIAOMI MI8 & OPPO R11s & FUJIFILM X100T &       &  \\
\cmidrule{2-14}          & \multicolumn{1}{l}{\# Images} & 15    & 60    & 100   & 55    & 25    & 30    & 36    & 67    & 36    & 50    & 39    & 71    &       &  \\
    \midrule
    \multirow{2}[2]{*}{/} & \multirow{2}[2]{*}{Noisy} & 33.41  & 34.51  & 35.97  & 38.94  & 39.17  & 36.94  & 38.21  & 37.30  & 42.89  & 34.49  & 37.84  & 39.39  & \multirow{2}[2]{*}{/} & \multirow{2}[2]{*}{/} \\
          &       & 0.8476  & 0.8712  & 0.9125  & 0.9482  & 0.9379  & 0.9185  & 0.9359  & 0.9143  & 0.9753  & 0.9251  & 0.9356  & 0.9307  &       &  \\
    \midrule
    \multirow{32}[32]{*}{\shortstack[l]{Traditional \\ denoisers}} & \multirow{2}[2]{*}{LCSD} & 36.20  & 37.69  & 37.97  & 40.65  & 41.45  & 39.22  & 39.84  & 39.53  & 44.32  & 35.18  & 38.69  & 38.75  & \multirow{2}[2]{*}{MATLAB + MEX} & \multirow{2}[2]{*}{6.45} \\
          &       & 0.9339  & 0.9626  & 0.9641  & 0.9732  & 0.9793  & 0.9633  & 0.9627  & 0.9625  & 0.9867  & 0.9560  & 0.9669  & 0.9440  &       &  \\
\cmidrule{2-16}          & \multirow{2}[2]{*}{LLRT} & 37.36  & 38.51  & 38.51  & 41.84  & 42.53  & 39.54  & 40.02  & 39.72  & 45.71  & 34.93  & 36.63  & 39.19  & \multirow{2}[2]{*}{MATLAB} & \multirow{2}[2]{*}{$>$1000} \\
          &       & 0.9559  & 0.9636  & 0.9707  & 0.9784  & 0.9816  & 0.9669  & 0.9676  & 0.9663  & 0.9899  & 0.9557  & 0.9687  & 0.9633  &       &  \\
\cmidrule{2-16}          & \multirow{2}[2]{*}{GID} & 37.02  & 38.41  & 38.37  & 40.86  & 41.60  & 39.52  & 40.12  & 40.16  & 44.94  & 36.11  & 40.50  & 42.31  & \multirow{2}[2]{*}{MATLAB} & \multirow{2}[2]{*}{38.88} \\
          &       & 0.9459  & 0.9633  & 0.9675  & 0.9743  & 0.9790  & 0.9653  & 0.9642  & 0.9670  & 0.9887  & 0.9599  & 0.9729  & 0.9750  &       &  \\
\cmidrule{2-16}          & \multirow{2}[2]{*}{MCWNNM} & 37.72  & 39.03  & 38.51  & 41.47  & 42.07  & 39.46  & 39.87  & 40.18  & 45.37  & 35.84  & 40.71  & 42.48  & \multirow{2}[2]{*}{MATLAB} & \multirow{2}[2]{*}{228.16} \\
          &       & 0.9568  & 0.9698  & 0.9671  & 0.9774  & 0.9795  & 0.9610  & 0.9567  & 0.9628  & 0.9894  & 0.9516  & 0.9734  & 0.9757  &       &  \\
\cmidrule{2-16}          & \multirow{2}[2]{*}{NLH} & 38.49  & 39.86  & 38.36  & 41.77  & 42.72  & 38.84  & 40.44  & 39.94  & \textbf{46.00}  & 35.73  & 40.69  & 42.80  & \multirow{2}[2]{*}{MATLAB + MEX} & \multirow{2}[2]{*}{51.86} \\
          &       & 0.9647  & 0.9759  & 0.9655  & 0.9790  & 0.9841  & 0.9593  & 0.9644  & 0.9630  & \textbf{0.9905}  & 0.9552  & 0.9740  & 0.9776  &       &  \\
\cmidrule{2-16}          & \multirow{2}[2]{*}{TWSC} & 37.90  & 39.66  & 38.62  & 41.65  & 42.52  & 39.71  & 40.27  & 40.12  & 45.48  & 35.41  & 40.71  & 42.22  & \multirow{2}[2]{*}{MATLAB} & \multirow{2}[2]{*}{368.68} \\
          &       & 0.9592  & 0.9759  & 0.9674  & 0.9767  & 0.9824  & 0.9651  & 0.9617  & 0.9619  & 0.9896  & 0.9385  & 0.9720  & 0.9728  &       &  \\
\cmidrule{2-16}          & \multirow{2}[2]{*}{M-SVD} & 37.62  & 39.27  & 38.56  & 41.81  & 42.54  & 39.94  & 40.66  & 40.35  & 45.69  & 36.24  & 40.69  & 42.52  & \multirow{2}[2]{*}{MATLAB} & \multirow{2}[2]{*}{75.88} \\
          &       & 0.9542  & 0.9705  & 0.9671  & 0.9788  & 0.9823  & 0.9671  & \textbf{0.9680}  & 0.9660  & 0.9897  & 0.9606  & 0.9735  & 0.9750  &       &  \\
\cmidrule{2-16}          & \multirow{2}[2]{*}{4DHOSVD} & 37.51  & 39.15  & 38.51  & 41.41  & 42.14  & 39.82  & 40.68  & 40.36  & 45.56  & 36.17  & 40.69  & 42.47  & \multirow{2}[2]{*}{MATLAB} & \multirow{2}[2]{*}{80.69} \\
          &       & 0.9559  & 0.9729  & 0.9673  & 0.9771  & 0.9797  & 0.9658  & 0.9664  & 0.9671  & 0.9901  & 0.9600  & 0.9736  & 0.9753  &       &  \\
\cmidrule{2-16}          & \multirow{2}[2]{*}{CMSt-SVD} & 37.95  & 39.75  & 38.82  & \textbf{41.99} & 42.75  & 40.08  & \textbf{40.84} & 40.53  & 45.89 & 36.40  & 40.85 & 42.68  & \multirow{2}[2]{*}{MATLAB + MEX} & \multirow{2}[2]{*}{6.18} \\
          &       & 0.9588  & 0.9756  & 0.9694  & \textbf{0.9794} & 0.9840  & 0.9674  & 0.9668  & 0.9674  & 0.9903 & 0.9616  & 0.9743  & 0.9770  &       &  \\
\cmidrule{2-16}          & \multirow{2}[2]{*}{CBM3D1} & 37.56  & 39.21  & 38.52  & 41.71  & 42.50  & 39.93  & 40.56  & 40.45  & 45.56  & 36.30  & 40.62  & 42.53  & \multirow{2}[2]{*}{MATLAB + MEX} & \multirow{2}[2]{*}{0.95} \\
          &       & 0.9562  & 0.9729  & 0.9676  & 0.9780  & 0.9827  & 0.9660  & 0.9666  & 0.9681  & 0.9897  & 0.9603  & 0.9725  & 0.9762  &       &  \\
\cmidrule{2-16}          & \multirow{2}[2]{*}{CBM3D2} & 37.70  & 39.41  & 38.69  & 41.69  & 42.54  & 39.97  & 40.77  & 40.55  & 45.70  & 36.38  & 40.75  & 42.65  & \multirow{2}[2]{*}{MATLAB + MEX} & \multirow{2}[2]{*}{4.38} \\
          &       & 0.9572  & 0.9740  & 0.9695  & 0.9780  & 0.9836  & 0.9669  & 0.9668  & 0.9693  & 0.9902  & 0.9610  & 0.9737  & 0.9767  &       &  \\
\cmidrule{2-16}          & \multirow{2}[2]{*}{CBM3D\_best} & 37.95  & 39.68  & 38.81  & {42.08} & 42.89  & {40.48} & {41.25} & {41.16} & {45.81} & {36.66} & {40.89} & {43.14} & \multirow{2}[2]{*}{MATLAB + MEX} & \multirow{2}[2]{*}{/} \\
          &       & 0.9621  & 0.9775  & 0.9712  & {0.9808} & 0.9851  & {0.9741} & {0.9758} & {0.9783} & {0.9904} & {0.9640} & 0.9742  & {0.9805} &       &  \\
    \midrule
    \midrule
    \multirow{52}[52]{*}{\shortstack[l]{DNN \\ methods}} & \multirow{2}[2]{*}{DnCNN} & 37.47  & 39.32  & 38.51  & 40.81  & 41.91  & 39.92  & 39.34  & 40.22  & 43.94  & 36.30  & 40.26  & 41.57  & \multirow{2}[2]{*}{MATLAB + MEX} & \multirow{2}[2]{*}{3.28} \\
          &       & 0.9537  & 0.9742  & 0.9663  & 0.9717  & 0.9790  & 0.9662  & 0.9639  & 0.9690  & 0.9840  & 0.9599  & 0.9736  & 0.9690  &       &  \\
\cmidrule{2-16}          & \multirow{2}[2]{*}{FFDNet} & 37.68  & 39.73  & 38.56  & 41.67  & 42.55  & 40.05  & 40.60  & 40.49  & 45.71  & 36.47  & 40.75  & 42.44  & \multirow{2}[2]{*}{MATLAB + MEX} & \multirow{2}[2]{*}{1.99} \\
          &       & 0.9563  & 0.9770  & 0.9658  & 0.9768  & 0.9824  & 0.9669  & 0.9645  & \textbf{0.9707}  & 0.9901  & 0.9607  & 0.9733  & 0.9759  &       &  \\
\cmidrule{2-16}          & \multirow{2}[2]{*}{CBDNet} & 36.36  & 37.46  & 36.93  & 39.43  & 41.26  & 37.11  & 38.56  & 37.62  & 42.17  & 35.09  & 37.68  & 40.21  & \multirow{2}[2]{*}{Python + GPU} & \multirow{2}[2]{*}{3.58} \\
          &       & 0.9294  & 0.9482  & 0.9457  & 0.9729  & 0.9788  & 0.9505  & 0.9586  & 0.9428  & 0.9819  & 0.9493  & 0.9601  & 0.9621  &       &  \\
\cmidrule{2-16}          & \multirow{2}[2]{*}{DBF} & 38.38  & 40.59  & 38.87  & 41.36  & 42.87  & 39.98  & 39.89  & 40.26  & 44.42  & 36.45  & 40.67  & 42.36  & \multirow{2}[2]{*}{Python + GPU} & \multirow{2}[2]{*}{0.32} \\
          &       & 0.9644  & 0.9811  & 0.9698  & 0.9765  & 0.9841  & 0.9669  & 0.9634  & 0.9660  & 0.9804  & 0.9631  & 0.9734  & 0.9751  &       &  \\
\cmidrule{2-16}          & \multirow{2}[2]{*}{DIDN} & 36.04  & \multirow{2}[2]{*}{N/A} & 37.32  & 40.66  & 41.58  & 38.09  & 38.22  & 38.50  & 43.89  & 35.25  & 39.66  & 41.17  & \multirow{2}[2]{*}{Python + GPU} & \multirow{2}[2]{*}{0.18} \\
          &       & 0.9456  &       & 0.9523  & 0.9739  & 0.9760  & 0.9500  & 0.9535  & 0.9493  & 0.9840  & 0.9485  & 0.9647  & 0.9655  &       &  \\
\cmidrule{2-16}          & \multirow{2}[2]{*}{FCCF} & \textbf{39.04} & \textbf{40.89} & 38.85  & \textbf{41.91} & \textbf{43.07} & 40.11  & 40.39  & 40.65  & 45.38  & \textbf{36.66 } & \textbf{40.93} & \textbf{42.78} & \multirow{2}[2]{*}{MATLAB + MEX} & \multirow{2}[2]{*}{36.36} \\
          &       & \textbf{0.9675} & \textbf{0.9819} & 0.9699  & \textbf{0.9794} & \textbf{0.9852} & 0.9656  & 0.9659  & 0.9680  & 0.9888  & \textbf{0.9640} & \textbf{0.9747} & \textbf{0.9782} &       &  \\
\cmidrule{2-16}          & \multirow{2}[2]{*}{NLNet} & 37.56  & 39.32  & 38.58  & 41.43  & 42.18  & 39.96  & 40.06  & 40.61  & 44.84  & 36.34  & 40.63  & 42.19  & \multirow{2}[2]{*}{MATLAB + MEX} & \multirow{2}[2]{*}{49.98} \\
          &       & 0.9554  & 0.9744  & 0.9684  & 0.9759  & 0.9811  & 0.9687  & 0.9637  & \textbf{0.9709}  & 0.9875  & 0.9617  & 0.9729  & 0.9739  &       &  \\
\cmidrule{2-16}          & \multirow{2}[2]{*}{UDnet} & 36.95  & 38.96  & 38.05  & 40.40  & 41.36  & 39.35  & 38.94  & 39.95  & 43.15  & 36.09  & 39.51  & 41.12  & \multirow{2}[2]{*}{MATLAB + MEX} & \multirow{2}[2]{*}{5.69} \\
          &       & 0.9425  & 0.9708  & 0.9622  & 0.9683  & 0.9774  & 0.9631  & 0.9591  & 0.9683  & 0.9811  & 0.9582  & 0.9660  & 0.9675  &       &  \\
\cmidrule{2-16}          & \multirow{2}[2]{*}{VDNet} & 35.86  & \multirow{2}[2]{*}{N/A} & 37.58  & 40.97  & 42.37  & 38.23  & 37.88  & 38.60  & 44.27  & 35.49  & 39.93  & 41.19  & \multirow{2}[2]{*}{Python + GPU} & \multirow{2}[2]{*}{0.21} \\
          &       & 0.9454  &       & 0.9574  & 0.9763  & 0.9831  & 0.9555  & 0.9562  & 0.9556  & 0.9844  & 0.9530  & 0.9699  & 0.9701  &       &  \\
\cmidrule{2-16}          & \multirow{2}[2]{*}{AINDNet} & 35.88  & 36.98  & 37.22  & 38.28  & 39.21  & 36.53  & 36.68  & 36.83  & 40.18  & 34.49  & 37.20  & 38.50  & \multirow{2}[2]{*}{Python + GPU} & \multirow{2}[2]{*}{1.29} \\
          &       & 0.9113  & 0.9305  & 0.9460  & 0.9748  & 0.9759  & 0.9544  & 0.9581  & 0.9527  & 0.9812  & 0.9523  & 0.9676  & 0.9656  &       &  \\
\cmidrule{2-16}          & \multirow{2}[2]{*}{CycleISP} & 35.40  & 36.87  & 37.61  & 41.24  & 41.79  & 38.43  & 38.69  & 38.60  & 44.71  & 35.50  & 39.85  & 41.58  & \multirow{2}[2]{*}{Python + GPU} & \multirow{2}[2]{*}{0.38} \\
          &       & 0.9158  & 0.9393  & 0.9546  & 0.9766  & 0.9770  & 0.9515  & 0.9605  & 0.9518  & 0.9874  & 0.9489  & 0.9682  & 0.9688  &       &  \\
\cmidrule{2-16}          & \multirow{2}[2]{*}{RIDNet} & 36.83  & 38.11  & 38.57  & 41.08  & 42.18  & 38.88  & 38.69  & 39.02  & 44.40  & 35.75  & 40.08  & 41.58  & \multirow{2}[2]{*}{Python + GPU} & \multirow{2}[2]{*}{1.06} \\
          &       & 0.9406  & 0.9609  & 0.9702  & 0.9770  & 0.9801  & 0.9621  & 0.9627  & 0.9588  & 0.9873  & 0.9570  & 0.9719  & 0.9750  &       &  \\
\cmidrule{2-16}          & \multirow{2}[2]{*}{SADNet} & 37.80  & \multirow{2}[2]{*}{N/A} & 39.30 & 40.74  & 42.03  & 39.55  & 39.91  & 39.72  & 44.69  & 36.38  & 40.72  & 41.91  & \multirow{2}[2]{*}{Python + GPU} & \multirow{2}[2]{*}{0.33} \\
          &       & 0.9521  &       & 0.9725 & 0.9758  & 0.9797  & 0.9646  & 0.9656  & 0.9626  & 0.9886  & 0.9620  & 0.9740  & 0.9752  &       &  \\
\cmidrule{2-16}          & \multirow{2}[2]{*}{DANet} & 37.17  & 38.63  & \textbf{41.76}  & 41.59  & 42.47  & 38.46  & 39.39  & 38.80  & 44.71  & 36.21  & 39.92  & 41.44  & \multirow{2}[2]{*}{Python + GPU} & \multirow{2}[2]{*}{0.12} \\
          &       & 0.9528  & 0.9694  & \textbf{0.9771}  & 0.9788  & 0.9833  & 0.9545  & 0.9661  & 0.9527  & 0.9874  & 0.9608  & 0.9673  & 0.9693  &       &  \\
\cmidrule{2-16}          & \multirow{2}[2]{*}{Self2Self} & 36.26 & 37.88 &	37.66 &	\multirow{2}[2]{*}{N/A} & \multirow{2}[2]{*}{N/A} &	\multirow{2}[2]{*}{N/A} &	\multirow{2}[2]{*}{N/A} & \multirow{2}[2]{*}{N/A} &	\multirow{2}[2]{*}{N/A} & \multirow{2}[2]{*}{N/A} &	\multirow{2}[2]{*}{N/A} & \multirow{2}[2]{*}{N/A}
 & \multirow{2}[2]{*}{Python + GPU} & \multirow{2}[2]{*}{$>$2000} \\
          &       & 0.9466 & 0.9672 & 0.9563 &  &	 &  &  &  &  &  &  &  &       &  \\
\cmidrule{2-16}          & \multirow{2}[2]{*}{MIRNet} & 36.06 & 37.25 &	37.49 &	40.72 & 41.68 &	38.28 &	38.60 &	38.55 &	43.66 &	35.42 &	39.73 &	41.05
 & \multirow{2}[2]{*}{Python + GPU} & \multirow{2}[2]{*}{0.18} \\
          &       & 0.9416 & 0.9556 & 0.9556 & 0.9747 &	0.9777 & 0.9539 & 0.9591 & 0.9522 & 0.9834 & 0.9525 & 0.9670 & 0.9679 &       &  \\
\cmidrule{2-16}          & \multirow{2}[2]{*}{DRUNet} & 38.30  & 40.33  & 38.93  & \textbf{41.94} & 42.62  & \textbf{40.35} & \textbf{40.80}  & \textbf{40.80} & 45.76  & 36.61  & \textbf{40.96} & \textbf{42.78} & \multirow{2}[2]{*}{Python + GPU} & \multirow{2}[2]{*}{0.18} \\
          &       & 0.9608  & 0.9791  & 0.9700  & 0.9779  & 0.9819  & \textbf{0.9694} & 0.9661  & 0.9700 & 0.9895  & 0.9607  & 0.9741  & 0.9770  &       &  \\
    \bottomrule
    \end{tabular}%
    }
  \label{Table_color_all}%
\end{table*}%

%% file: Fig_CC15.tex
\begin{figure*}[htbp]
\graphicspath{{Figs/combined_CC15/}}
\centering
\subfigure[Mean]{
\label{Fig4}
\includegraphics[width=1.1in]{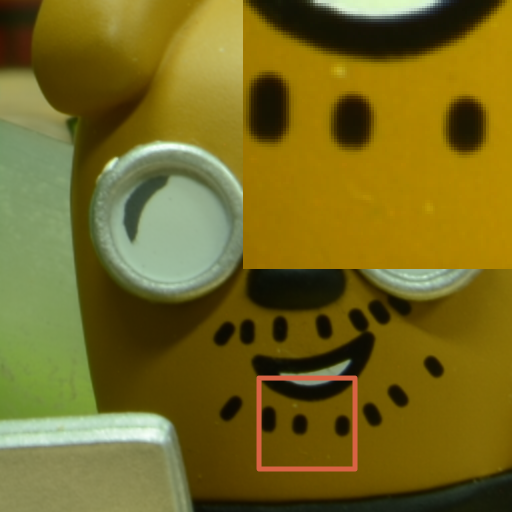}}
\subfigure[Noisy (33.26)]{
\label{Fig4}
\includegraphics[width=1.1in]{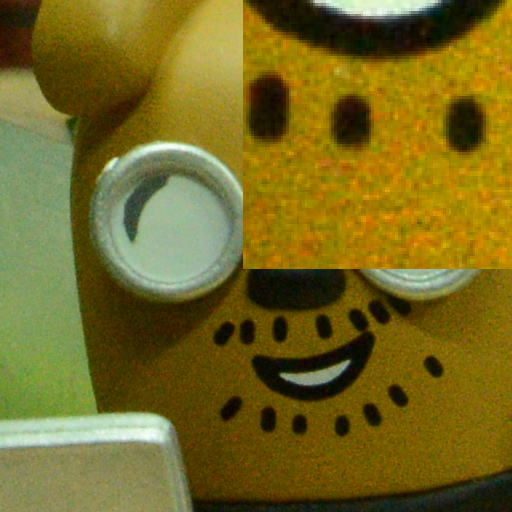}}
\subfigure[NI (38.78)]{
\label{Fig4}
\includegraphics[width=1.1in]{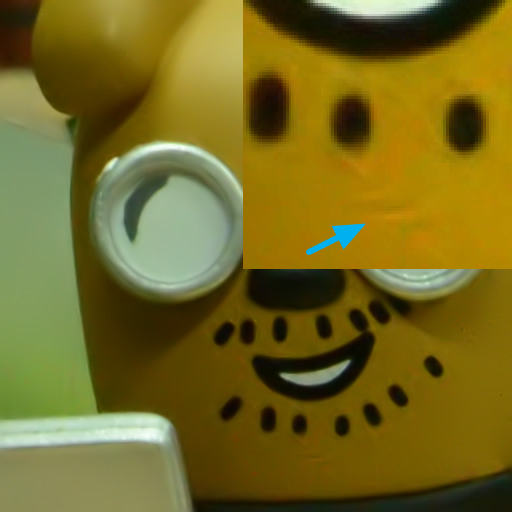}}
\subfigure[DeNoiseAI (36.63)]{
\label{Fig4}
\includegraphics[width=1.1in]{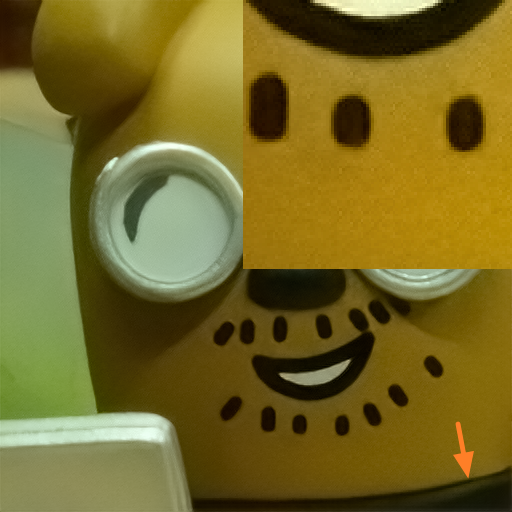}}
\subfigure[DBF (41.42)]{
\label{Fig4}
\includegraphics[width=1.1in]{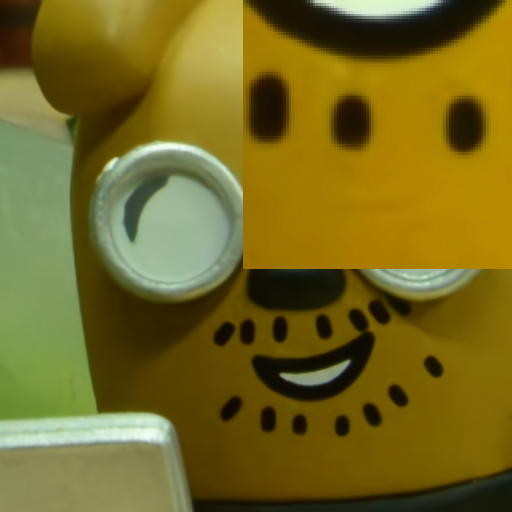}}
\subfigure[DRUNet (40.85)]{
\label{Fig4}
\includegraphics[width=1.1in]{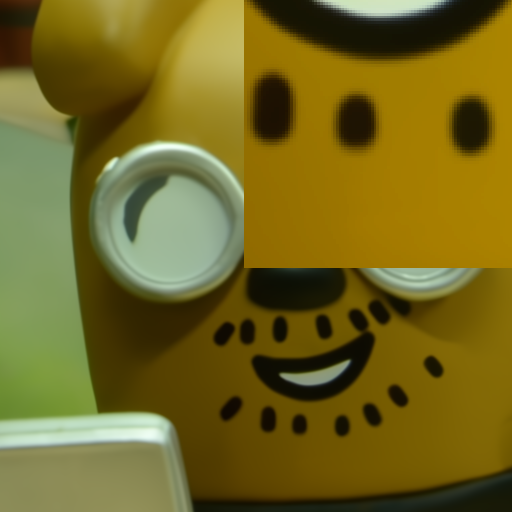}}\\

\subfigure[FFDNet (40.15)]{
\label{Fig4}
\includegraphics[width=1.1in]{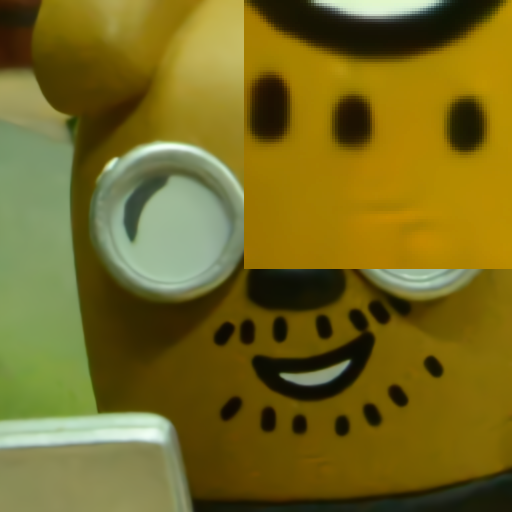}}
\subfigure[FCCF (41.21)]{
\label{Fig4}
\includegraphics[width=1.1in]{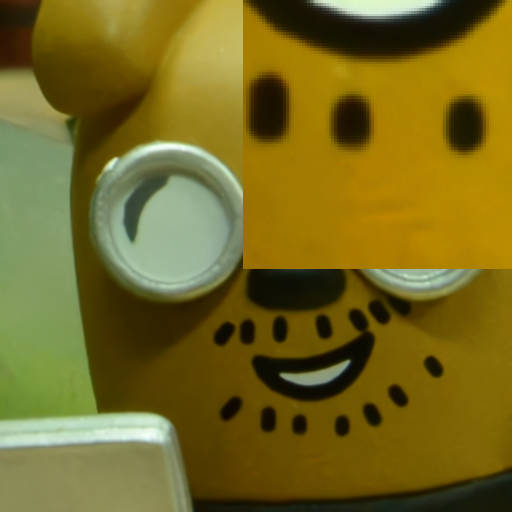}}
\subfigure[CBM3D2 (39.59)]{
\label{Fig4}
\includegraphics[width=1.1in]{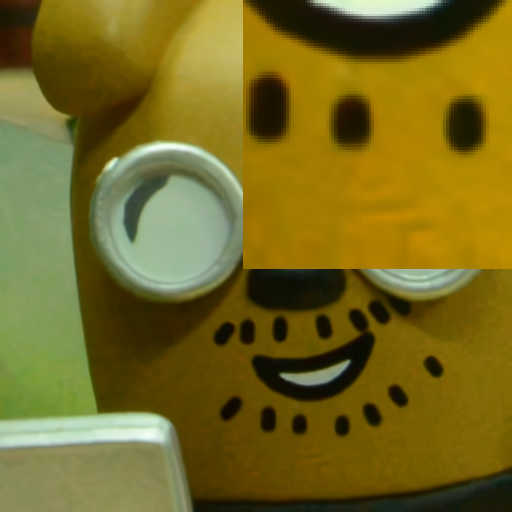}}
\subfigure[CMSt-SVD (39.98)]{
\label{Fig4}
\includegraphics[width=1.1in]{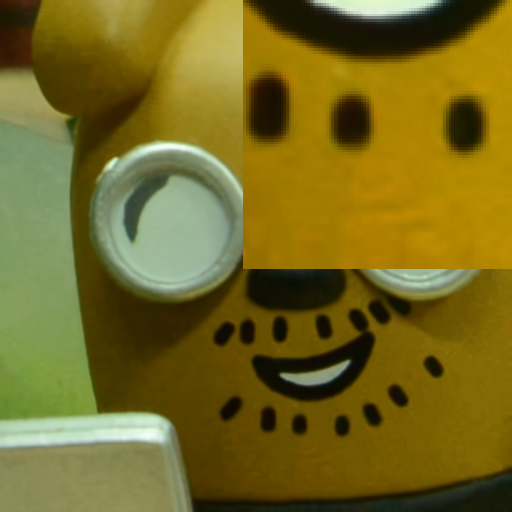}}
\subfigure[TWSC (40.27)]{
\label{Fig4}
\includegraphics[width=1.1in]{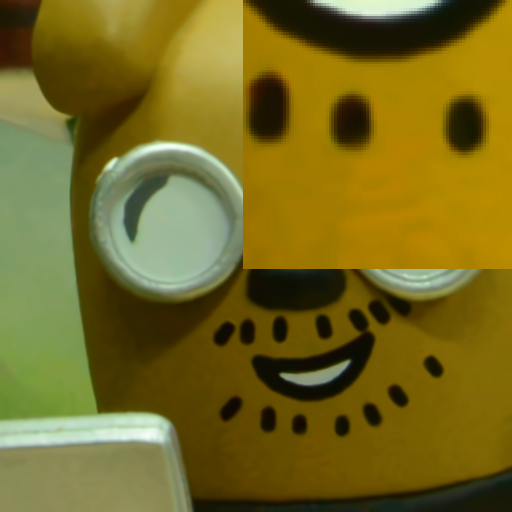}}
\subfigure[NLH (40.21)]{
\label{Fig4}
\includegraphics[width=1.1in]{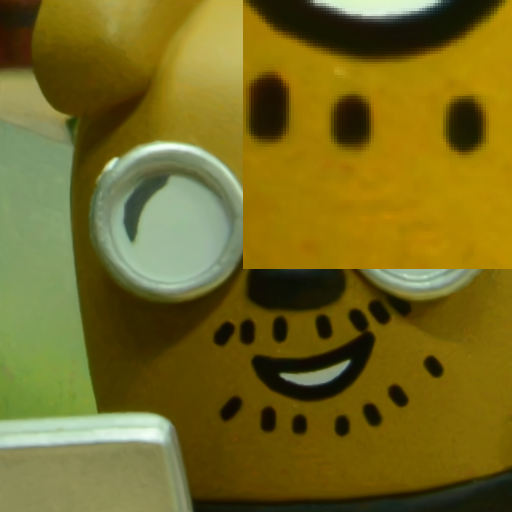}}

\caption{Visual evaluations of compared methods (PSNR) on CC15 dataset.}
\label{Fig_CC15}
\end{figure*}

%% file: Fig_Xu_100.tex
\begin{figure*}[htbp]
\graphicspath{{Figs/combined_Xu_100/}}
\centering
\subfigure[Mean]{
\label{Fig4}
\includegraphics[width=1.1in]{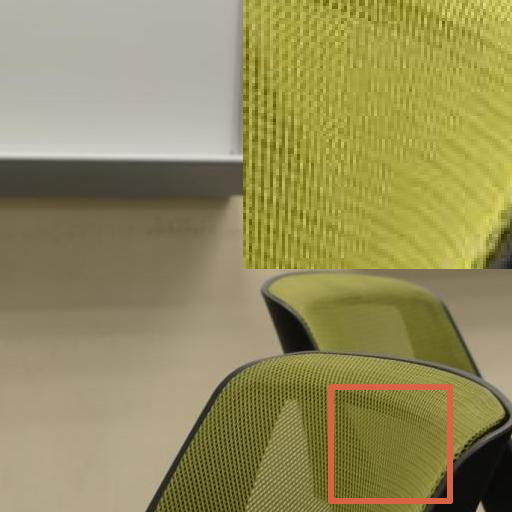}}
\subfigure[Noisy (36.68)]{
\label{Fig4}
\includegraphics[width=1.1in]{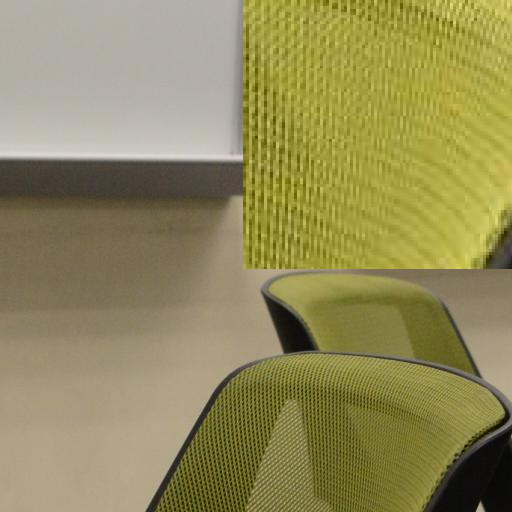}}
\subfigure[NI (36.78)]{
\label{Fig4}
\includegraphics[width=1.1in]{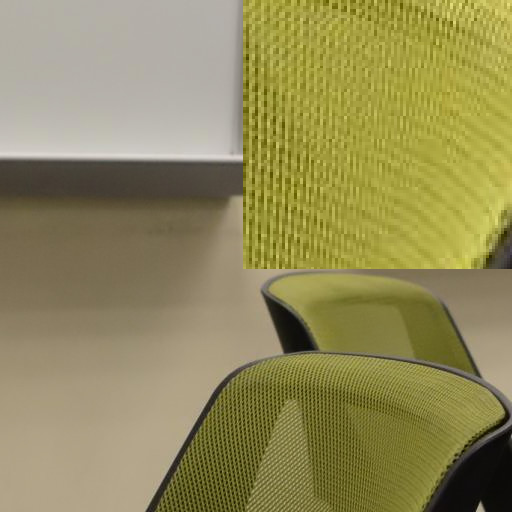}}
\subfigure[DeNoiseAI (36.01)]{
\label{Fig4}
\includegraphics[width=1.1in]{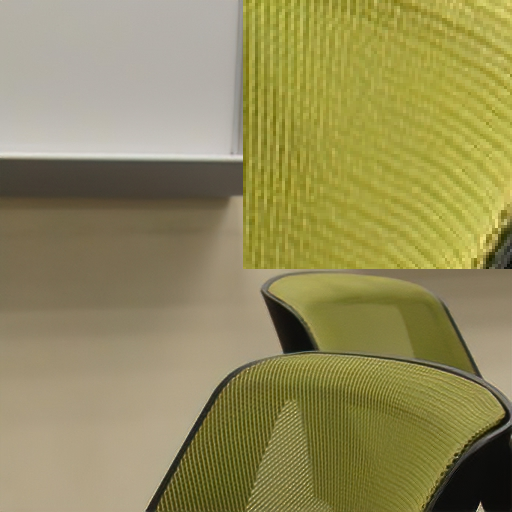}}
\subfigure[DBF (36.62)]{
\label{Fig4}
\includegraphics[width=1.1in]{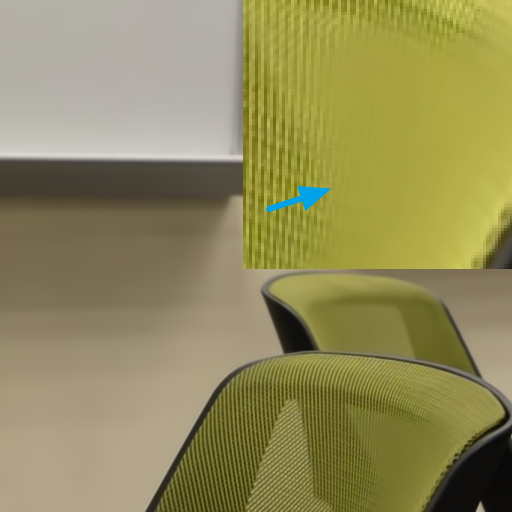}}
\subfigure[DRUNet (37.57)]{
\label{Fig4}
\includegraphics[width=1.1in]{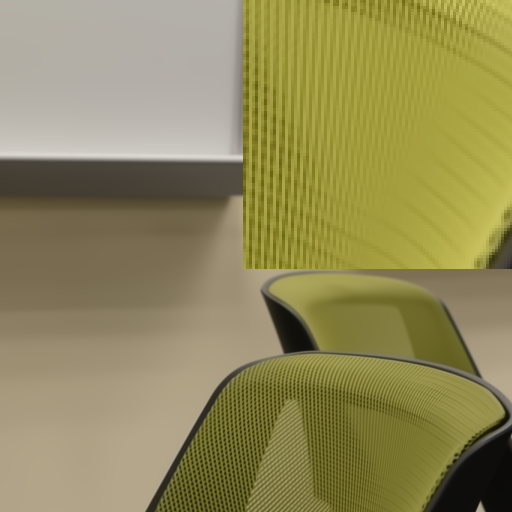}}\\
\subfigure[FFDNet (36.95)]{
\label{Fig4}
\includegraphics[width=1.1in]{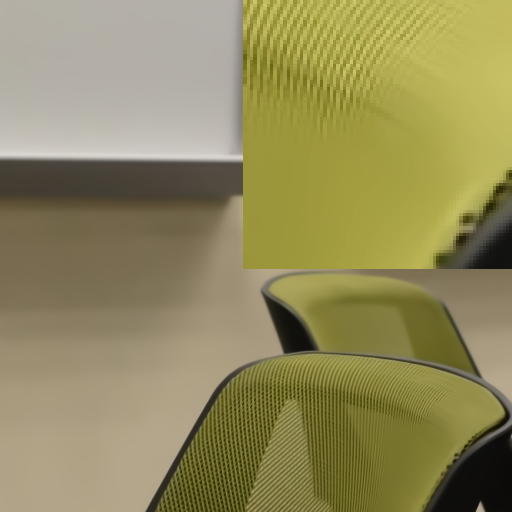}}
\subfigure[FCCF (37.48)]{
\label{Fig4}
\includegraphics[width=1.1in]{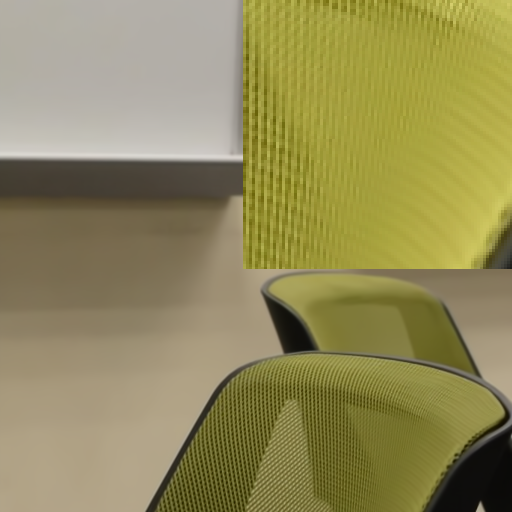}}
\subfigure[CBM3D2 (37.81)]{
\label{Fig4}
\includegraphics[width=1.1in]{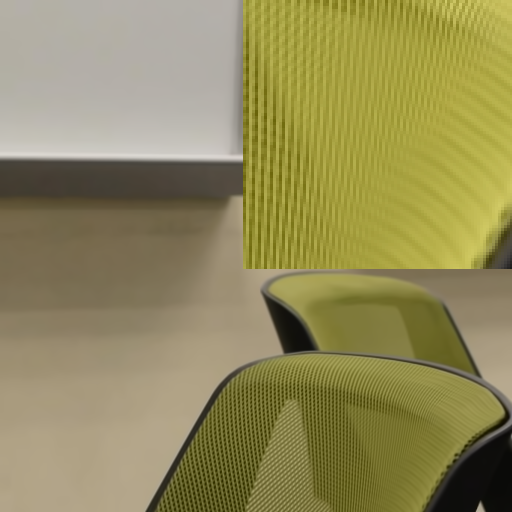}}
\subfigure[CMSt-SVD (37.80)]{
\label{Fig4}
\includegraphics[width=1.1in]{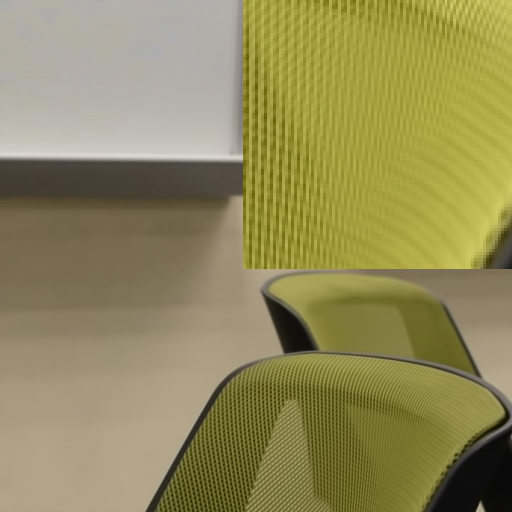}}
\subfigure[TWSC (37.16)]{
\label{Fig4}
\includegraphics[width=1.1in]{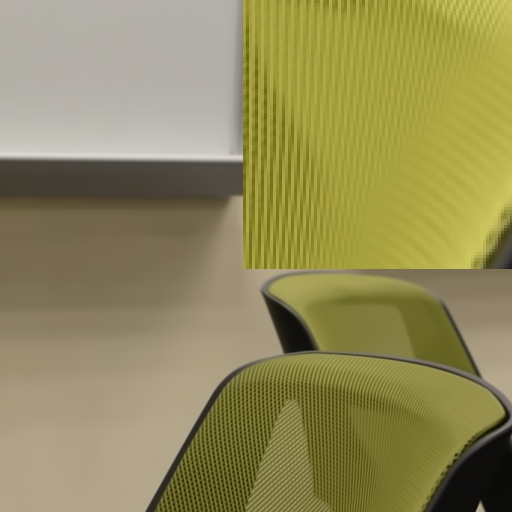}}
\subfigure[NLH (37.14)]{
\label{Fig4}
\includegraphics[width=1.1in]{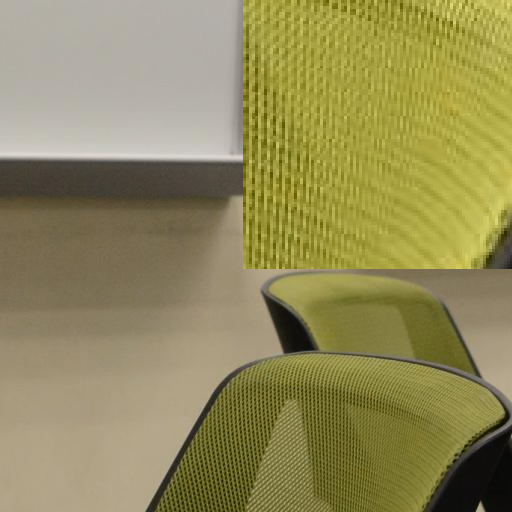}}

\caption{Visual evaluation of compared methods (PSNR) on PolyU dataset.}
\label{Fig_Xu_100}
\end{figure*}

%% file: Fig_my_own.tex
\begin{figure*}[htbp]
\graphicspath{{Figs/combined_my_own/}}
\centering
\subfigure[Mean]{
\label{Fig4}
\includegraphics[width=1.1in]{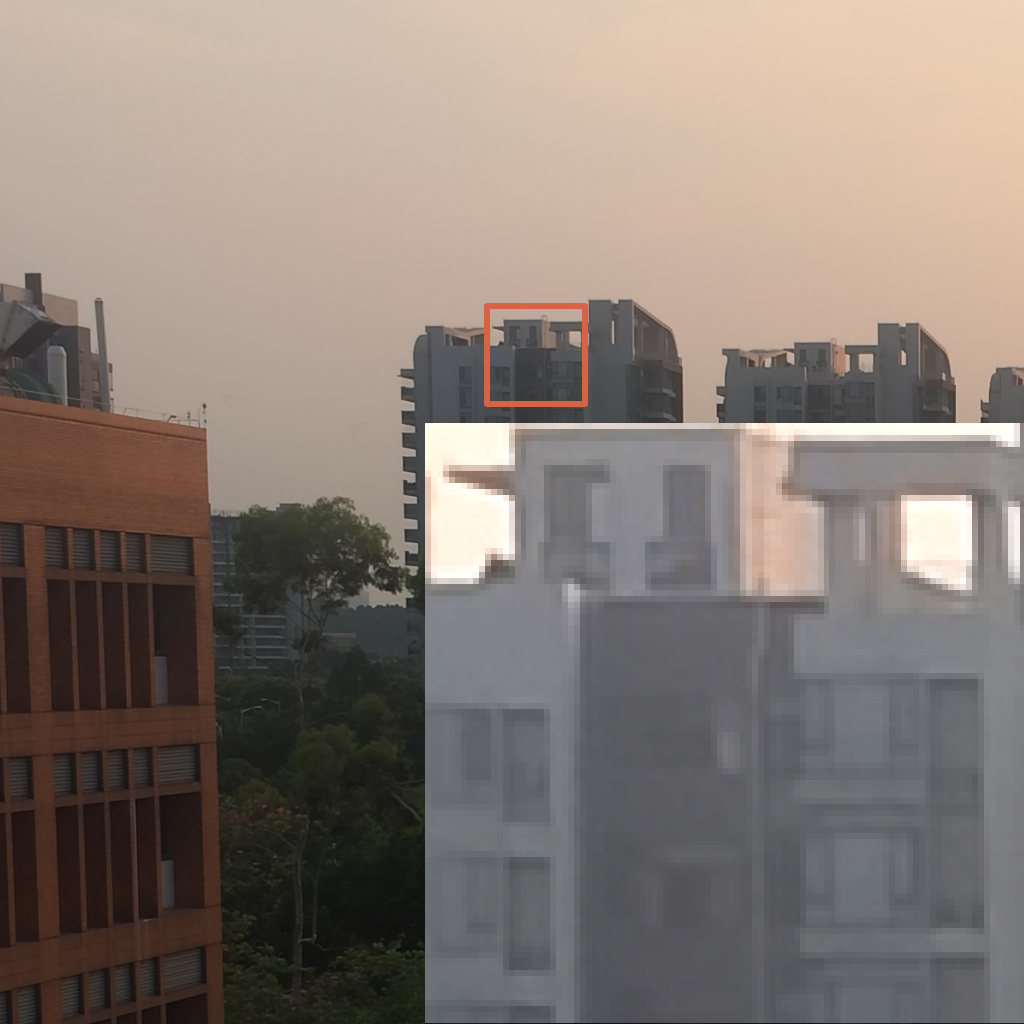}}
\subfigure[Noisy (41.23)]{
\label{Fig4}
\includegraphics[width=1.1in]{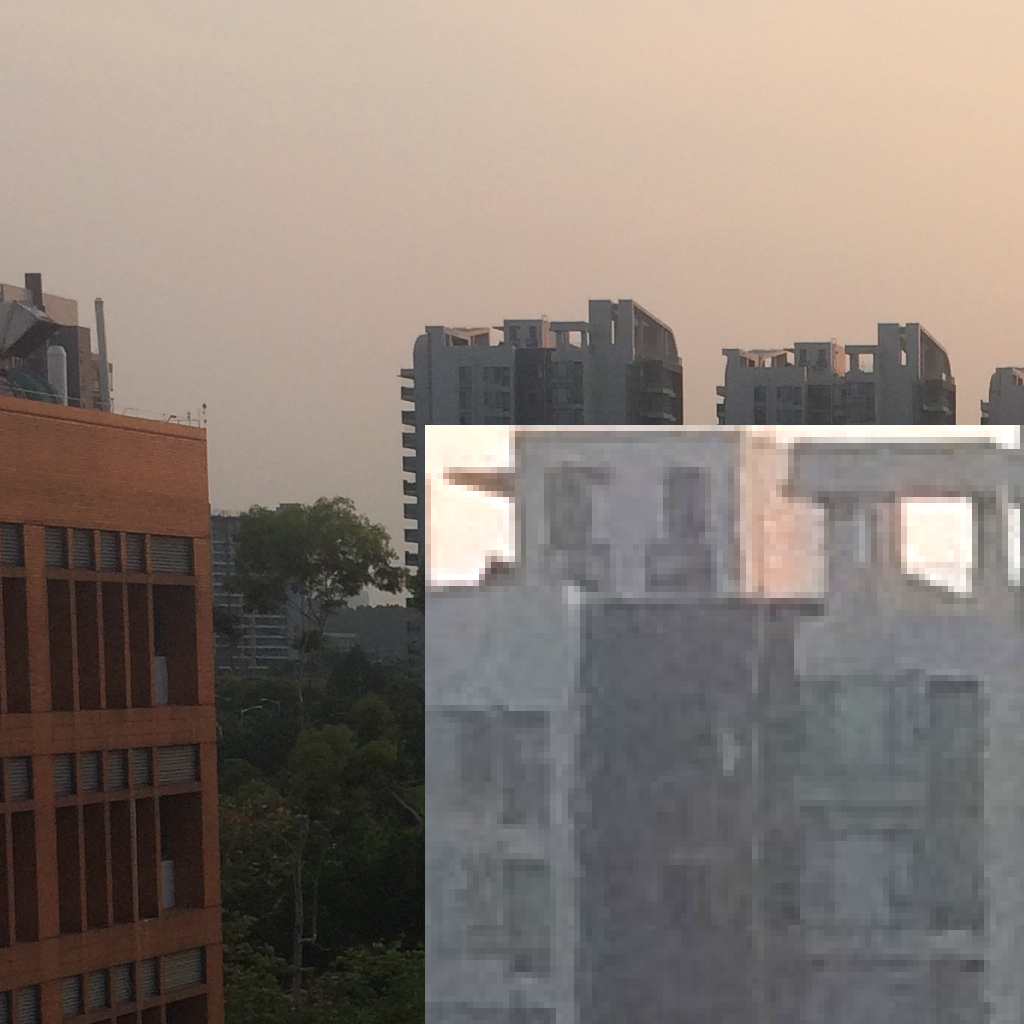}}
\subfigure[NI (42.89)]{
\label{Fig4}
\includegraphics[width=1.1in]{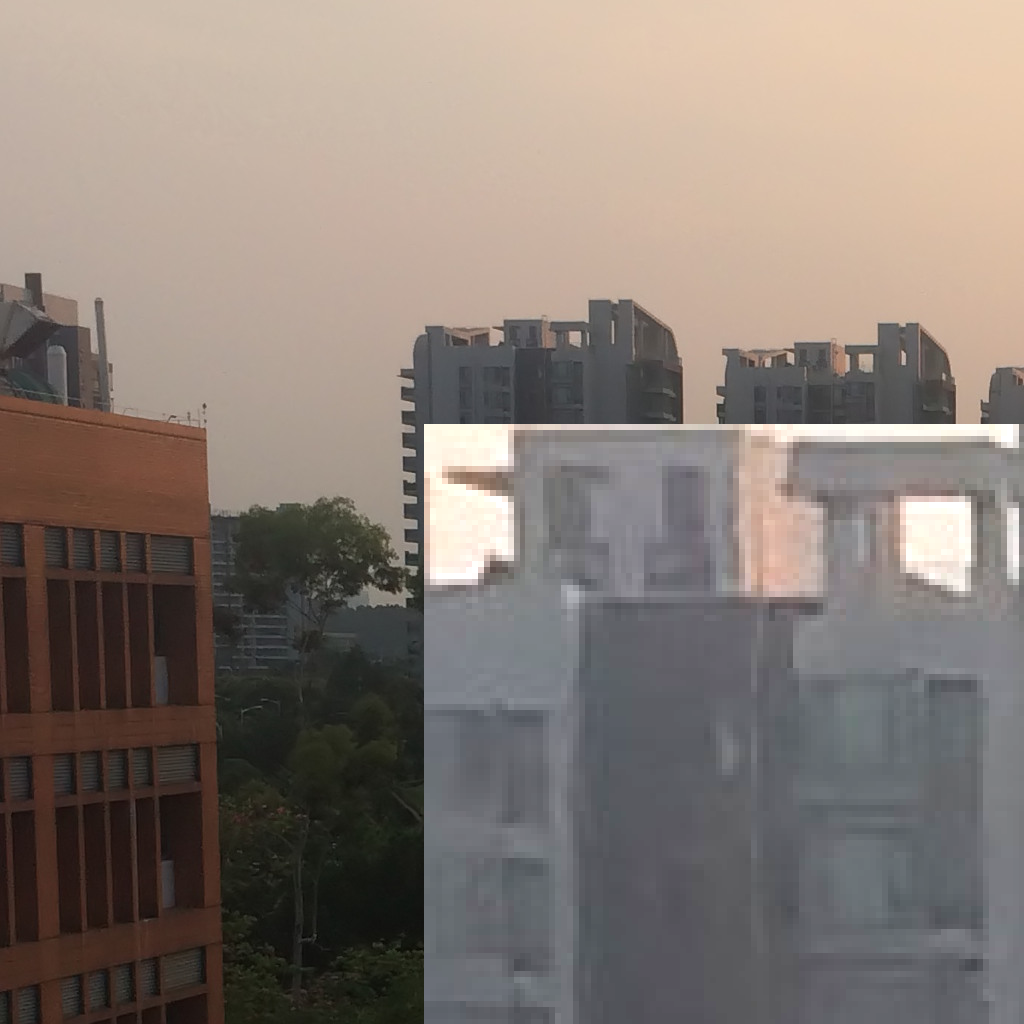}}
\subfigure[DeNoiseAI (41.06)]{
\label{Fig4}
\includegraphics[width=1.1in]{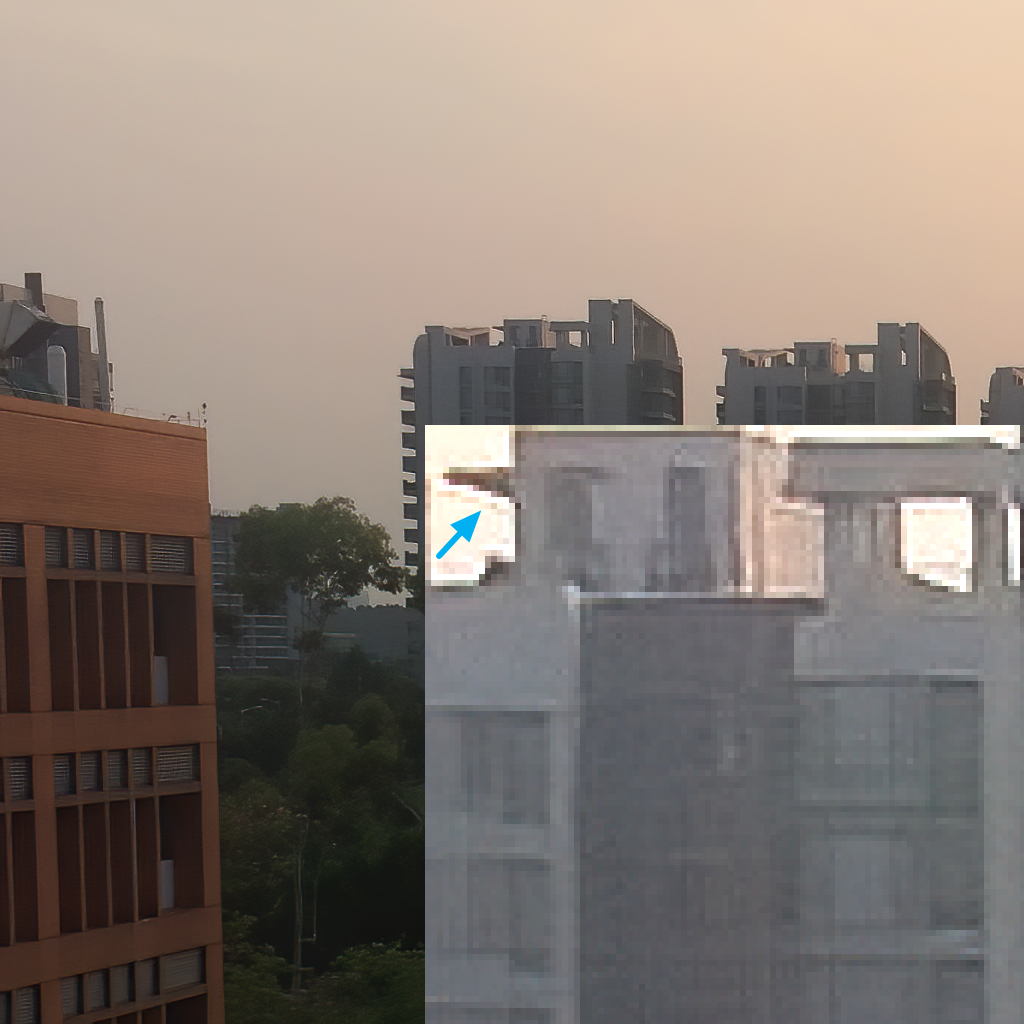}}
\subfigure[DBF (42.66)]{
\label{Fig4}
\includegraphics[width=1.1in]{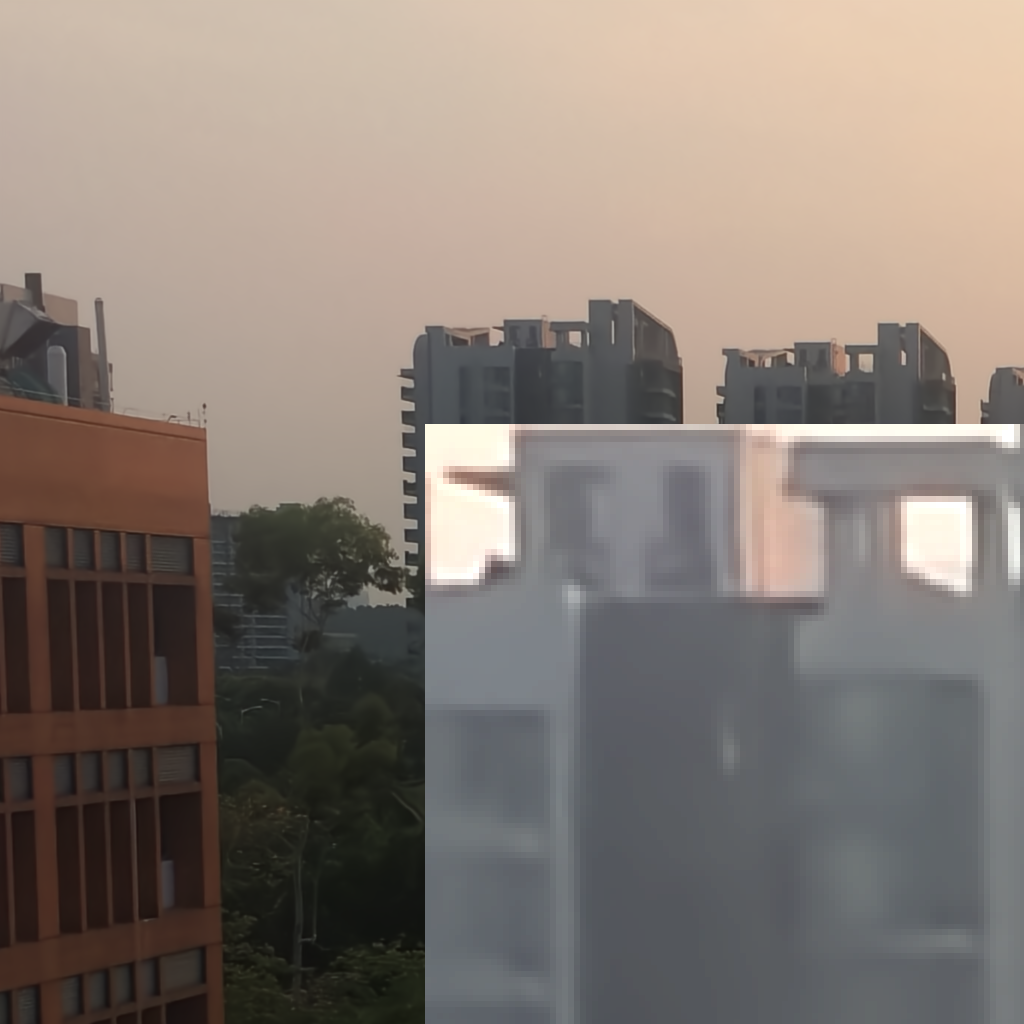}}
\subfigure[DRUNet (44.47)]{
\label{Fig4}
\includegraphics[width=1.1in]{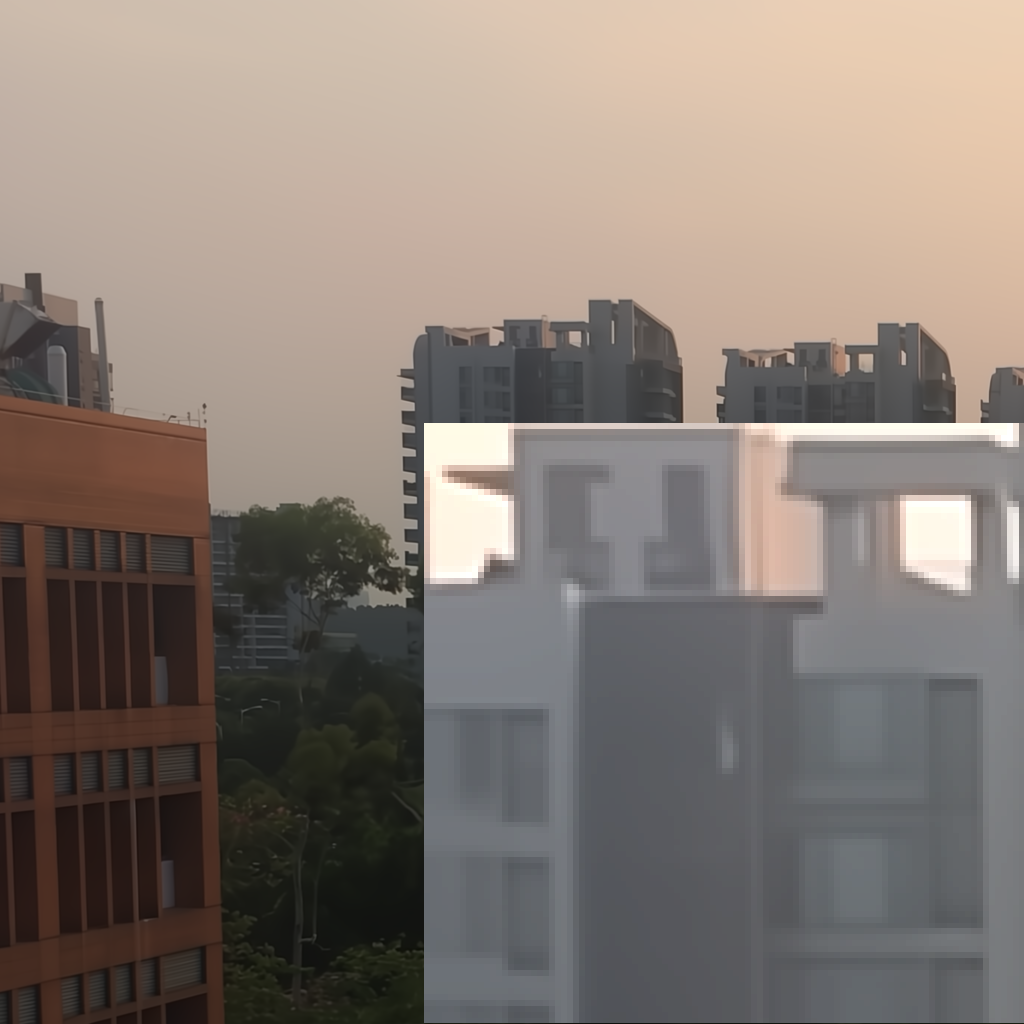}}\\
\subfigure[FFDNet (44.77)]{
\label{Fig4}
\includegraphics[width=1.1in]{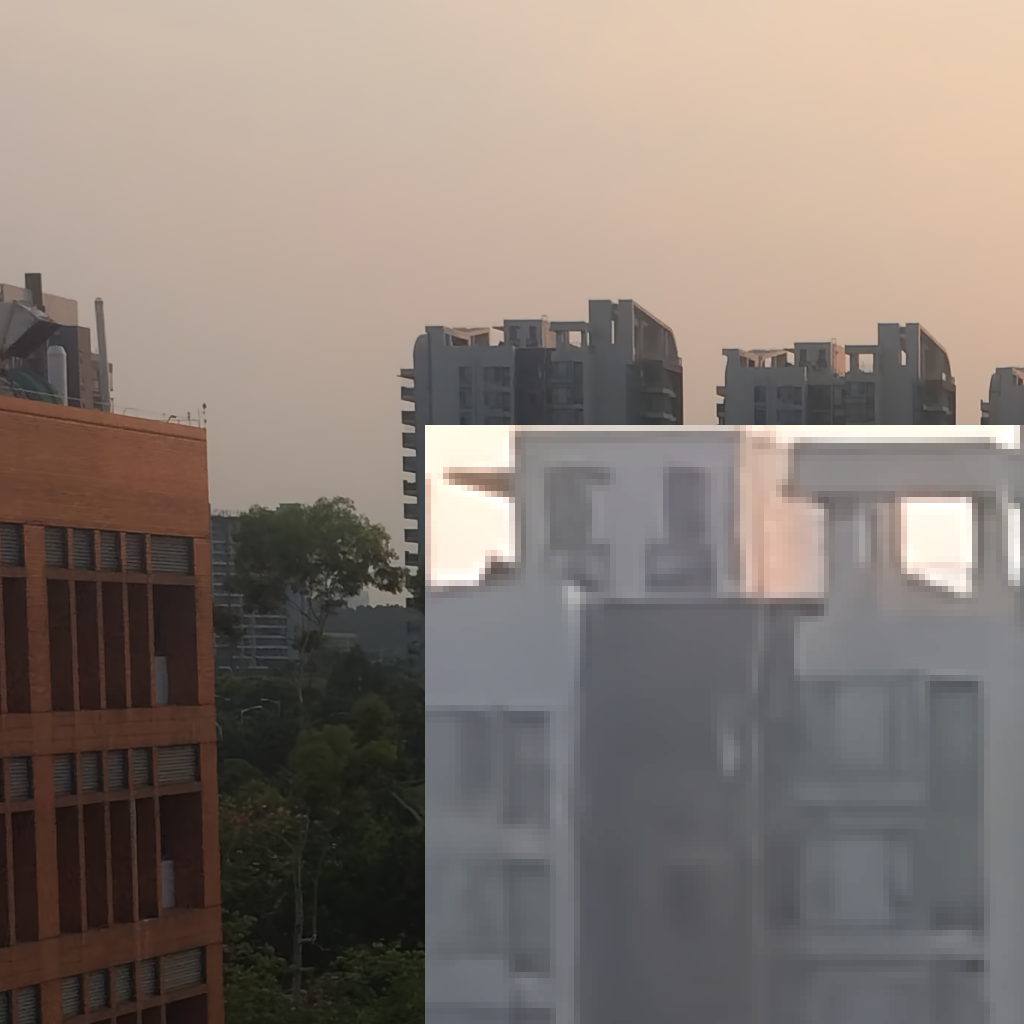}}
\subfigure[FCCF (43.40)]{
\label{Fig4}
\includegraphics[width=1.1in]{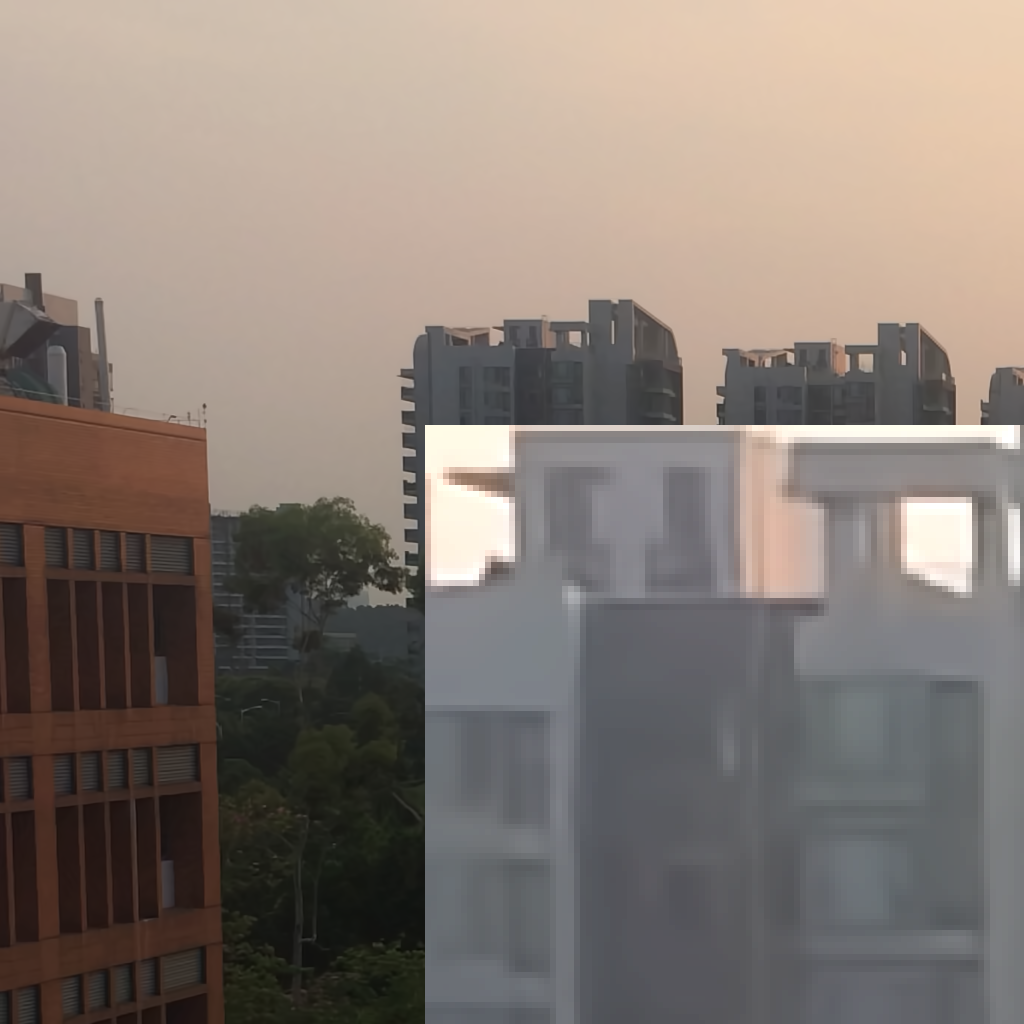}}
\subfigure[CBM3D2 (44.54)]{
\label{Fig4}
\includegraphics[width=1.1in]{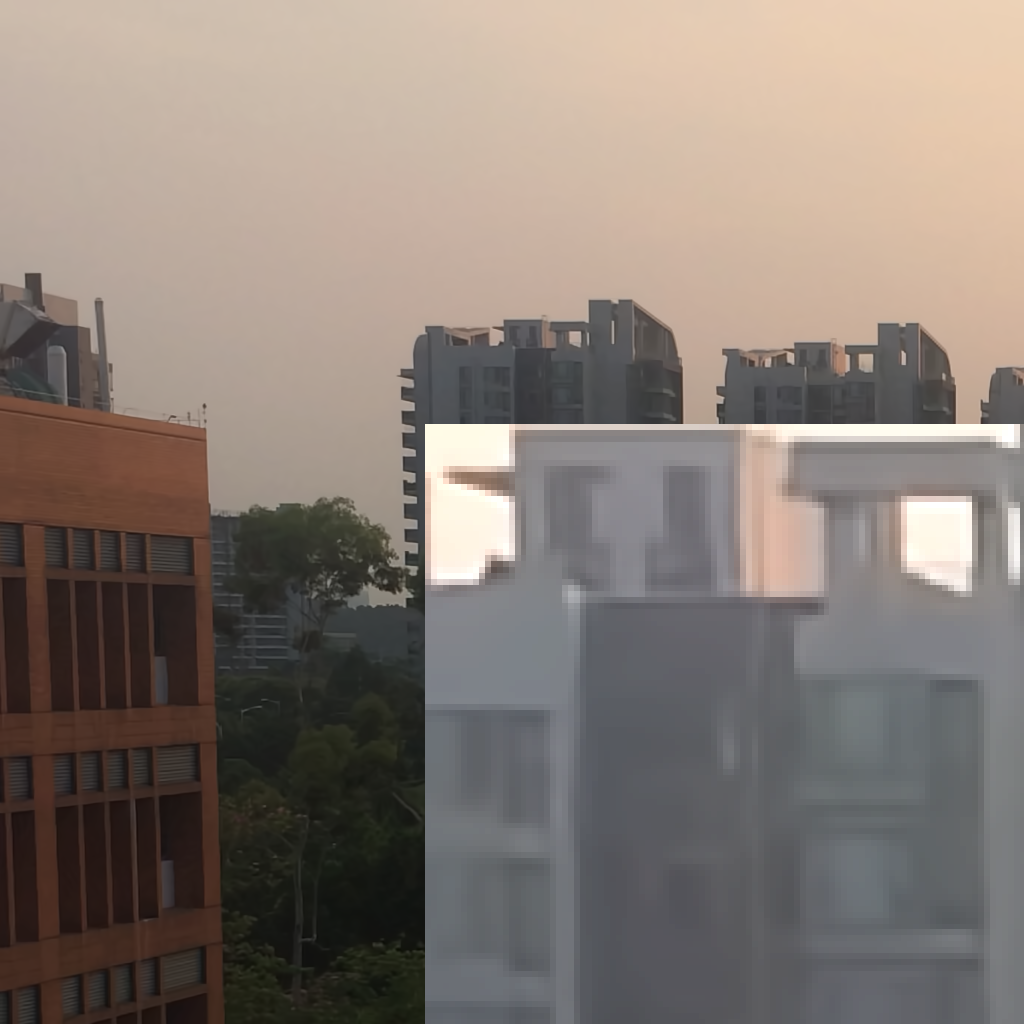}}
\subfigure[CMSt-SVD (44.81)]{
\label{Fig4}
\includegraphics[width=1.1in]{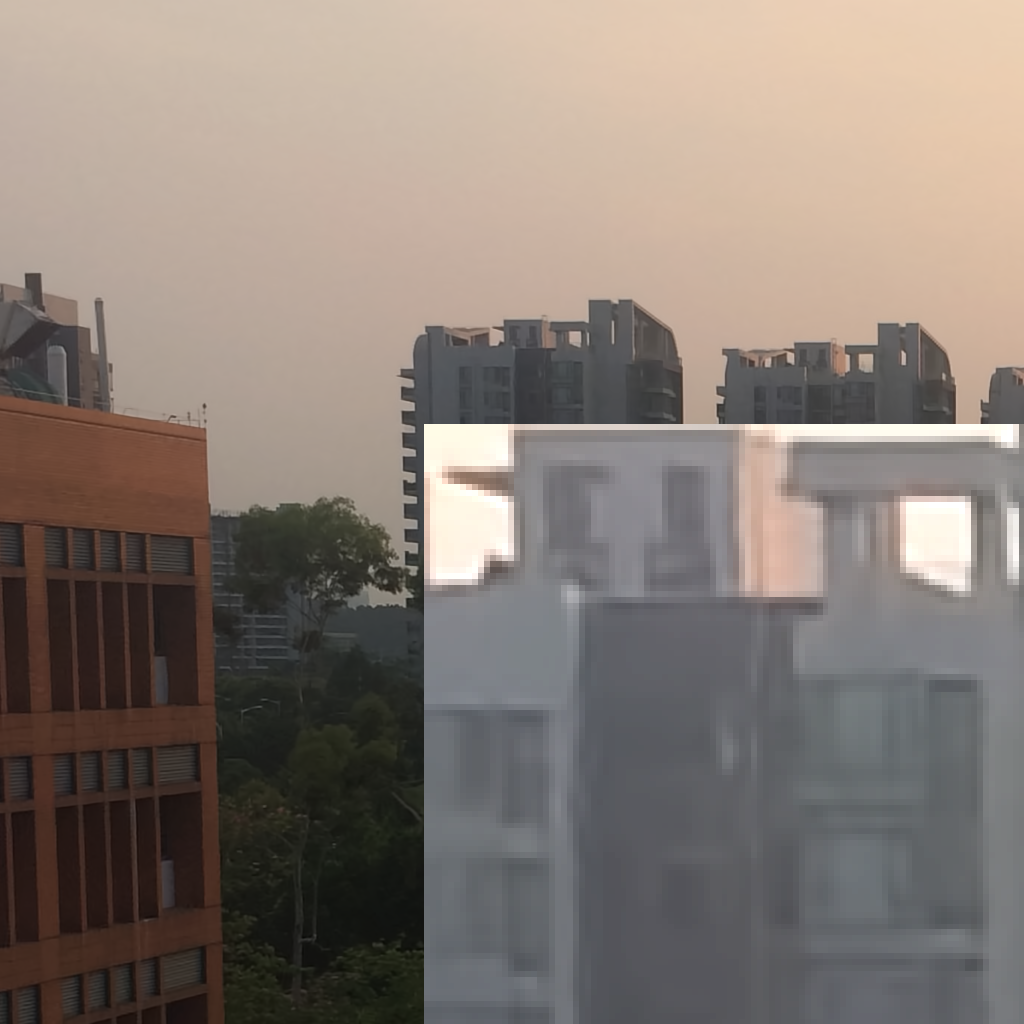}}
\subfigure[TWSC (42.60)]{
\label{Fig4}
\includegraphics[width=1.1in]{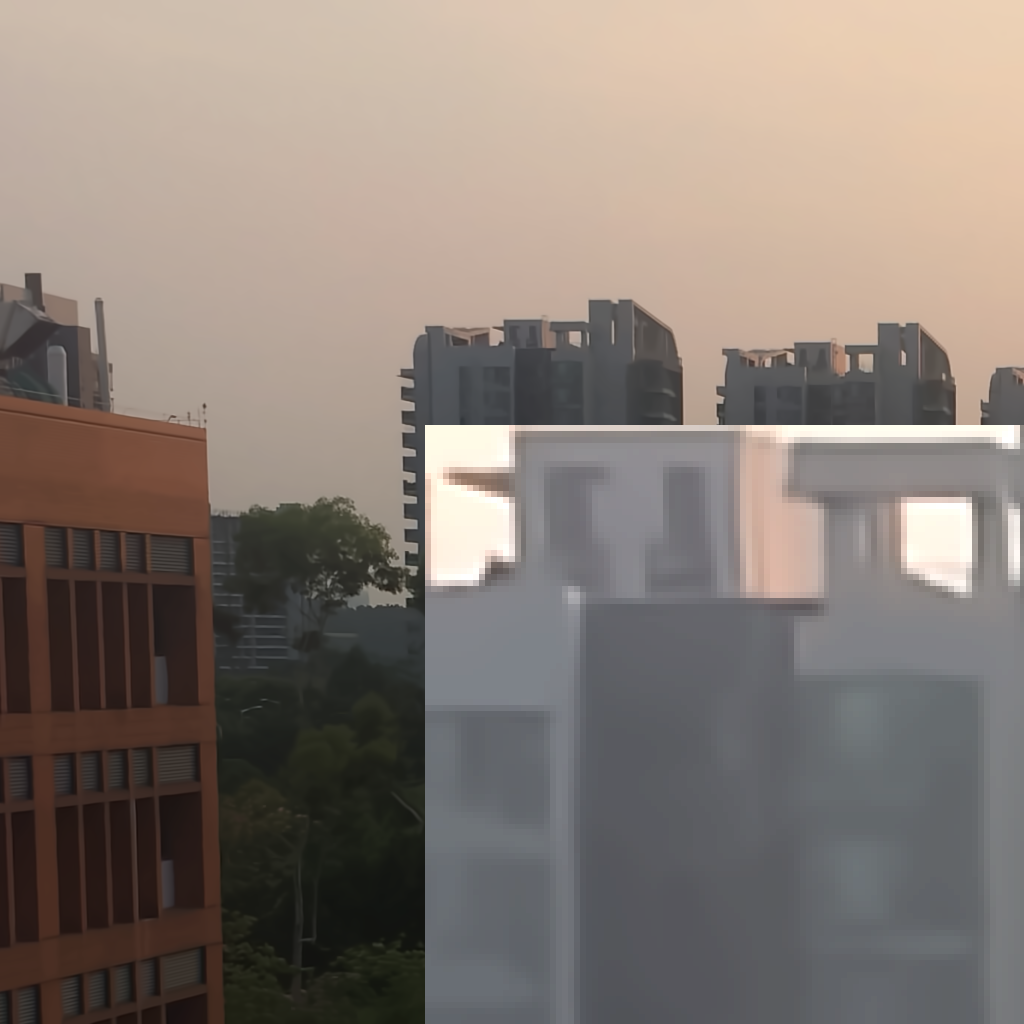}}
\subfigure[NLH (44.47)]{
\label{Fig4}
\includegraphics[width=1.1in]{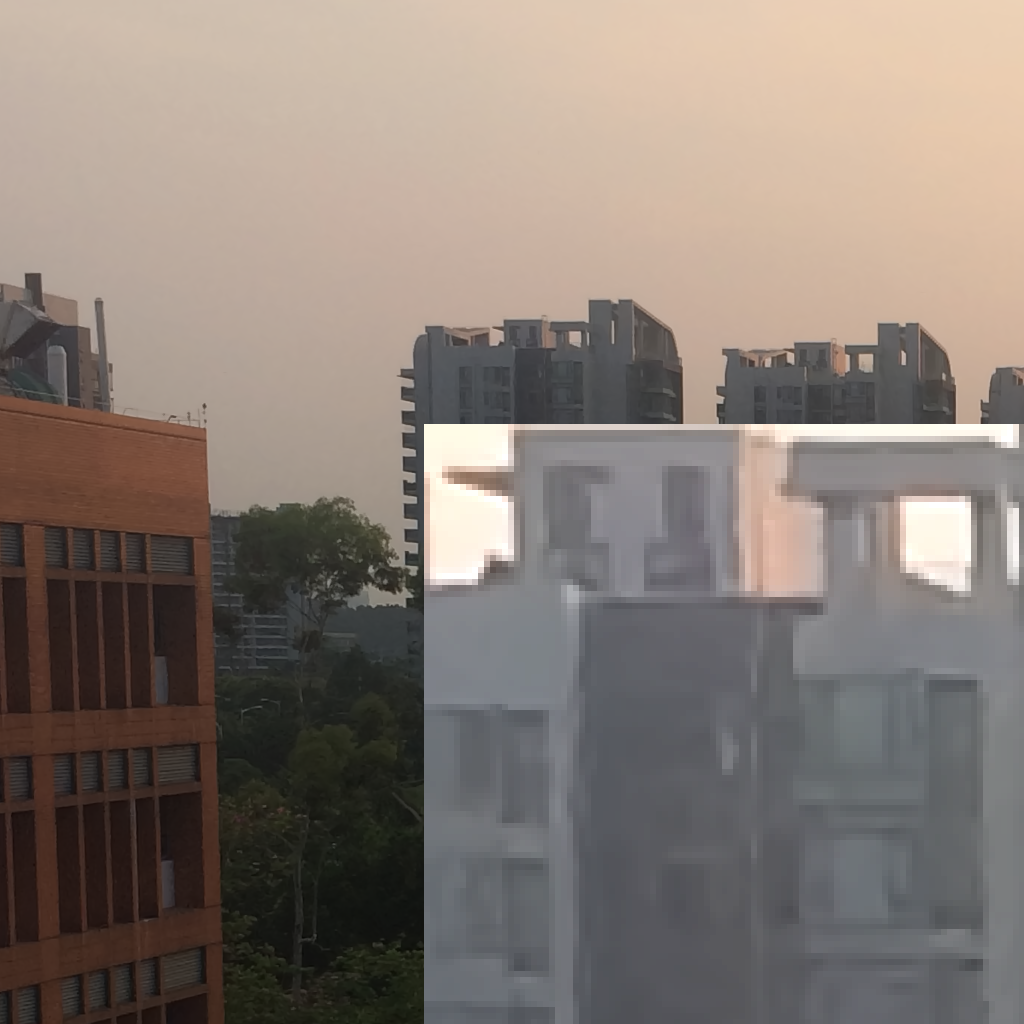}}

\caption{Visual evaluation of compared methods (PSNR) on our IOCI's IPHONE 5S dataset.}
\label{Fig_my_own}
\end{figure*}

%% file: Table_video_psnr_ssim.tex
\begin{table}[htbp]
\scriptsize
  \centering
  \caption{PSNR and SSIM values of compared methods on real-world color video dataset. Average time per frame is also included. The best results are bolded. 'FC' and 'BC' represent front camera and back camera, respectively.}
  \scalebox{0.89}{
  \renewcommand{\arraystretch}{0.99}
    \begin{tabular}{cccccc}
    \toprule
    \multirow{2}[4]{*}{Dataset} & \multicolumn{5}{c}{Method} \\
\cmidrule{2-6}          & Noisy & CVMSt-SVD & VBM4D1 & FastDVDNet & VNLNet \\
    \midrule
    \multirow{2}[4]{*}{HUAWEI\_BC} & 38.11  & 40.80  & 41.19  & \textbf{41.26} & \textbf{41.35} \\
\cmidrule{2-6}          & 0.9593  & 0.9834  & 0.9830  & 0.9857  & \textbf{0.9868} \\
    \midrule
    \multirow{2}[4]{*}{HUAWEI\_FC} & 35.58  & \textbf{38.71} & \textbf{38.76} & 38.03  & \textbf{38.79} \\
\cmidrule{2-6}          & 0.9413  & 0.9780  & 0.9785  & 0.9776  & \textbf{0.9809} \\
    \midrule
    \multirow{2}[4]{*}{OPPO\_BC} & 32.06  & \textbf{33.44} & 33.26  & 33.05  & \textbf{33.56} \\
\cmidrule{2-6}          & 0.9071  & 0.9508  & 0.9456  & 0.9476  & \textbf{0.9544} \\
    \midrule
    \multirow{2}[4]{*}{OPPO\_FC} & 36.90  & 39.66  & 39.56  & 39.06  & \textbf{40.11} \\
\cmidrule{2-6}          & 0.9447  & 0.9791  & 0.9785  & 0.9751  & \textbf{0.9823} \\
    \midrule
    Implementation & /     & MEX   & MEX   & GPU   & GPU \\
    \midrule
    Time/Frame (s) & /     & 1.86  & 1.29  & \textbf{0.08}  & 0.66 \\
    \bottomrule
    \end{tabular}%
    }
  \label{Table_video_psnr_ssim}%
\end{table}%

%% file: Fig_color_video.tex
\begin{figure}[htbp]
\graphicspath{{Figs/combined_real_video/combine/}}
\centering
\subfigure[Mean]{
\label{Fig4}
\includegraphics[width=1.08in]{Clean}}
\subfigure[Noisy]{
\label{Fig4}
\includegraphics[width=1.08in]{Noisy}}
\subfigure[CVMSt-SVD]{
\label{Fig4}
\includegraphics[width=1.08in]{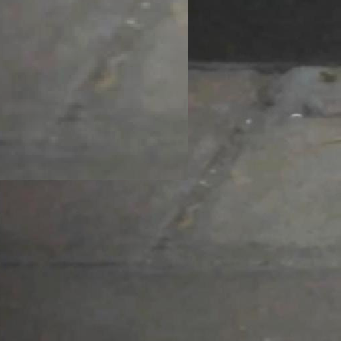}}\\
\subfigure[VBM4D1]{
\label{Fig4}
\includegraphics[width=1.08in]{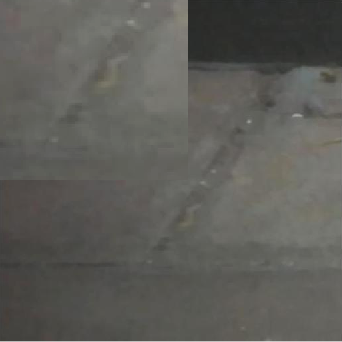}}
\subfigure[FastDVDNet]{
\label{Fig4}
\includegraphics[width=1.08in]{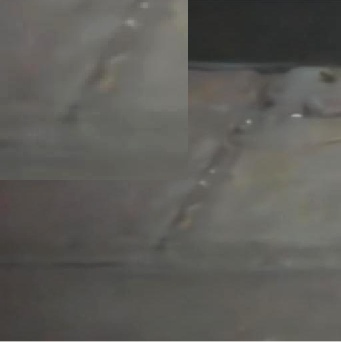}}
\subfigure[Vnlnet]{
\label{Fig4}
\includegraphics[width=1.08in]{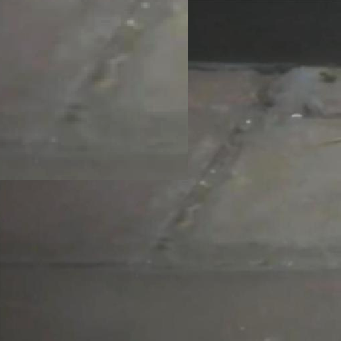}}\\

\caption{Visual evaluation of color video denoising methods on our IOCV dataset.}
\label{Fig_color_video}
\end{figure}

%% file: Table_color_video_rating.tex
% Table generated by Excel2LaTeX from sheet 'Investigation'
%\begin{table}[htbp]
%\vspace{-0.1cm}
%\scriptsize
%  \centering
%  \caption{Human rating results of sequences generated by different methods on our IOCV dataset. The top two results are bolded.}
%  \scalebox{0.88}{
%  \renewcommand{\arraystretch}{0.89}
%    \begin{tabular}{ccccccc}
%    \multicolumn{7}{c}{Ratings} \\
%    \midrule
%    \# Image & Mean  & Noisy & CVMSt-SVD & VBM4D & Fastdvdnet & Vnlnet \\
%    \midrule
%    1     & \textbf{9}     & 0     & 1     & 1     & 5     & 3 \\
%    \midrule
%    2     & \textbf{8}     & 1     & 0     & 3     & \textbf{7}     & 6 \\
%    \midrule
%    3     & \textbf{10}    & 0     & 0     & \textbf{8}     & 3     & 5 \\
%    \midrule
%    4     & \textbf{8}     & 0     & 0     & 4     & 7     & \textbf{8} \\
%    \midrule
%    5     & \textbf{10}    & 0     & 0     & 4     & 4     & \textbf{7} \\
%    \midrule
%    6     & \textbf{10}    & 0     & 0     & 5     & 6     & \textbf{7} \\
%    \midrule
%    7     & \textbf{10}    & 0     & 0     & 5     & \textbf{8}     & 5 \\
%    \midrule
%    8     & \textbf{10}    & 2     & 2     & \textbf{8}     & 5     & 6 \\
%    \midrule
%    9     & \textbf{9}     & 0     & 2     & 4     & 2     & \textbf{7} \\
%    \midrule
%    10    & \textbf{4}     & 4     & 4     & 4     & 5     & \textbf{8} \\
%    \midrule
%    Average & 8.80  & 0.70  & 0.90  & 4.60  & 5.20  & 6.20  \\
%    \bottomrule
%    \end{tabular}%
%    }
%  \label{Table_color_video_rating}%
%\end{table}%

\begin{table}[htbp]
\vspace{-0.1cm}
\scriptsize
  \centering
  \caption{Human rating results of sequences generated by different methods on our IOCV dataset. The top two results are bolded.}
  \scalebox{0.88}{
  \renewcommand{\arraystretch}{0.89}
    \begin{tabular}{ccccccc}
    \multicolumn{7}{c}{Ratings} \\
    \midrule
    \# Image & Mean  & Noisy & CVMSt-SVD & VBM4D1 & FastDVDNet & VNLNet \\
    \midrule
    1     & \textbf{9} & 0     & 1     & 2     & \textbf{5} & 3 \\
    \midrule
    2     & \textbf{8} & 1     & 0     & 4     & \textbf{7} & 6 \\
    \midrule
    3     & \textbf{10} & 0     & 0     & \textbf{8} & 3     & 5 \\
    \midrule
    4     & \textbf{8} & 0     & 0     & 4     & 7     & \textbf{8} \\
    \midrule
    5     & \textbf{10} & 0     & 0     & 4     & 4     & \textbf{7} \\
    \midrule
    6     & \textbf{10} & 0     & 0     & 5     & 6     & \textbf{7} \\
    \midrule
    7     & \textbf{10} & 0     & 0     & 5     & \textbf{8} & 5 \\
    \midrule
    8     & \textbf{10} & 2     & 2     & \textbf{8} & 5     & 6 \\
    \midrule
    9     & \textbf{9} & 0     & 2     & 4     & 2     & \textbf{7} \\
    \midrule
    10    & 4     & 4     & 4     & 4     & 5     & \textbf{8} \\
    \midrule
    Average & \textbf{8.80 } & 0.70  & 0.90  & 4.80  & 5.20  & \textbf{6.20 } \\
    \bottomrule
    \end{tabular}%
    }
  \label{Table_color_video_rating}%
\end{table}%

%% file: Table_msi_psnr_ssim.tex
\begin{table*}[ht]
\tiny
  \centering
  \caption{Comparison of quantitative results and computational time (minutes) under i.i.d Gaussian noise $\sigma = \{10, 30, 50, 100\}$ on CAVE dataset.}
  \renewcommand{\arraystretch}{0.8}
    \begin{tabular}{cccccccccccccccccc}
    \toprule
    \multirow{2}[4]{*}{Methods} & \multicolumn{4}{c}{$\sigma$ = 10} & \multicolumn{4}{c}{$\sigma$ = 30} & \multicolumn{4}{c}{$\sigma$ = 50} & \multicolumn{4}{c}{$\sigma$ = 100} & \multirow{2}[4]{*}{Time (m)} \\
\cmidrule{2-17}          & PSNR  & SSIM  & ERGAS & SAM   & PSNR  & SSIM  & ERGAS & SAM   & PSNR  & SSIM  & ERGAS & SAM   & PSNR  & SSIM  & ERGAS & SAM   &  \\
    \midrule
    Noisy & 28.13  & 0.4371 & 236.40  & 0.7199 & 18.59  & 0.1069 & 676.01  & 1.0085 & 14.15 & 0.0475  & 1126.7 & 1.1461 & 8.13  & 0.0136  & 2253.4 & 1.3074  & / \\
    \midrule
    PARAFAC & 35.39  & 0.8759 & 108.78  & 0.2363 & 33.65  & 0.8294 & 125.13  & 0.3291 & 31.52 & 0.7393  & 154.19 & 0.4351 & 27.13  & 0.4637  & 250.68 & 0.6681  & 3.9 \\
    \midrule
    LRTA  & 41.36  & 0.9499 & 49.53  & 0.1718 & 36.06  & 0.8775 & 90.50  & 0.2446 & 33.52 & 0.8201  & 121.15 & 0.2897 & 30.06  & 0.7138  & 180.03 & 0.3649  & \textbf{0.5} \\
    \midrule
    LRMR  & 39.27  & 0.9094 & 64.81  & 0.3343 & 31.36  & 0.6451 & 157.65  & 0.6021 & 26.67 & 0.4000  & 264.28 & 0.7534 & 26.67  & 0.1850  & 469.26 & 0.9306  & 4.1 \\
    \midrule
    LLRGTV & 38.96  & 0.9384 & 71.54  & 0.2292 & 33.56  & 0.7862 & 122.45  & 0.4756 & 30.33 & 0.6375  & 172.22 & 0.6386 & 25.62  & 0.3917  & 288.78 & 0.8593  & 3.8 \\
    \midrule
    4DHOSVD1 & 45.43  & 0.9811 & 30.83  & 0.1084 & 39.78  & 0.9336 & 59.12  & 0.2272 & 36.82 & 0.8722  & 83.36 & 0.3385 & 32.66  & 0.7307  & 134.34 & 0.5599  & 4.9 \\
    \midrule
    BM4D1 & 42.76  & 0.9716  & 42.19  & 0.1585  & 37.13  & 0.9235  & 82.33  & 0.2910  & 34.55  & 0.8803  & 109.43  & 0.3591  & 31.17  & 0.7763  & 158.83  & 0.4751  & 2.1 \\
    \midrule
    BM4D2 & 44.61  & 0.9784 & 33.32  & 0.1289 & 38.80  & 0.9283 & 65.23  & 0.2579 & 35.90 & 0.8685  & 91.19 & 0.3557 & 31.84  & 0.7197  & 144.91 & 0.5160  & 8.8 \\
    \midrule
    TDL   & 44.41  & 0.9797 & 34.32  & 0.1048 & 39.07  & 0.9493 & 63.18  & 0.1493 & 36.46 & 0.9171  & 85.24 & 0.2008 & 32.92  & 0.8284  & 128.15 & 0.3132  & 0.9 \\
    \midrule
    ITSReg & 45.77  & 0.9802 & 30.53  & 0.1086 & 40.51  & 0.9488 & 53.05  & 0.1374 & 37.75 & 0.9271  & 70.16 & 0.1619 & 33.01  & 0.8648  & 120.77 & 0.2376  & 5.2 \\
    \midrule
    FastHyDe & 42.87  & 0.9763 & 42.90  & 0.1398 & 37.22  & 0.9151 & 81.44  & 0.3123 & 34.56 & 0.8471  & 108.75 & 0.4395 & 30.62  & 0.6717  & 168.92 & 0.6096  & 0.8 \\
    \midrule
    HSI-SDeCNN & 41.40  & 0.9624 & 52.04  & 0.1523 & 37.57  & 0.9247  & 78.28 & 0.1959  & 35.49  & 0.8947  & 98.54 & 0.2141  & 32.51  & 0.8309  & 136.16 & 0.2748  & 0.98 \\
    \midrule
    QRNN3D & 40.36  & 0.9672  & 65.31  & 0.1926  & 38.48  & 0.9496  & 76.21  & 0.2320  & 37.02  & 0.9314  & 87.15  & 0.2741  & 30.33  & 0.7128  & 172.68  & 0.5161  & 0.01  \\
    \midrule
    LLRT  & \textbf{46.60} & 0.9868 & \textbf{26.75} & \textbf{0.0842} & 41.49  & 0.9681 & 48.50 & \textbf{0.1221} & 38.65 & 0.9482  & 67.56 & \textbf{0.1551} & \textbf{35.39} & \textbf{0.9154} & \textbf{99.37} & \textbf{0.1962} & 43.8 \\
    \midrule
    MSt-SVD & 45.20  & 0.9814 & 32.05  & 0.1064 & 40.23  & 0.9530  & 56.26  & 0.1737 & 37.73 & 0.9285  & 75.03 & 0.2231 & 34.20  & 0.8800  & 115.24 & 0.3142  & 1.5 \\
    \midrule
    NGMeet & \textbf{46.58} & \textbf{0.9881} & 27.27  & 0.0853 & \textbf{41.60} & \textbf{0.9692} & \textbf{48.14} & 0.1238 & \textbf{38.88} & \textbf{0.9507} & \textbf{65.64} & 0.1616  & 34.97  & 0.9072  & 101.98 & 0.2506  & 4.6 \\
    \midrule
    LTDL  & 45.90  & 0.9859  & 29.06  & 0.0925 & 41.20  & 0.9644  & 49.73  & 0.1301 & 38.63  & 0.9459  & 66.68  & 0.1650  & 34.87  & 0.9015  & 103.65  & 0.2353  & 60.2 \\
    \bottomrule
    \end{tabular}%
  \label{Table_msi_psnr_ssim}%
\end{table*}%

%% file: Table_ICVL.tex
%\begin{table}[htbp]
%  \centering
%  \caption{Average PSNR/SSIM of representative denoising methods under i.i.d Gaussian noise $\sigma \in \{10,20,40\}$ with ICVL dataset.}
%    \begin{tabular}{cccc}
%    \toprule
%    Methods & $\sigma = 10$    & $\sigma = 20$    & $\sigma = 40$ \\
%    \midrule
%    MSt-SVD & 46.58/0.9868 & 42.98/0.9746 & 39.32/0.9525 \\
%    \midrule
%    QRNN3D & 47.37/0.9908 & 45.08/0.9842 & 42.38/0.9731 \\
%    \bottomrule
%    \end{tabular}%
%  \label{Table_ICVL}%
%\end{table}%

% Table generated by Excel2LaTeX from sheet 'ICVL'
\begin{table}[htbp]
\scriptsize
  \centering
  \caption{Average PSNR/SSIM and computational time (minutes) of MSt-SVD and QRNN3D under i.i.d Gaussian noise $\sigma \in \{10,20,40\}$ on the first 20 MSI data of ICVL dataset.}
    \begin{tabular}{ccccc}
    \toprule
    Methods & $\sigma = 10$    & $\sigma = 20$    & $\sigma = 40$    & Time (m) \\
    \midrule
    MSt-SVD & 46.58/0.9868 & 42.98/0.9746 & 39.32/0.9525 & 10.1 \\
    \midrule
    QRNN3D & \textbf{47.37/0.9908} & \textbf{45.08/0.9842} & \textbf{42.38/0.9731} & \textbf{0.16} \\
    \bottomrule
    \end{tabular}%
  \label{Table_ICVL}%
\end{table}%

%% file: Fig_CAVE.tex
\begin{figure}[htbp]
\graphicspath{{Figs/combine_CAVE/}}
\centering
\subfigure[Clean]{
\label{Fig4}
\includegraphics[width=0.78in]{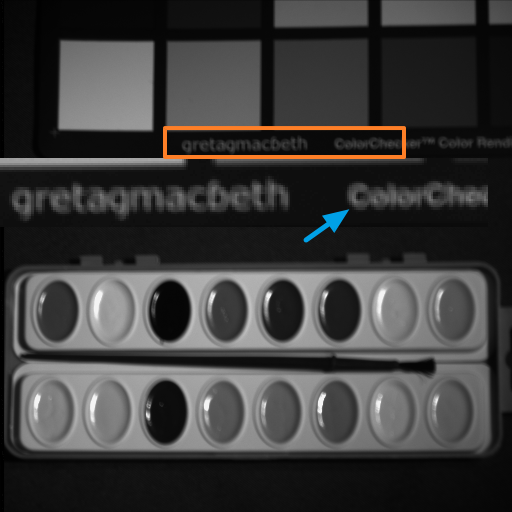}}
\subfigure[Noisy]{
\label{Fig4}
\includegraphics[width=0.78in]{Noisy}}
\subfigure[BM4D2]{
\label{Fig4}
\includegraphics[width=0.78in]{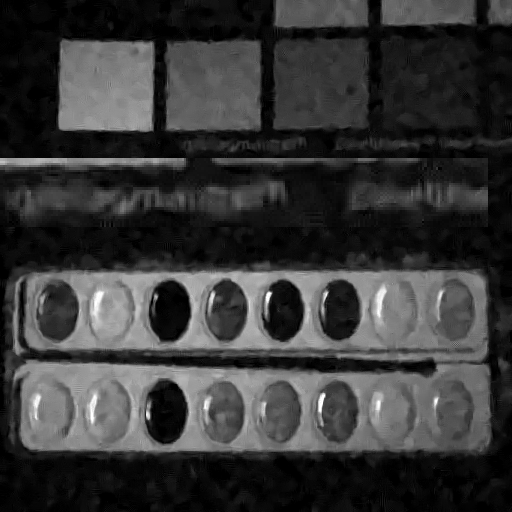}}
\subfigure[NGMeet]{
\label{Fig4}
\includegraphics[width=0.78in]{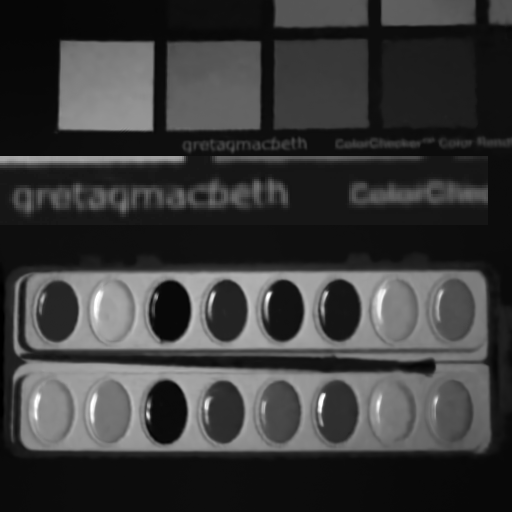}}\\

\subfigure[4DHOSVD1]{
\label{Fig4}
\includegraphics[width=0.78in]{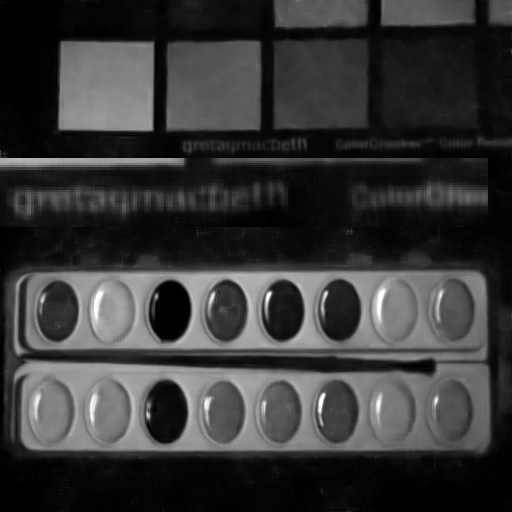}}
\subfigure[MSt-SVD]{
\label{Fig4}
\includegraphics[width=0.78in]{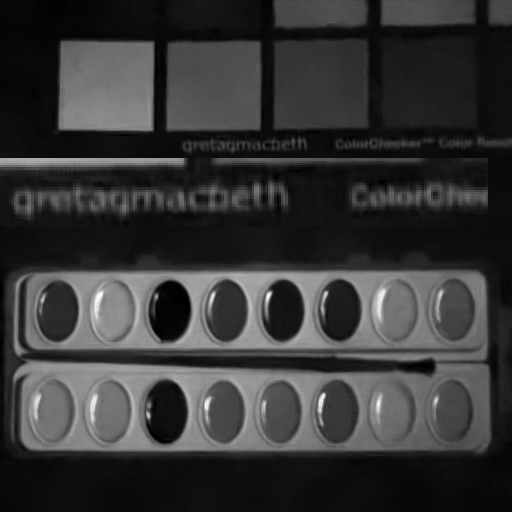}}
\subfigure[LTDL]{
\label{Fig4}
\includegraphics[width=0.78in]{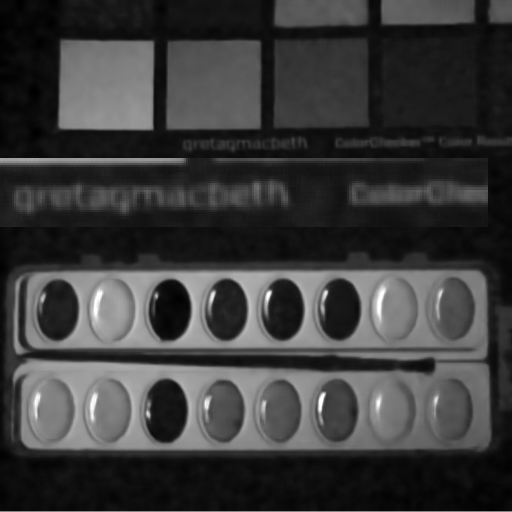}}
\subfigure[LLRT]{
\label{Fig4}
\includegraphics[width=0.78in]{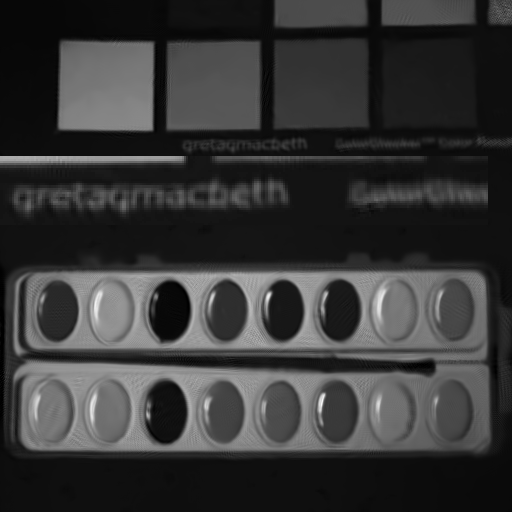}}

\caption{Visual evaluation of MSI denoising methods on CAVE dataset under noise level $\sigma = 100$.}
\label{Fig_CAVE}
\end{figure}

%% file: Fig_HHD.tex
\begin{figure}[htbp]
\graphicspath{{Figs/combine_HHD/Combine_new/}}
\centering
%\subfigure[Noisy]{
%\label{Fig4}
%\includegraphics[width=1.08in]{imga4_combined}}
%\subfigure[BM4D1]{
%\label{Fig4}
%\includegraphics[width=1.08in]{imga4_BM4D1_combined}}
%\subfigure[MSt-SVD]{
%\label{Fig4}
%\includegraphics[width=1.08in]{imga4_MStSVD_combined}}\\

%\subfigure[Noisy]{
%\label{Fig4}
%\includegraphics[width=1.08in]{imgc6_combined_marked}}
%\subfigure[BM4D1]{
%\label{Fig4}
%\includegraphics[width=1.08in]{imgc6_BM4D1_combined_marked}}
%\subfigure[MSt-SVD]{
%\label{Fig4}
%\includegraphics[width=1.08in]{imgc6_MStSVD_combined}}

\subfigure[Noisy]{
\label{Fig4}
\includegraphics[width=1.08in]{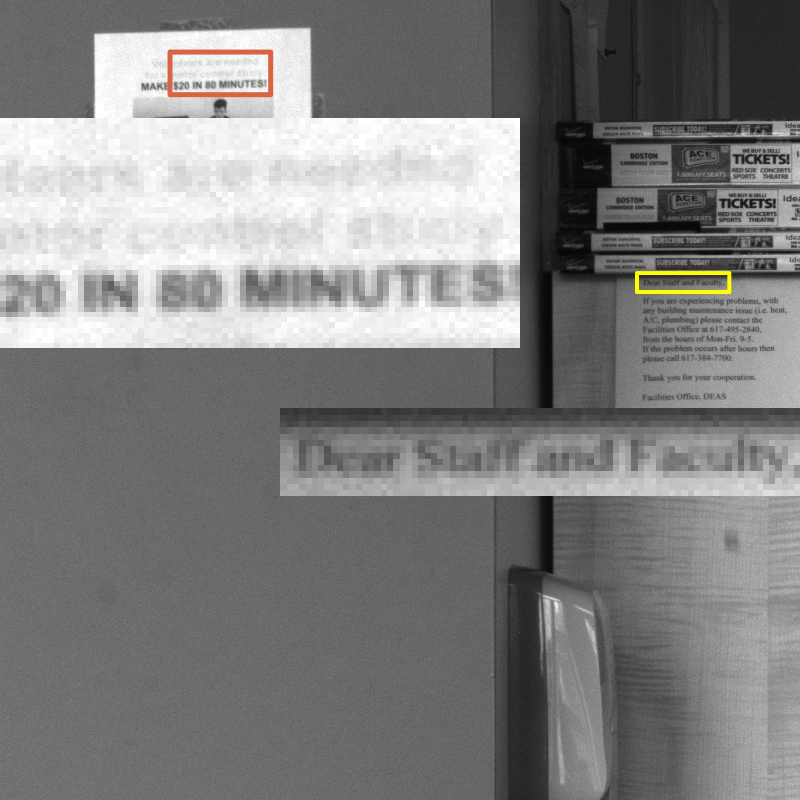}}
\subfigure[BM4D1]{
\label{Fig4}
\includegraphics[width=1.08in]{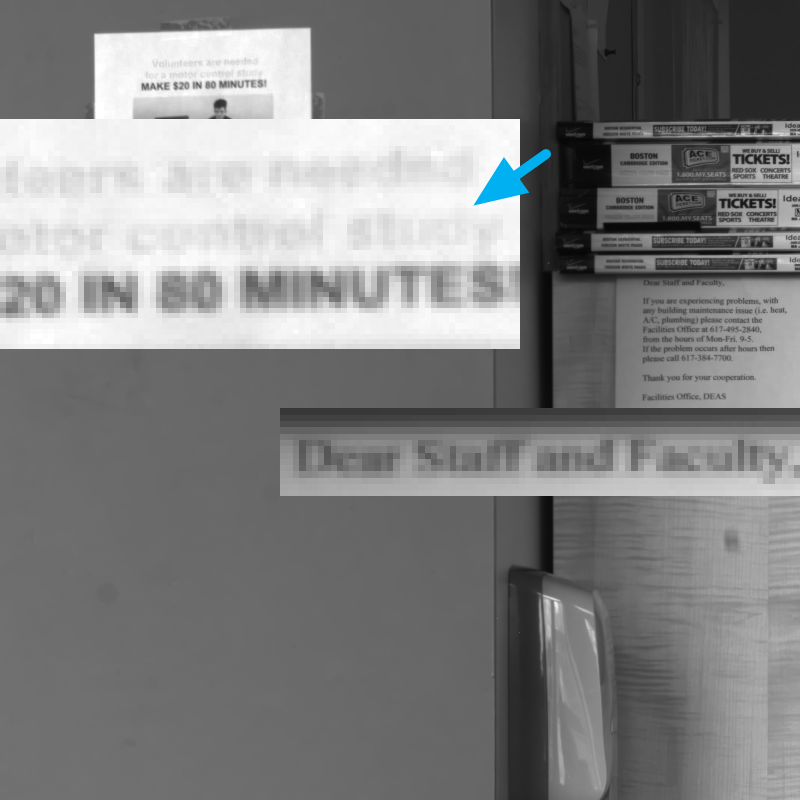}}
\subfigure[TDL]{
\label{Fig4}
\includegraphics[width=1.08in]{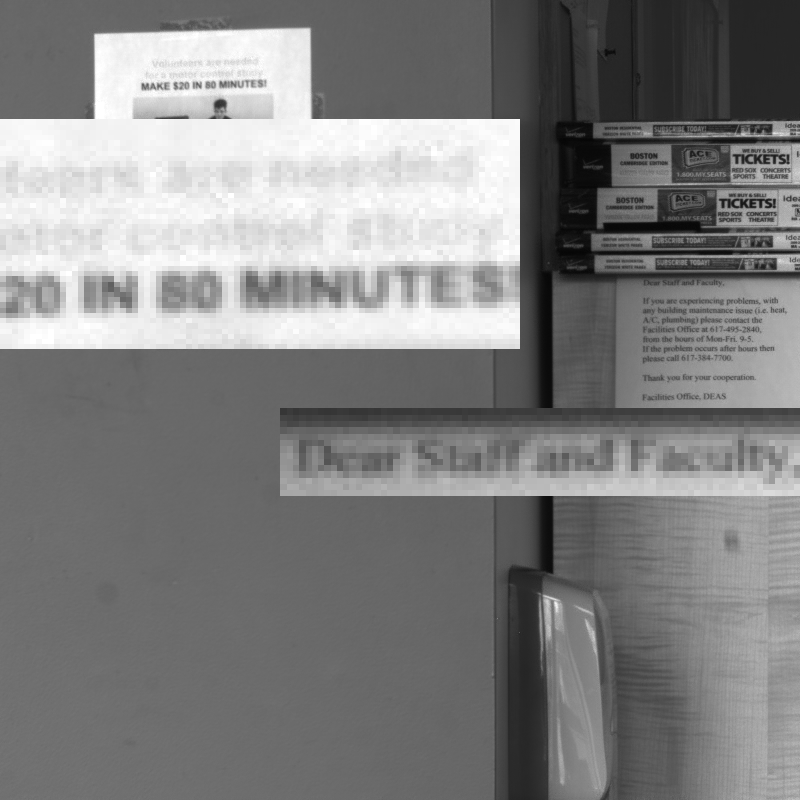}}\\
%\subfigure[FastHyDe]{
%\label{Fig4}
%\includegraphics[width=1.08in]{imgc6_FastHyDe_combine_new}}
\subfigure[QRNN3D]{
\label{Fig4}
\includegraphics[width=1.08in]{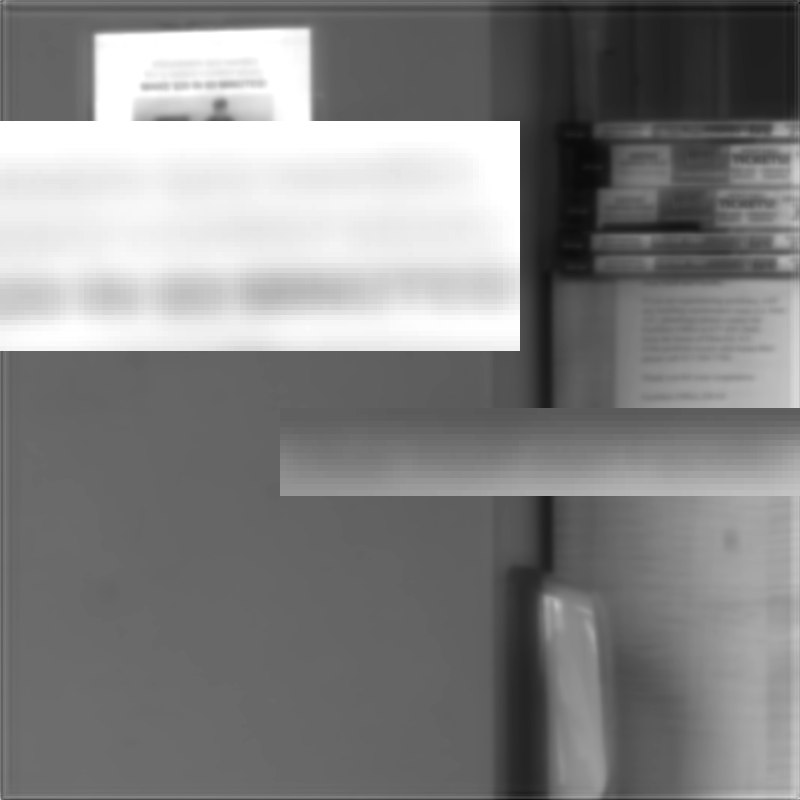}}
\subfigure[NGMeet]{
\label{Fig4}
\includegraphics[width=1.08in]{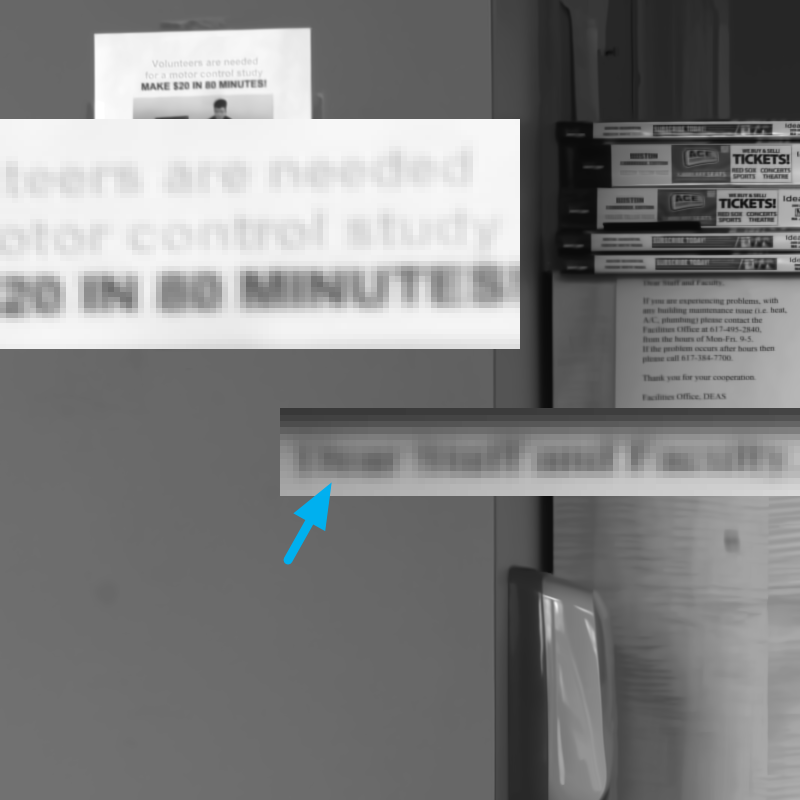}}
\subfigure[MSt-SVD]{
\label{Fig4}
\includegraphics[width=1.08in]{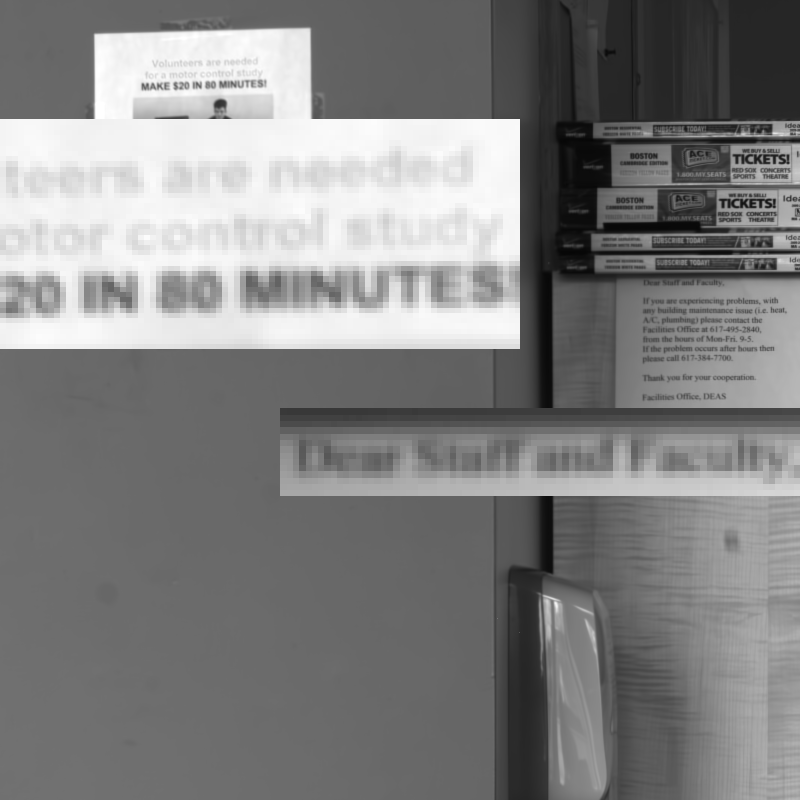}}

\caption{Visual evaluation of compared MSI denoising methods on the real-world HHD dataset.}
\label{Fig_HHD}
\end{figure}

%% file: Table_MRI.tex
\begin{table*}[htbp]
\scriptsize
  \centering
  \caption{Average PSNR/SSIM and computational time (s) of different methods with T1w, T2w, and PDw data corrupted by Rician noise. The standard deviations $\sigma$ of the noise is expressed as percentage relative to the maximum intensity value of the noise-free data \cite{maggioni2012nonlocal}.}
  \renewcommand{\arraystretch}{0.98}
  \scalebox{0.83}{
    \begin{tabular}{ccccccccccccc}
    \toprule
    Noise & Noisy & ONLM  & AONLM & ODCT  & PRINLM & BM4D1 & BM4D2 & NLPCA & PRI-NLPCA & ILR-HOSVD & 4DHOSVD & MSt-SVD \\
    \midrule
    1     & 40.0/0.971 & 41.7/0.991 & 41.9/0.991 & 43.3/0.992 & 43.3/0.993 & 43.2/0.992 & 43.4/0.992 & 44.1/0.994 & 44.3/0.994 & \textbf{44.4/0.994} & 43.9/0.993 & 44.0/0.993 \\
    \midrule
    3     & 30.5/0.822 & 36.4/0.969 & 36.5/0.971 & 36.4/0.970 & 36.9/0.975 & 36.6/0.972 & 37.2/0.975 & 37.7/0.977 & \textbf{38.1/0.981} & 38.0/0.978 & 37.5/0.976 & 37.5/0.975 \\
    \midrule
    5     & 26.1/0.676 & 33.3/0.940 & 33.8/0.947 & 33.6/0.949 & 34.0/0.955 & 33.7/0.950 & 34.6/0.959 & 33.8/0.940 & \textbf{35.3/0.967} & 35.2/0.963 & 34.6/0.959 & 34.6/0.958 \\
    \midrule
    7     & 23.2/0.564 & 31.2/0.911 & 31.9/0.921 & 31.7/0.927 & 32.0/0.934 & 31.8/0.926 & 32.8/0.943 & 31.9/0.912 & \textbf{33.4/0.951} & 33.4/0.948 & 32.9/0.941 & 32.8/0.940 \\
    \midrule
    9     & 21.0/0.479 & 29.4/0.880 & 30.5/0.894 & 30.2/0.905 & 30.5/0.911 & 30.3/0.899 & 31.5/0.928 & 31.2/0.908 & \textbf{31.9/0.934} & \textbf{32.0/0.933} & 31.5/0.922 & 31.3/0.924 \\
    \midrule
    11    & 19.3/0.414 & 27.8/0.847 & 29.3/0.865 & 29.1/0.883 & 29.3/0.888 & 29.2/0.873 & 30.5/0.912 & 30.0/0.883 & 30.7/0.916 & \textbf{30.9/0.917} & 30.4/0.903 & 30.2/0.906 \\
    \midrule
    13    & 17.8/0.361 & 26.3/0.811 & 28.3/0.836 & 28.1/0.861 & 28.2/0.864 & 28.1/0.844 & 29.6/0.895 & 29.0/0.855 & 29.6/0.898 & \textbf{29.9/0.902} & 29.4/0.883 & 29.3/0.888 \\
    \midrule
    15    & 16.6/0.319 & 24.9/0.771 & 27.4/0.807 & 27.3/0.840 & 27.4/0.841 & 27.3/0.814 & 28.8/0.879 & 28.1/0.826 & 28.8/0.879 & \textbf{29.1/0.886} & 28.6/0.863 & 28.5/0.871 \\
    \midrule
    17    & 15.5/0.284 & 23.4/0.729 & 26.5/0.775 & 26.5/0.820 & 26.6/0.817 & 26.5/0.785 & \textbf{28.1/0.862} & 27.3/0.798 & 28.0/0.860 & 27.7/0.838 & 27.8/0.842 & 27.7/0.853 \\
    \midrule
    19    & 14.6/0.254 & 22.0/0.684 & 25.7/0.744 & 25.8/0.799 & 25.9/0.797 & 25.7/0.755 & \textbf{27.4/0.845} & 26.6/0.772 & 27.3/0.841 & 27.0/0.819 & 27.1/0.822 & 27.1/0.835 \\
    \midrule
    Implementation & /     & Mex   & Mex   & Mex   & Mex   & Mex   & Mex   & MATLAB & MEX   & MATLAB & MATLAB & MEX \\
    \midrule
    Time (s) & /     & 139.9 & 868.1 & \textbf{10.5} & 18.6  & 69.9  & 211.3 & 638.8 & 868.9 & 1896.3 & 706.9 & 99.6 \\
    \bottomrule
    \end{tabular}%
    }
  \label{Table_MRI}%
\end{table*}%

%% file: Fig_illus_PSNR_SSIM_MRI.tex
\begin{figure}[htbp]
\graphicspath{{Figs/Fig_others/}}
\centering
\subfigure[PSNR]{
\label{Fig4}
\includegraphics[width=1.69in]{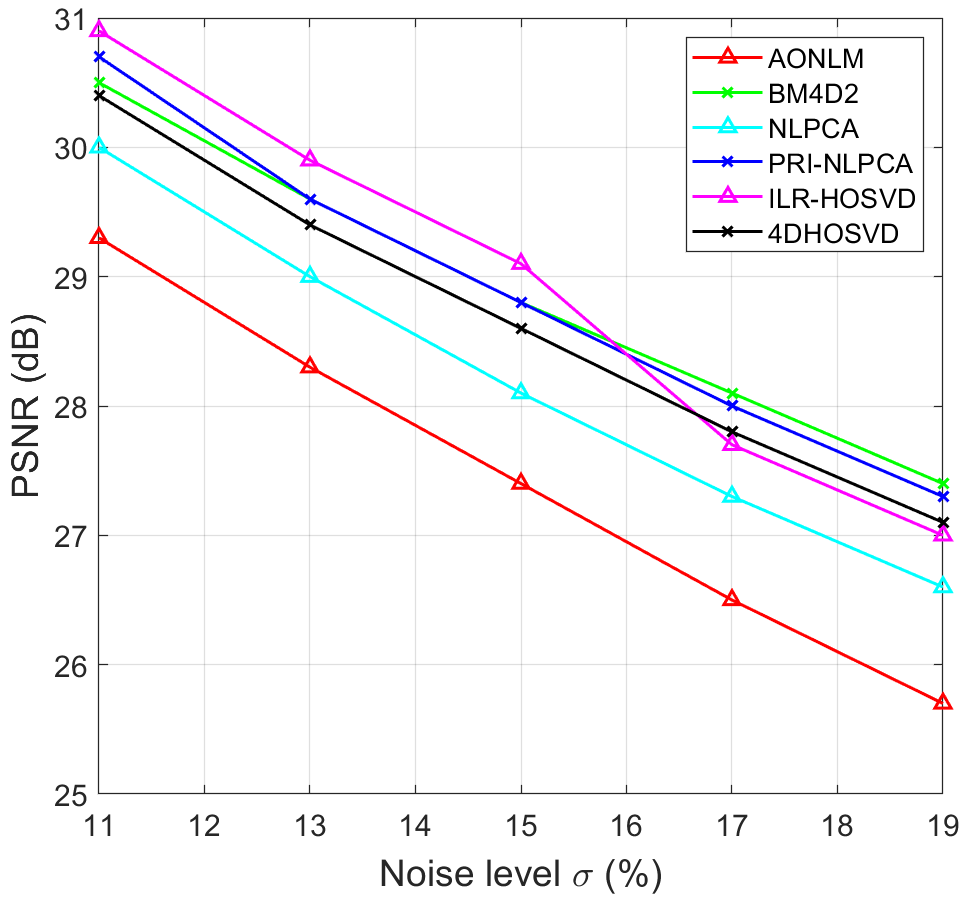}}
\subfigure[SSIM]{
\label{Fig4}
\includegraphics[width=1.69in]{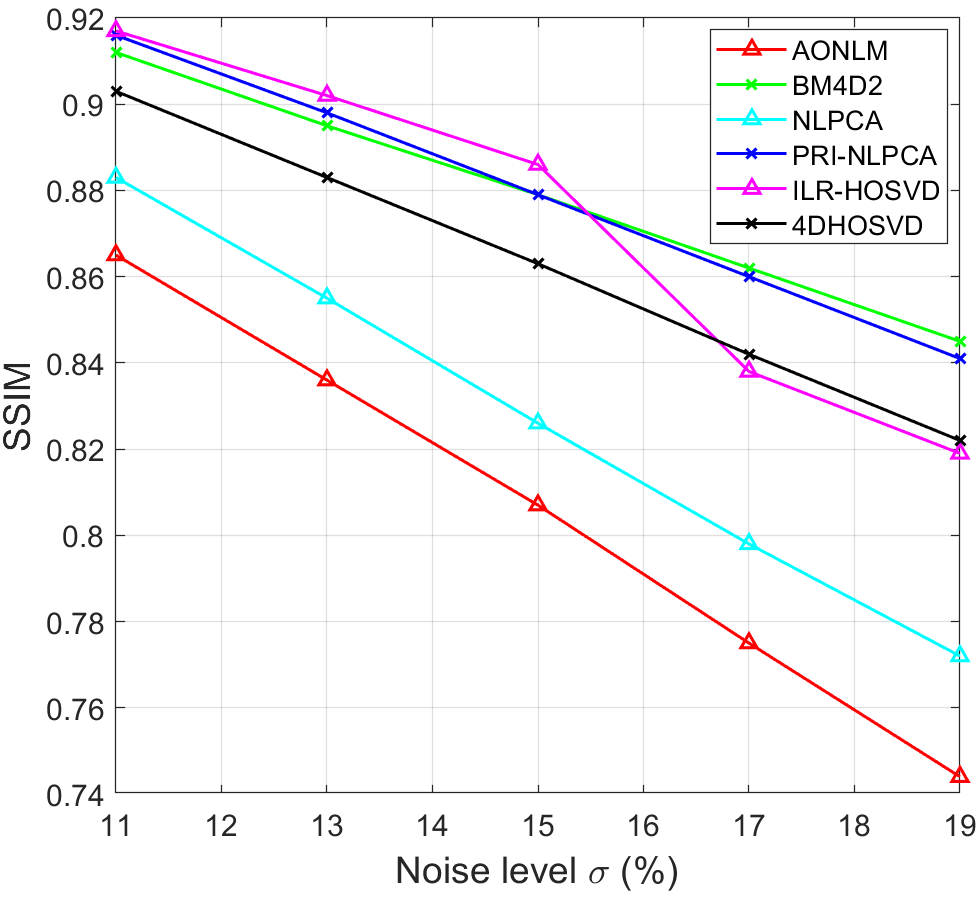}}

\caption{Average PSNR and SSIM values of representative compared methods with T1w, T2w and PDw data at high noise levels ($\sigma \geq 11\%$).}
\label{Fig_illus_PSNR_SSIM_MRI}
\end{figure}

%% file: Fig_Brainweb.tex
\begin{figure*}[htbp]
\graphicspath{{Figs/combine_Brainweb/}}
\centering
\subfigure[Clean]{
\label{Fig4}
\includegraphics[width=0.81in]{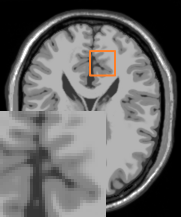}}
\subfigure[Noisy]{
\label{Fig4}
\includegraphics[width=0.81in]{Noisy}}
\subfigure[NLPCA]{
\label{Fig4}
\includegraphics[width=0.81in]{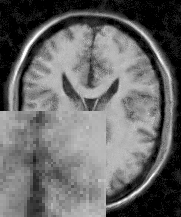}}
\subfigure[PRINLM]{
\label{Fig4}
\includegraphics[width=0.81in]{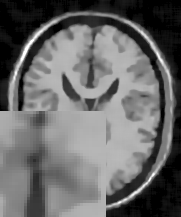}}
\subfigure[PRI-NLPCA]{
\label{Fig4}
\includegraphics[width=0.81in]{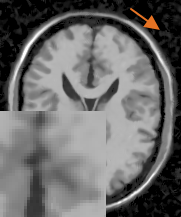}}
\subfigure[ILR-HOSVD]{
\label{Fig4}
\includegraphics[width=0.81in]{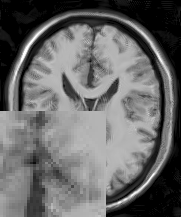}}
\subfigure[MSt-SVD]{
\label{Fig4}
\includegraphics[width=0.81in]{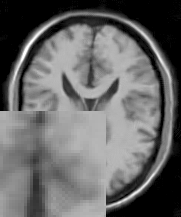}}
\subfigure[BM4D2]{
\label{Fig4}
\includegraphics[width=0.81in]{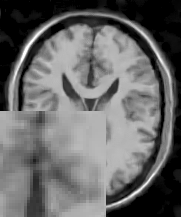}}

\caption{Visual evaluation of compared methods on synthetic Brainweb T1w data with noise level $\sigma$ = 19\%.}
\label{Fig_Brainweb}
\end{figure*}

%% file: Fig_real_MRI_0112.tex
\begin{figure}[htbp]
\graphicspath{{Figs/combine_real_MRI/}}
\centering
\subfigure[MRI\_0112]{
\label{Fig4}
\includegraphics[width=0.78in]{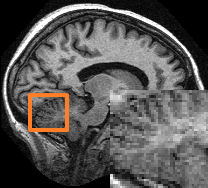}}
\subfigure[NLPCA]{
\label{Fig4}
\includegraphics[width=0.78in]{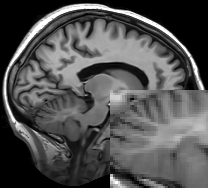}}
\subfigure[PRINLM]{
\label{Fig4}
\includegraphics[width=0.78in]{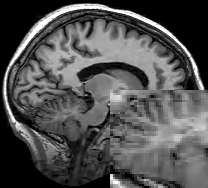}}
\subfigure[PRI-NLPCA]{
\label{Fig4}
\includegraphics[width=0.78in]{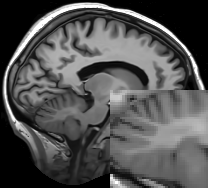}}\\

\subfigure[MSt-SVD]{
\label{Fig4}
\includegraphics[width=0.78in]{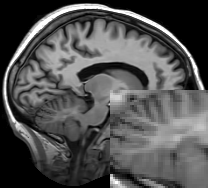}}
\subfigure[HOSVDR\_S]{
\label{Fig4}
\includegraphics[width=0.78in]{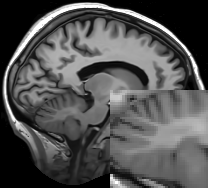}}
\subfigure[BM4D1]{
\label{Fig4}
\includegraphics[width=0.78in]{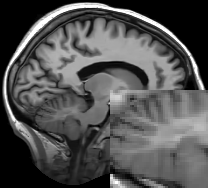}}
\subfigure[BM4D2]{
\label{Fig4}
\includegraphics[width=0.78in]{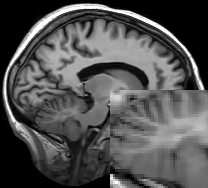}}

\caption{Visual evaluation of compared methods on Real OAS1\_0112 T1w data with estimated noise level $\sigma$ = 3\%.}
\label{Fig_real_MRI_0112}
\end{figure}

%% file: Fig_real_MRI.tex
\begin{figure}[htbp]
\graphicspath{{Figs/combine_real_MRI/}}
\centering
\subfigure[MRI\_0092]{
\label{Fig4}
\includegraphics[width=0.78in]{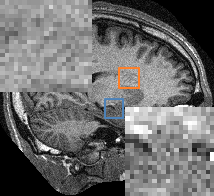}}
\subfigure[NLPCA]{
\label{Fig4}
\includegraphics[width=0.78in]{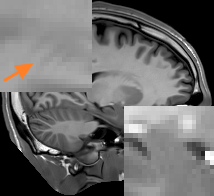}}
\subfigure[PRINLM]{
\label{Fig4}
\includegraphics[width=0.78in]{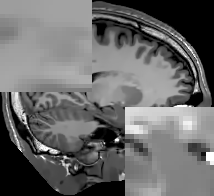}}
\subfigure[PRI-NLPCA]{
\label{Fig4}
\includegraphics[width=0.78in]{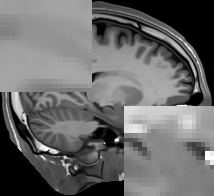}}\\

\subfigure[MSt-SVD]{
\label{Fig4}
\includegraphics[width=0.78in]{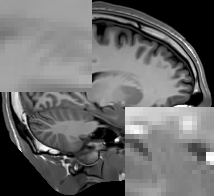}}
\subfigure[HOSVDR\_S]{
\label{Fig4}
\includegraphics[width=0.78in]{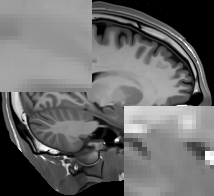}}
\subfigure[BM4D1]{
\label{Fig4}
\includegraphics[width=0.78in]{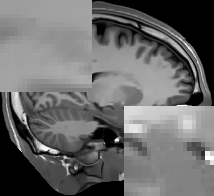}}
\subfigure[BM4D2]{
\label{Fig4}
\includegraphics[width=0.78in]{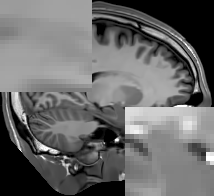}}

\caption{Visual evaluation of compared methods on Real OAS1\_0092 T1w data with estimated noise level $\sigma$ = 4.5\%.}
\label{Fig_real_MRI}
\end{figure}

%% file: Fig_CBM3D_noise_level_influence.tex
\begin{figure}[htbp]
\graphicspath{{Figs/Fig_discussion/}}
\centering
\subfigure[PSNR]{
\label{Fig4}
\includegraphics[width=1.65in]{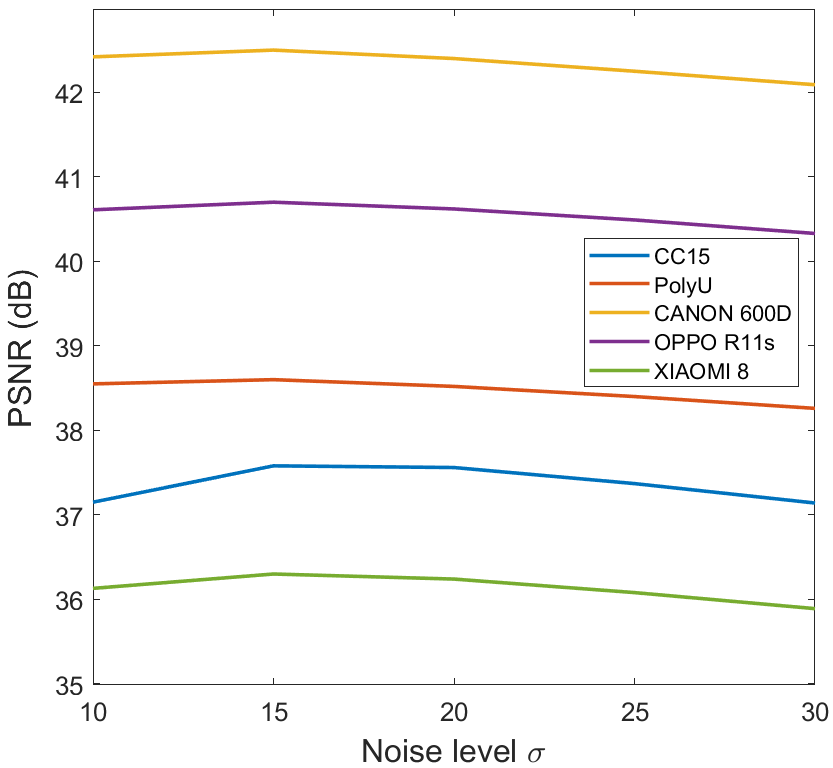}}
\subfigure[SSIM]{
\label{Fig4}
\includegraphics[width=1.699in]{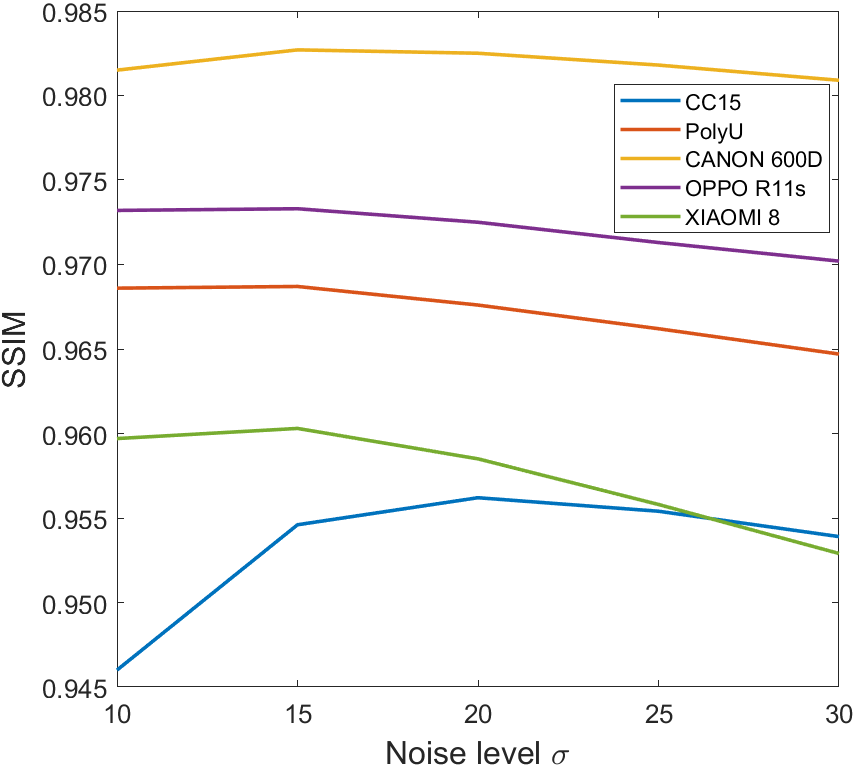}}

\caption{PSNR and SSIM change of CBM3D1 with noise level $\sigma \in [10, 30]$ on five real-world color image datasets.}
\label{Fig_CBM3D_noise_level_influence}
\end{figure}

%% file: Fig_PSNR_diff_Fujifilm_test.tex
\begin{figure}[htbp]

\graphicspath{{Figs/Fig_discussion/}}

%\centering
%\includegraphics[width=2.36in]{PSNR_diff}
\centering
\subfigure[PSNR]{
\label{Fig4}
\includegraphics[width=1.66in]{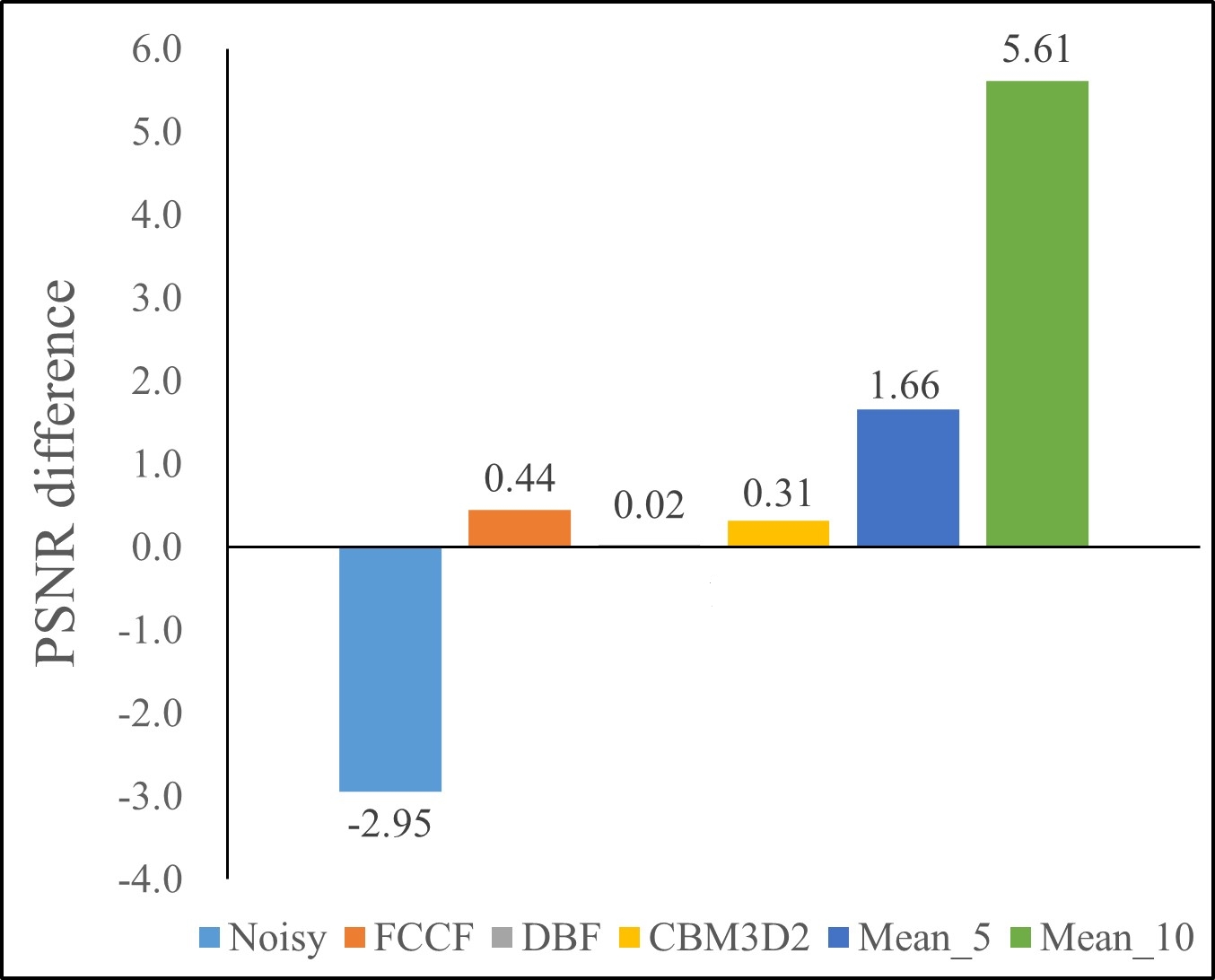}}
\subfigure[SSIM]{
\label{Fig4}
\includegraphics[width=1.66in]{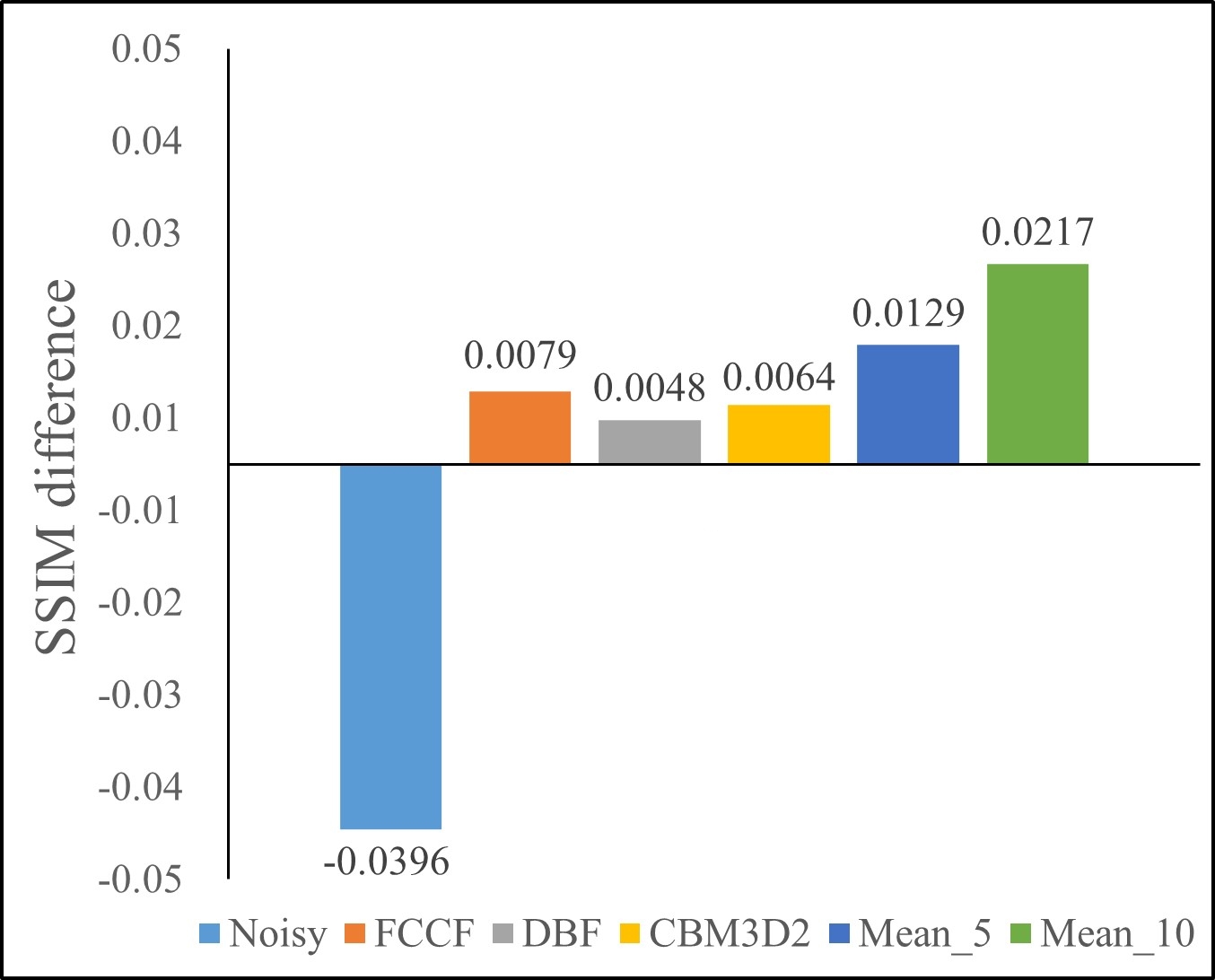}}

\caption{PSNR and SSIM difference of six implementations compared with 'Mean\_3' on FUJI dataset.}

\label{Fig_PSNR_diff_Fujifilm_test}
\end{figure}

%% file: Fig_discussion_visual_enhancement.tex
\begin{figure}[htbp]
\graphicspath{{Figs/Fig_discussion/}}
\centering
\subfigure[Noisy (RENOIR)]{
\label{Fig4}
\includegraphics[width=1.08in]{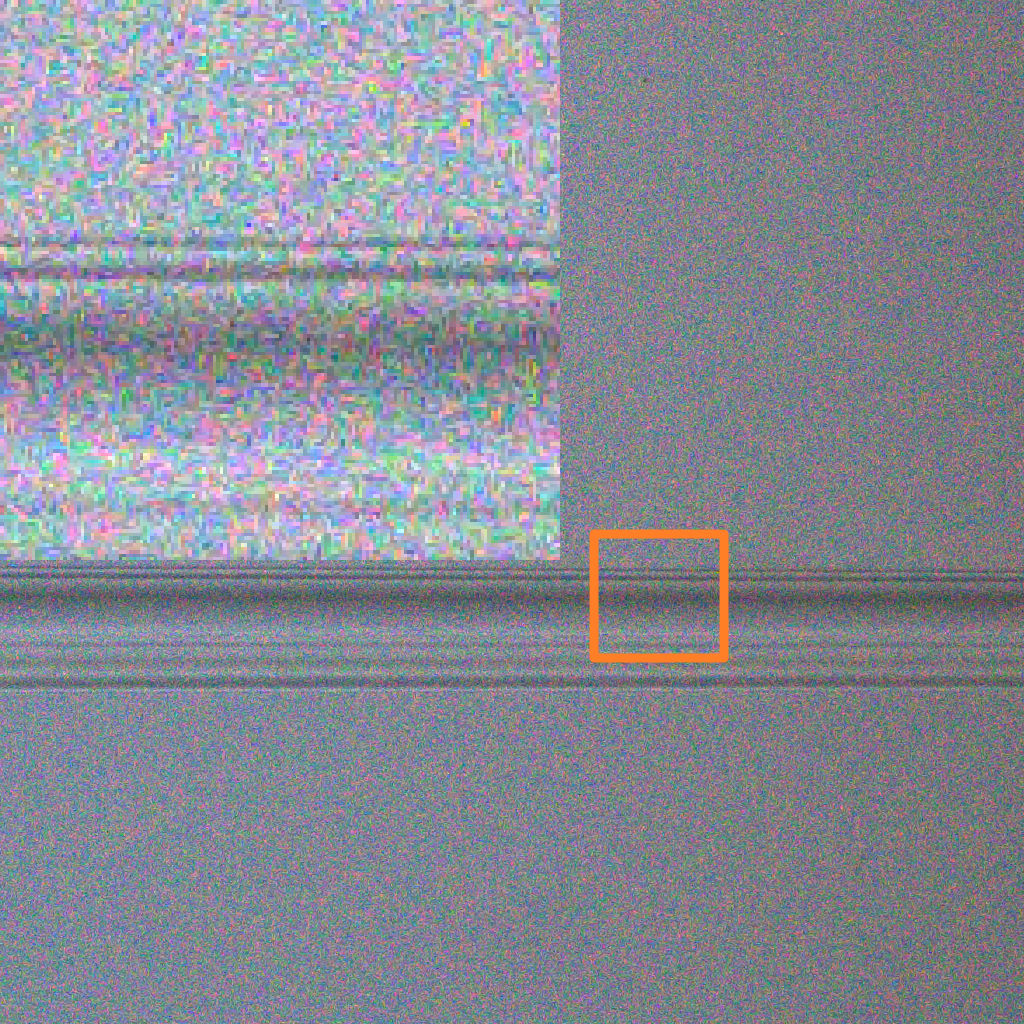}}
\subfigure[CMSt-SVD]{
\label{Fig4}
\includegraphics[width=1.08in]{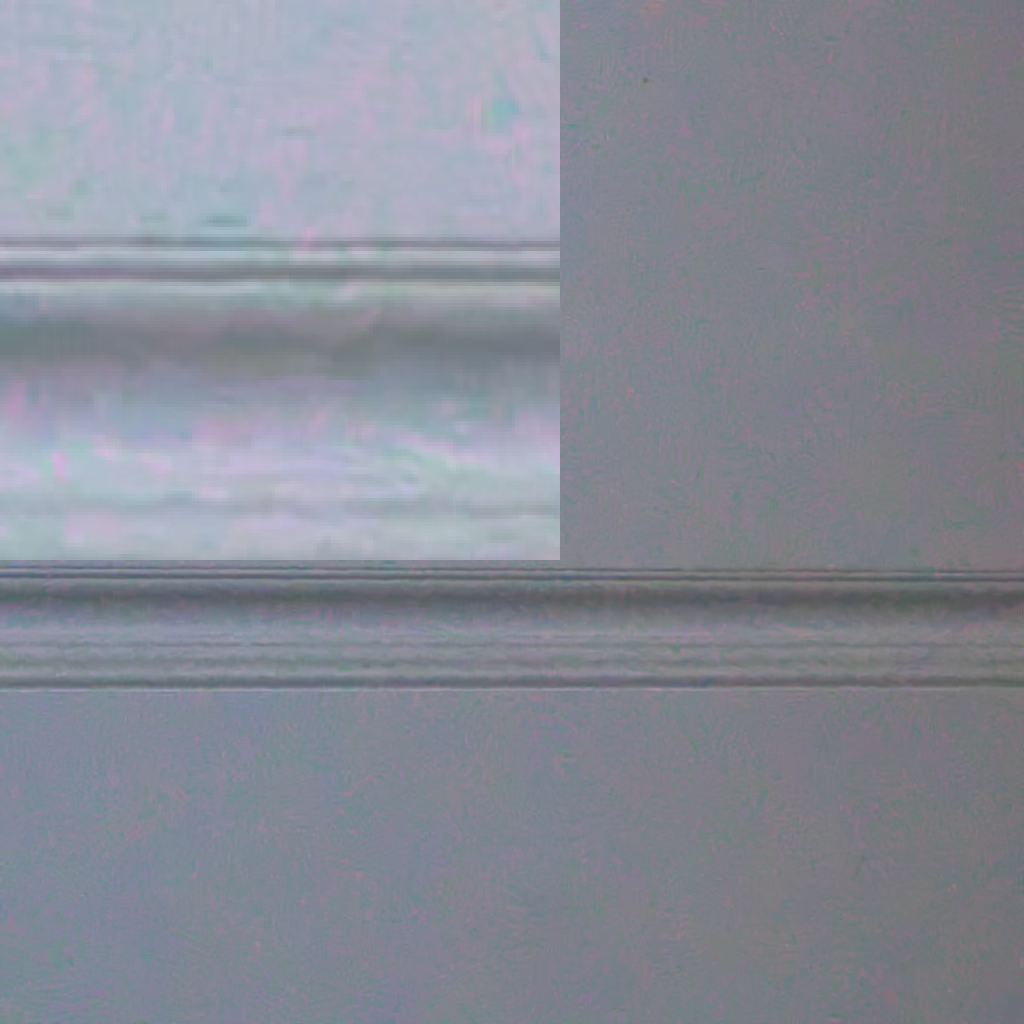}}
\subfigure[CMSt-SVD\_R]{
\label{Fig4}
\includegraphics[width=1.08in]{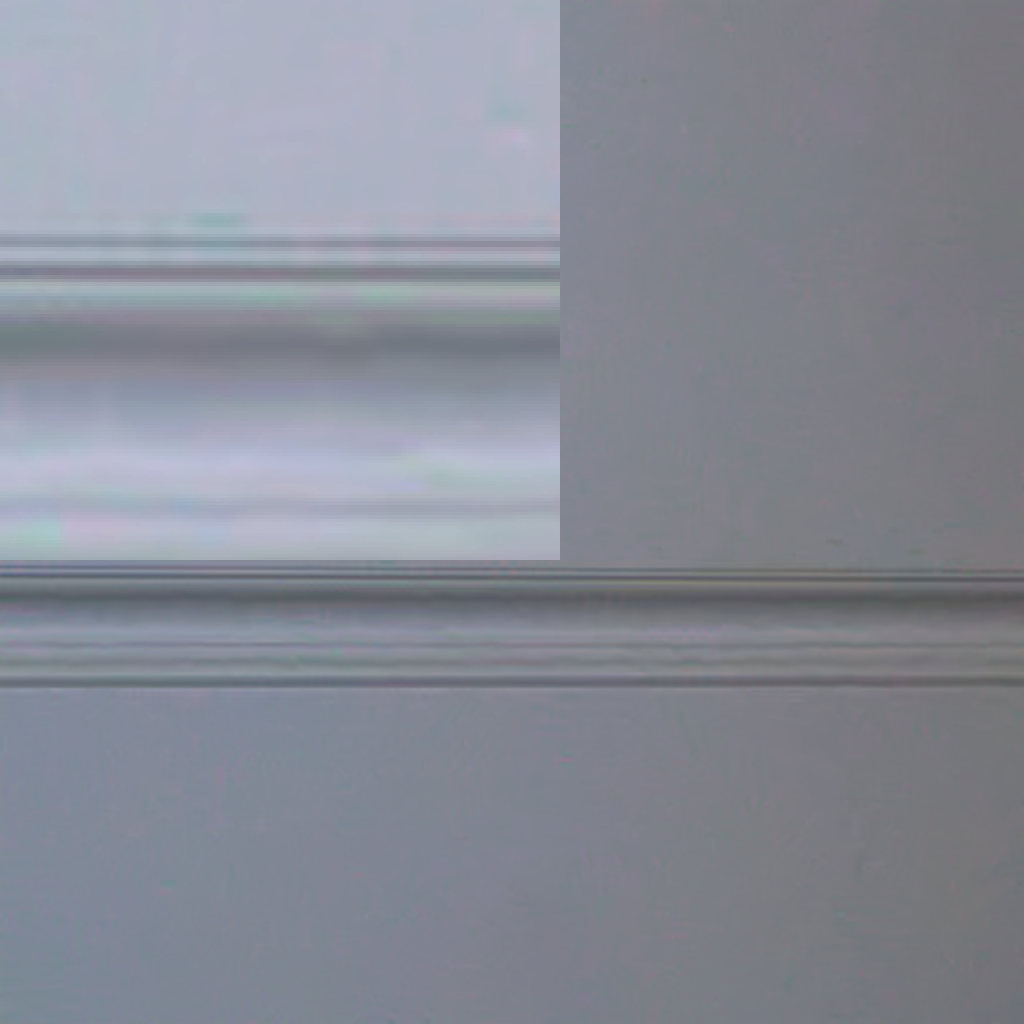}}

\subfigure[Noisy (DND)]{
\label{Fig4}
\includegraphics[width=1.08in]{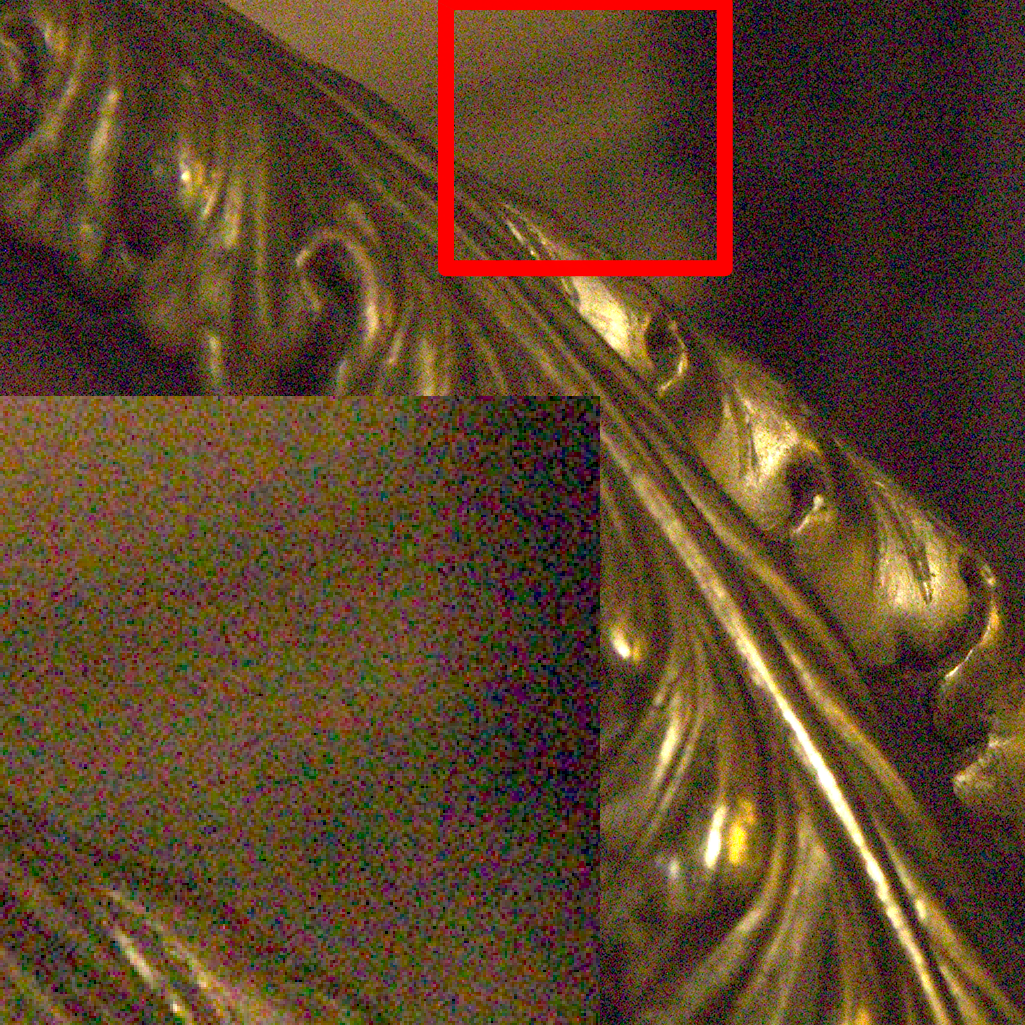}}
\subfigure[CMSt-SVD]{
\label{Fig4}
\includegraphics[width=1.08in]{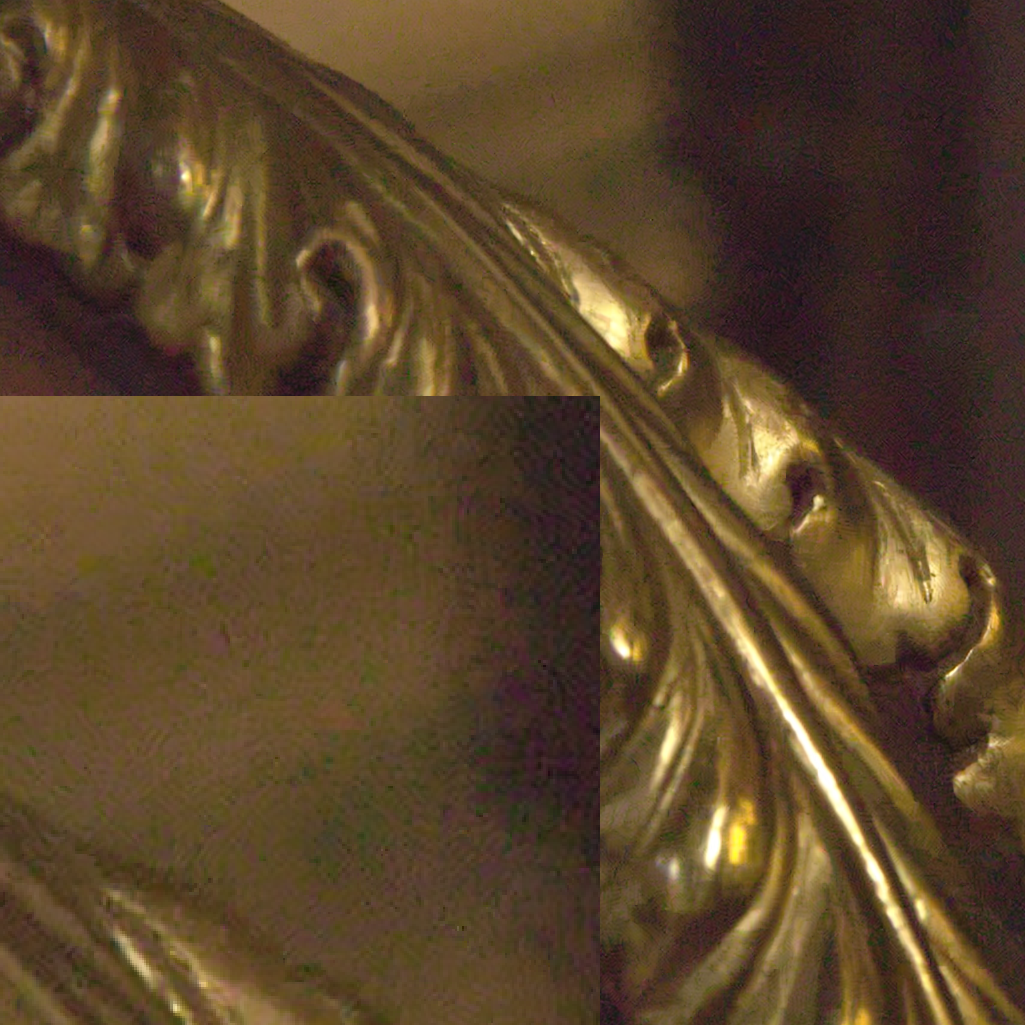}}
\subfigure[CMSt-SVD\_R]{
\label{Fig4}
\includegraphics[width=1.08in]{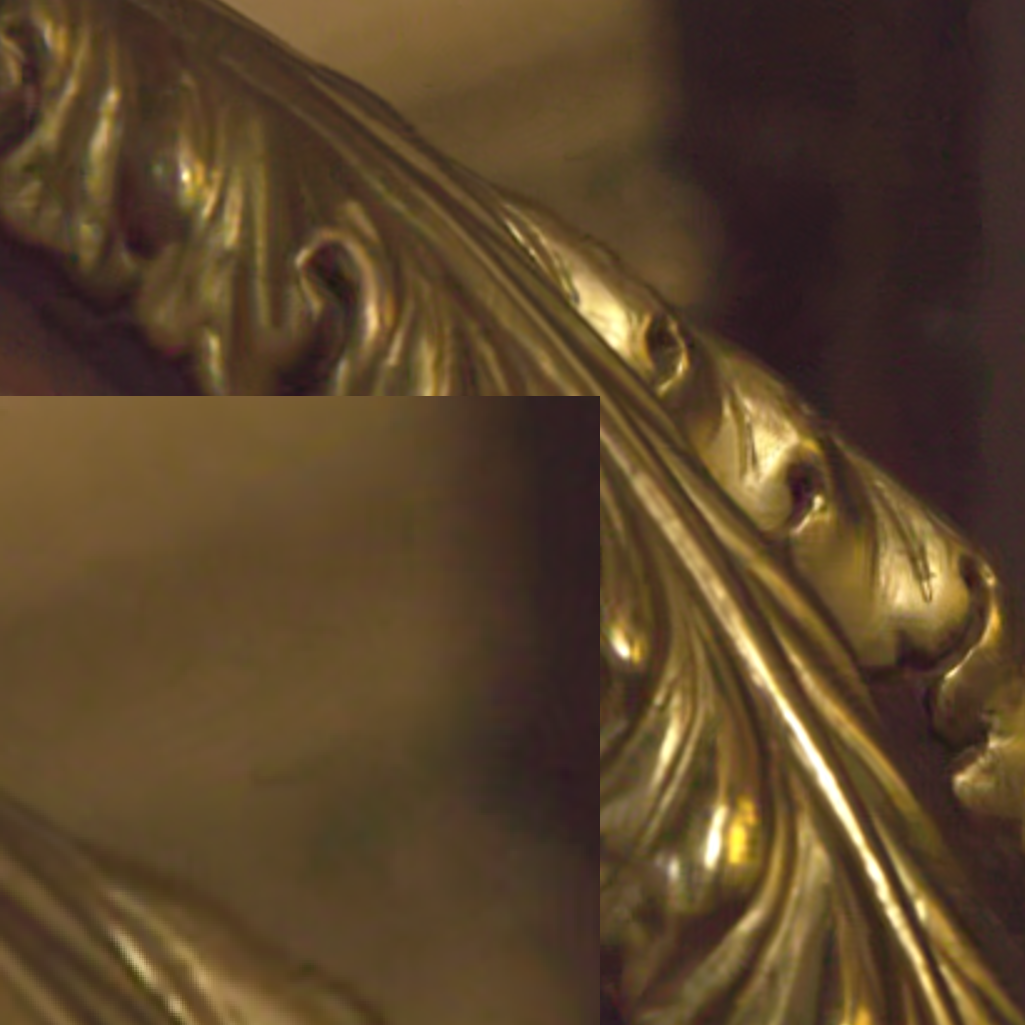}}

\caption{Visual evaluation of CMSt-SVD with and without the image resizing strategy on RENOIR and DND datasets.}
\label{Fig_discussion_visual_enhancement}
\end{figure}

%% file: Algorithm_M_SVD.tex
\begin{algorithm}[ht]
\caption{M-SVD}
{\bf Input:} Noisy image $\mathcal{A}$, number of similar patches $K$.\\
{\bf Output:} Estimated clean image $\hat{\mathcal{A}}_c$.\\
{\bf Step 1} (Grouping): For every reference patch of $\mathcal{A}$, stack $K$ similar patches in a group $\mathcal{G} \in \mathbb{R}^{ps \times ps \times 3 \times K}$.\\
{\bf Step 2} (Collaborative filtering):\\
 \hspace*{0.2in}(1) Obtain the group and patch level transform matrices $\mathbf{U}$ and $\mathbf{V}$ by performing SVD on $\mathbf{G}_{opp_{(3)}}$ and $\mathbf{G}_{(4)}^T$, respectively.\\
 \hspace*{0.2in}(2) Apply the hard-threshold technique to $\mathbf{C} = \mathbf{U}^T \mathbf{G}_{(4)} \mathbf{V}$ via $\mathbf{C}_{trun} = \text{hard-threshold}(\mathbf{C})$.\\
 \hspace*{0.2in}(3) Take the inverse transform to obtain estimated clean group via $\hat{\mathbf{G}}_{c} = \mathbf{U} \mathbf{C}_{trun} \mathbf{V}^T$.\\
{\bf Step 4} (Aggregation): Averagely write back all image patches in $\hat{\mathbf{G}}_{c}$ to their original locations according to Eq. (\ref{aggregation}).
\label{m-svd}
\end{algorithm}

%% file: Fig_oversmooth.tex
\begin{figure}[htbp]
\graphicspath{{Figs/Fig_appendix/}}
\centering
\subfigure[Origin]{
\label{Fig4}
\includegraphics[width=1.10in]{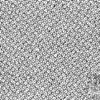}}
\subfigure[T-HOSVD]{
\label{Fig4}
\includegraphics[width=1.10in]{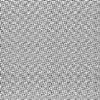}}
\subfigure[Origin - T-HOSVD]{
\label{Fig4}
\includegraphics[width=1.10in]{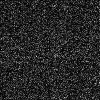}}

\caption{Illustration of over-smooth effects produced by truncated HOSVD (T-HOSVD) when all the patches in a group are the same.}
\label{Fig_oversmooth}
\end{figure}

%% file: ms.bbl
\begin{thebibliography}{100}
\providecommand{\url}[1]{#1}
\csname url@samestyle\endcsname
\providecommand{\newblock}{\relax}
\providecommand{\bibinfo}[2]{#2}
\providecommand{\BIBentrySTDinterwordspacing}{\spaceskip=0pt\relax}
\providecommand{\BIBentryALTinterwordstretchfactor}{4}
\providecommand{\BIBentryALTinterwordspacing}{\spaceskip=\fontdimen2\font plus
\BIBentryALTinterwordstretchfactor\fontdimen3\font minus
  \fontdimen4\font\relax}
\providecommand{\BIBforeignlanguage}[2]{{%
\expandafter\ifx\csname l@#1\endcsname\relax
\typeout{** WARNING: IEEEtran.bst: No hyphenation pattern has been}%
\typeout{** loaded for the language `#1'. Using the pattern for}%
\typeout{** the default language instead.}%
\else
\language=\csname l@#1\endcsname
\fi
#2}}
\providecommand{\BIBdecl}{\relax}
\BIBdecl

\bibitem{tirer2020back}
T.~Tirer and R.~Giryes, ``Back-projection based fidelity term for ill-posed
  linear inverse problems,'' \emph{IEEE Trans. Image Process.}, vol.~29, pp.
  6164--6179, 2020.

\bibitem{isard1998condensation}
M.~Isard and A.~Blake, ``Condensation conditional density propagation for
  visual tracking,'' \emph{Int. J. Comput. Vis.}, vol.~29, no.~1, pp. 5--28,
  1998.

\bibitem{pal1993review}
N.~R. Pal and S.~K. Pal, ``A review on image segmentation techniques,''
  \emph{Pattern Recognit.}, vol.~26, no.~9, pp. 1277--1294, 1993.

\bibitem{raczkowska2019influence}
M.~K. Raczkowska, P.~Koziol, S.~Urbaniak-Wasik, C.~Paluszkiewicz, W.~M.
  Kwiatek, and T.~P. Wrobel, ``Influence of denoising on classification results
  in the context of hyperspectral data: High definition ft-ir imaging,''
  \emph{Anal. Chim. Acta}, vol. 1085, pp. 39--47, 2019.

\bibitem{wiener1950extrapolation}
N.~Wiener, \emph{Extrapolation, interpolation, and smoothing of stationary time
  series: with engineering applications}.\hskip 1em plus 0.5em minus
  0.4em\relax MIT press, 1950.

\bibitem{dabov2007image}
K.~Dabov, A.~Foi, V.~Katkovnik, and K.~Egiazarian, ``Image denoising by sparse
  3-d transform-domain collaborative filtering,'' \emph{IEEE Trans. Image
  Process.}, vol.~16, no.~8, pp. 2080--2095, 2007.

\bibitem{buades2005review}
A.~Buades, B.~Coll, and J.-M. Morel, ``A review of image denoising algorithms,
  with a new one,'' \emph{Multiscale Model. Simul.}, vol.~4, no.~2, pp.
  490--530, 2005.

\bibitem{yaroslavsky2001transform}
L.~P. Yaroslavsky, K.~O. Egiazarian, and J.~T. Astola, ``Transform domain image
  restoration methods: review, comparison, and interpretation,'' in \emph{Proc.
  Nonlinear Image Process. Pattern Anal. XII}, vol. 4304, 2001, pp. 155--169.

\bibitem{dabov2009bm3d}
K.~Dabov, A.~Foi, V.~Katkovnik, and K.~Egiazarian, ``Bm3d image denoising with
  shape-adaptive principal component analysis,'' in \emph{Proc. Workshop on
  SPARS}, 2009, pp. 1--6.

\bibitem{katkovnik2010local}
V.~Katkovnik, A.~Foi, K.~Egiazarian, and J.~Astola, ``From local kernel to
  nonlocal multiple-model image denoising,'' \emph{Int. J. Comput. Vis.},
  vol.~86, no.~1, pp. 1--32, 2010.

\bibitem{dabov2007color}
K.~Dabov, A.~Foi, V.~Katkovnik, and K.~Egiazarian, ``Color image denoising via
  sparse 3d collaborative filtering with grouping constraint in
  luminance-chrominance space,'' in \emph{Proc. IEEE Int. Conf. Image
  Process.}, 2007, pp. 313--316.

\bibitem{danielyan2010denoising}
A.~Danielyan, A.~Foi, V.~Katkovnik, and K.~Egiazarian, ``Denoising of
  multispectral images via nonlocal groupwise spectrum-pca,'' in \emph{Proc.
  Conf. Colour Graph., Imag. Vis.}, vol. 2010, no.~1, 2010, pp. 261--266.

\bibitem{maggioni2012nonlocal}
M.~Maggioni, V.~Katkovnik, K.~Egiazarian, and A.~Foi, ``Nonlocal
  transform-domain filter for volumetric data denoising and reconstruction,''
  \emph{IEEE Trans. Image Process.}, vol.~22, no.~1, pp. 119--133, 2012.

\bibitem{shao2013heuristic}
L.~Shao, R.~Yan, X.~Li, and Y.~Liu, ``From heuristic optimization to dictionary
  learning: A review and comprehensive comparison of image denoising
  algorithms,'' \emph{IEEE Trans. Cybern.}, vol.~44, no.~7, pp. 1001--1013,
  2013.

\bibitem{tian2019deep}
C.~Tian, L.~Fei, W.~Zheng, Y.~Xu, W.~Zuo, and C.-W. Lin, ``Deep learning on
  image denoising: An overview,'' \emph{Neural Networks}, vol. 131, pp.
  251--275, 2020.

\bibitem{kostadin2007video}
D.~Kostadin, F.~Alessandro, and E.~Karen, ``Video denoising by sparse 3d
  transform-domain collaborative filtering,'' in \emph{Proc. 15th Eur. Signal
  Process. Conf.}, 2007, pp. 145--149.

\bibitem{maggioni2012video}
M.~Maggioni, G.~Boracchi, A.~Foi, and K.~Egiazarian, ``Video denoising,
  deblocking, and enhancement through separable 4-d nonlocal spatiotemporal
  transforms,'' \emph{IEEE Trans. Image Process.}, vol.~21, no.~9, pp.
  3952--3966, 2012.

\bibitem{plotz2017benchmarking}
T.~Plotz and S.~Roth, ``Benchmarking denoising algorithms with real
  photographs,'' in \emph{Proc. IEEE Conf. Comput. Vis. Pattern Recognit.},
  2017, pp. 1586--1595.

\bibitem{kong2019color}
Z.~Kong and X.~Yang, ``Color image and multispectral image denoising using
  block diagonal representation,'' \emph{IEEE Trans. Image Process.}, vol.~28,
  no.~9, pp. 4247--4259, 2019.

\bibitem{liu2013single}
X.~Liu, M.~Tanaka, and M.~Okutomi, ``Single-image noise level estimation for
  blind denoising,'' \emph{IEEE Trans. Image Process.}, vol.~22, no.~12, pp.
  5226--5237, 2013.

\bibitem{chen2015efficient}
G.~Chen, F.~Zhu, and P.~Ann~Heng, ``An efficient statistical method for image
  noise level estimation,'' in \emph{Proc. IEEE Int. Conf. Comput. Vis.}, 2015,
  pp. 477--485.

\bibitem{Collection_denoising_methods}
B.~Wen, ``Reproducible denoising methods,'' [Online]. Available:
  \url{https://github.com/wenbihan/reproducible-image-denoising-state-of-the-art}.

\bibitem{milanfar2012tour}
P.~Milanfar, ``A tour of modern image filtering: New insights and methods, both
  practical and theoretical,'' \emph{IEEE Sig. Process. Mag.}, vol.~30, no.~1,
  pp. 106--128, 2012.

\bibitem{thanh2019review}
D.~Thanh, P.~Surya \emph{et~al.}, ``A review on ct and x-ray images denoising
  methods,'' \emph{Informatica}, vol.~43, no.~2, 2019.

\bibitem{mohan2014survey}
J.~Mohan, V.~Krishnaveni, and Y.~Guo, ``A survey on the magnetic resonance
  image denoising methods,'' \emph{Biomed. Sign. Process. Control}, vol.~9, pp.
  56--69, 2014.

\bibitem{schmidhuber2015deep}
J.~Schmidhuber, ``Deep learning in neural networks: An overview,'' \emph{Neural
  networks}, vol.~61, pp. 85--117, 2015.

\bibitem{kong2018brief}
Z.~Kong and X.~Yang, ``A brief review of real-world color image denoising,''
  \emph{arXiv preprint arXiv:1809.03298}, 2018.

\bibitem{yue2019high}
H.~Yue, J.~Liu, J.~Yang, T.~Q. Nguyen, and F.~Wu, ``High iso jpeg image
  denoising by deep fusion of collaborative and convolutional filtering,''
  \emph{IEEE Trans. Image Process.}, vol.~28, no.~9, pp. 4339--4353, 2019.

\bibitem{ronneberger2015u}
O.~Ronneberger, P.~Fischer, and T.~Brox, ``U-net: Convolutional networks for
  biomedical image segmentation,'' in \emph{Proc. Int. Conf. Medical Image
  Comput. Comput.-Assisted Intervention}, 2015, pp. 234--241.

\bibitem{he2016deep}
K.~He, X.~Zhang, S.~Ren, and J.~Sun, ``Deep residual learning for image
  recognition,'' in \emph{Proc. IEEE Conf. Comput. Vis. Pattern Recognit.},
  2016, pp. 770--778.

\bibitem{zhang2020plug}
K.~Zhang, Y.~Li, W.~Zuo, L.~Zhang, L.~Van~Gool, and R.~Timofte, ``Plug-and-play
  image restoration with deep denoiser prior,'' \emph{arXiv preprint
  arXiv:2008.13751}, 2020.

\bibitem{Tassano_2020_CVPR}
M.~Tassano, J.~Delon, and T.~Veit, ``Fastdvdnet: Towards real-time deep video
  denoising without flow estimation,'' in \emph{Proc. IEEE Conf. Comput. Vis.
  Pattern Recognit.}, June 2020.

\bibitem{davy2018non}
A.~Davy, T.~Ehret, J.-M. Morel, P.~Arias, and G.~Facciolo, ``Non-local video
  denoising by cnn,'' \emph{arXiv preprint arXiv:1811.12758}, 2018.

\bibitem{kolda2009tensor}
T.~G. Kolda and B.~W. Bader, ``Tensor decompositions and applications,''
  \emph{SIAM Rev.}, vol.~51, no.~3, pp. 455--500, 2009.

\bibitem{ramanath2005color}
R.~Ramanath, W.~E. Snyder, Y.~Yoo, and M.~S. Drew, ``Color image processing
  pipeline,'' \emph{IEEE Sig. Process. Mag.}, vol.~22, no.~1, pp. 34--43, 2005.

\bibitem{yan2013restoration}
M.~Yan, ``Restoration of images corrupted by impulse noise and mixed gaussian
  impulse noise using blind inpainting,'' \emph{SIAM J. Imag. Sci.}, vol.~6,
  no.~3, pp. 1227--1245, 2013.

\bibitem{huang2017mixed}
T.~Huang, W.~Dong, X.~Xie, G.~Shi, and X.~Bai, ``Mixed noise removal via
  laplacian scale mixture modeling and nonlocal low-rank approximation,''
  \emph{IEEE Trans. Image Process.}, vol.~26, no.~7, pp. 3171--3186, 2017.

\bibitem{zhang2014novel}
Z.~Zhang, G.~Ely, S.~Aeron, N.~Hao, and M.~Kilmer, ``Novel methods for
  multilinear data completion and de-noising based on tensor-svd,'' in
  \emph{Proc. IEEE Conf. Comput. Vis. Pattern Recognit.}, 2014, pp. 3842--3849.

\bibitem{chen2017denoising}
Y.~Chen, X.~Cao, Q.~Zhao, D.~Meng, and Z.~Xu, ``Denoising hyperspectral image
  with non-iid noise structure,'' \emph{IEEE Trans. Cybern.}, vol.~48, no.~3,
  pp. 1054--1066, 2017.

\bibitem{awate2007feature}
S.~P. Awate and R.~T. Whitaker, ``Feature-preserving mri denoising: a
  nonparametric empirical bayes approach,'' \emph{IEEE Trans. Med. Imag.},
  vol.~26, no.~9, pp. 1242--1255, 2007.

\bibitem{romano2017little}
Y.~Romano, M.~Elad, and P.~Milanfar, ``The little engine that could:
  Regularization by denoising (red),'' \emph{SIAM J. Imag. Sci.}, vol.~10,
  no.~4, pp. 1804--1844, 2017.

\bibitem{xu2015patch}
J.~Xu, L.~Zhang, W.~Zuo, D.~Zhang, and X.~Feng, ``Patch group based nonlocal
  self-similarity prior learning for image denoising,'' in \emph{Proc. IEEE
  Conf. Comput. Vis.}, 2015, pp. 244--252.

\bibitem{donoho1995noising}
D.~L. Donoho, ``De-noising by soft-thresholding,'' \emph{IEEE Trans. Inf.
  Theory}, vol.~41, no.~3, pp. 613--627, 1995.

\bibitem{huang2005color}
K.-q. Huang, Z.-y. Wu, G.~S. Fung, and F.~H. Chan, ``Color image denoising with
  wavelet thresholding based on human visual system model,'' \emph{Signal
  Process.: Image Commun.}, vol.~20, no.~2, pp. 115--127, 2005.

\bibitem{othman2006noise}
H.~Othman and S.-E. Qian, ``Noise reduction of hyperspectral imagery using
  hybrid spatial-spectral derivative-domain wavelet shrinkage,'' \emph{IEEE
  Trans. Geosci. Remote Sens.}, vol.~44, no.~2, pp. 397--408, 2006.

\bibitem{yaroslavsky1996local}
L.~P. Yaroslavsky, ``Local adaptive image restoration and enhancement with the
  use of dft and dct in a running window,'' in \emph{Proc. SPIE}, vol. 2825,
  1996, pp. 2--13.

\bibitem{foi2007pointwise}
A.~Foi, V.~Katkovnik, and K.~Egiazarian, ``Pointwise shape-adaptive dct for
  high-quality denoising and deblocking of grayscale and color images,''
  \emph{IEEE Trans. Image Process.}, vol.~16, no.~5, pp. 1395--1411, 2007.

\bibitem{dai2013multichannel}
J.~Dai, O.~C. Au, L.~Fang, C.~Pang, F.~Zou, and J.~Li, ``Multichannel nonlocal
  means fusion for color image denoising,'' \emph{IEEE Trans. Circuits Syst.
  Video Technol.}, vol.~23, no.~11, pp. 1873--1886, 2013.

\bibitem{elad2006image}
M.~Elad and M.~Aharon, ``Image denoising via sparse and redundant
  representations over learned dictionaries,'' \emph{IEEE Trans. Image
  Process.}, vol.~15, no.~12, pp. 3736--3745, 2006.

\bibitem{mairal2007sparse}
J.~Mairal, M.~Elad, and G.~Sapiro, ``Sparse representation for color image
  restoration,'' \emph{IEEE Trans. Image Process.}, vol.~17, no.~1, pp. 53--69,
  2007.

\bibitem{fu2015adaptive}
Y.~Fu, A.~Lam, I.~Sato, and Y.~Sato, ``Adaptive spatial-spectral dictionary
  learning for hyperspectral image denoising,'' in \emph{Proc. IEEE Int. Conf.
  Comput. Vis.}, 2015, pp. 343--351.

\bibitem{liu2010high}
C.~Liu and W.~T. Freeman, ``A high-quality video denoising algorithm based on
  reliable motion estimation,'' in \emph{Proc. Eur. Conf. Comput. Vis.}, 2010,
  pp. 706--719.

\bibitem{zhang2010two}
L.~Zhang, W.~Dong, D.~Zhang, and G.~Shi, ``Two-stage image denoising by
  principal component analysis with local pixel grouping,'' \emph{Pattern
  Recognit.}, vol.~43, no.~4, pp. 1531--1549, 2010.

\bibitem{dong2012nonlocal}
W.~Dong, G.~Shi, and X.~Li, ``Nonlocal image restoration with bilateral
  variance estimation: a low-rank approach,'' \emph{IEEE Trans. Image
  Process.}, vol.~22, no.~2, pp. 700--711, 2012.

\bibitem{phophalia20173d}
A.~Phophalia and S.~K. Mitra, ``3d mr image denoising using rough set and
  kernel pca method,'' \emph{Magn. Reson. Imaging}, vol.~36, pp. 135--145,
  2017.

\bibitem{zoran2011learning}
D.~Zoran and Y.~Weiss, ``From learning models of natural image patches to whole
  image restoration,'' in \emph{Proc. IEEE Int. Conf. Comput. Vis.}, 2011, pp.
  479--486.

\bibitem{hurault2018epll}
S.~Hurault, T.~Ehret, and P.~Arias, ``Epll: an image denoising method using a
  gaussian mixture model learned on a large set of patches,'' \emph{Image
  Processing On Line}, vol.~8, pp. 465--489, 2018.

\bibitem{zhang2013hyperspectral}
H.~Zhang, W.~He, L.~Zhang, H.~Shen, and Q.~Yuan, ``Hyperspectral image
  restoration using low-rank matrix recovery,'' \emph{IEEE Trans. Geosci.
  Remote Sens.}, vol.~52, no.~8, pp. 4729--4743, 2013.

\bibitem{xu2018external}
J.~Xu, L.~Zhang, and D.~Zhang, ``External prior guided internal prior learning
  for real-world noisy image denoising,'' \emph{IEEE Trans. Image Process.},
  vol.~27, no.~6, pp. 2996--3010, 2018.

\bibitem{buades2016patch}
A.~Buades, J.-L. Lisani, and M.~Miladinovi{\'c}, ``Patch-based video denoising
  with optical flow estimation,'' \emph{IEEE Trans. Image Process.}, vol.~25,
  no.~6, pp. 2573--2586, 2016.

\bibitem{rizkinia2016local}
M.~Rizkinia, T.~Baba, K.~Shirai, and M.~Okuda, ``Local spectral component
  decomposition for multi-channel image denoising,'' \emph{IEEE Trans. Image
  Process.}, vol.~25, no.~7, pp. 3208--3218, 2016.

\bibitem{gu2014weighted}
S.~Gu, L.~Zhang, W.~Zuo, and X.~Feng, ``Weighted nuclear norm minimization with
  application to image denoising,'' in \emph{Proc. IEEE Conf. Comput. Vis.
  Pattern Recognit.}, 2014, pp. 2862--2869.

\bibitem{xu2017multi}
J.~Xu, L.~Zhang, D.~Zhang, and X.~Feng, ``Multi-channel weighted nuclear norm
  minimization for real color image denoising,'' in \emph{Proc. IEEE Int. Conf.
  Comput. Vis.}, 2017, pp. 1096--1104.

\bibitem{xu2018trilateral}
J.~Xu, L.~Zhang, and D.~Zhang, ``A trilateral weighted sparse coding scheme for
  real-world image denoising,'' in \emph{Proc. Eur. Conf. Comput. Vis.}, 2018,
  pp. 20--36.

\bibitem{arias2018video}
P.~Arias and J.-M. Morel, ``Video denoising via empirical bayesian estimation
  of space-time patches,'' \emph{J. Math. Imag. Vis.}, vol.~60, no.~1, pp.
  70--93, 2018.

\bibitem{hou2020nlh}
Y.~Hou, J.~Xu, M.~Liu, G.~Liu, L.~Liu, F.~Zhu, and L.~Shao, ``Nlh: A blind
  pixel-level non-local method for real-world image denoising,'' \emph{IEEE
  Trans. Image Process.}, vol.~29, pp. 5121--5135, 2020.

\bibitem{chan2005salt}
R.~H. Chan, C.-W. Ho, and M.~Nikolova, ``Salt-and-pepper noise removal by
  median-type noise detectors and detail-preserving regularization,''
  \emph{IEEE Trans. Image Process.}, vol.~14, no.~10, pp. 1479--1485, 2005.

\bibitem{lebrun2015multiscale}
M.~Lebrun, M.~Colom, and J.-M. Morel, ``Multiscale image blind denoising,''
  \emph{IEEE Trans. Image Process.}, vol.~24, no.~10, pp. 3149--3161, 2015.

\bibitem{zhuang2018fast}
L.~Zhuang and J.~M. Bioucas-Dias, ``Fast hyperspectral image denoising and
  inpainting based on low-rank and sparse representations,'' \emph{IEEE J. Sel.
  Topics Appl. Earth Observ. Remote Sens.}, vol.~11, no.~3, pp. 730--742, 2018.

\bibitem{manjon2012new}
J.~V. Manj{\'o}n, P.~Coup{\'e}, A.~Buades, D.~L. Collins, and M.~Robles, ``New
  methods for mri denoising based on sparseness and self-similarity,''
  \emph{Med. Image Anal.}, vol.~16, no.~1, pp. 18--27, 2012.

\bibitem{coupe2008optimized}
P.~Coup{\'e}, P.~Yger, S.~Prima, P.~Hellier, C.~Kervrann, and C.~Barillot, ``An
  optimized blockwise nonlocal means denoising filter for 3-d magnetic
  resonance images,'' \emph{IEEE Trans. Med. Imag.}, vol.~27, no.~4, pp.
  425--441, 2008.

\bibitem{manjon2010adaptive}
J.~V. Manj{\'o}n, P.~Coup{\'e}, L.~Mart{\'\i}-Bonmat{\'\i}, D.~L. Collins, and
  M.~Robles, ``Adaptive non-local means denoising of mr images with spatially
  varying noise levels,'' \emph{J. Magn. Reson. Imag.}, vol.~31, no.~1, pp.
  192--203, 2010.

\bibitem{aja2008noise}
S.~Aja-Fern{\'a}ndez, C.~Alberola-L{\'o}pez, and C.-F. Westin, ``Noise and
  signal estimation in magnitude mri and rician distributed images: a lmmse
  approach,'' \emph{IEEE Trans. Image Process.}, vol.~17, no.~8, pp.
  1383--1398, 2008.

\bibitem{manjon2015mri}
J.~V. Manj{\'o}n, P.~Coup{\'e}, and A.~Buades, ``Mri noise estimation and
  denoising using non-local pca,'' \emph{Med. Image Anal.}, vol.~22, no.~1, pp.
  35--47, 2015.

\bibitem{renard2008denoising}
N.~Renard, S.~Bourennane, and J.~Blanc-Talon, ``Denoising and dimensionality
  reduction using multilinear tools for hyperspectral images,'' \emph{IEEE
  Geosci. Remote. Sens. Lett.}, vol.~5, no.~2, pp. 138--142, 2008.

\bibitem{liu2012denoising}
X.~Liu, S.~Bourennane, and C.~Fossati, ``Denoising of hyperspectral images
  using the parafac model and statistical performance analysis,'' \emph{IEEE
  Trans. Geosci. Remote Sens.}, vol.~50, no.~10, pp. 3717--3724, 2012.

\bibitem{rajwade2012image}
A.~Rajwade, A.~Rangarajan, and A.~Banerjee, ``Image denoising using the higher
  order singular value decomposition,'' \emph{IEEE Trans. Pattern Anal. Mach.
  Intell.}, vol.~35, no.~4, pp. 849--862, 2012.

\bibitem{zhang2015denoising}
X.~Zhang, Z.~Xu, N.~Jia, W.~Yang, Q.~Feng, W.~Chen, and Y.~Feng, ``Denoising of
  3d magnetic resonance images by using higher-order singular value
  decomposition,'' \emph{Med. Image Anal.}, vol.~19, no.~1, pp. 75--86, 2015.

\bibitem{zhang2017denoise}
X.~Zhang, J.~Peng, M.~Xu, W.~Yang, Z.~Zhang, H.~Guo, W.~Chen, Q.~Feng, E.~X.
  Wu, and Y.~Feng, ``Denoise diffusion-weighted images using higher-order
  singular value decomposition,'' \emph{Neuroimage}, vol. 156, pp. 128--145,
  2017.

\bibitem{peng2014decomposable}
Y.~Peng, D.~Meng, Z.~Xu, C.~Gao, Y.~Yang, and B.~Zhang, ``Decomposable nonlocal
  tensor dictionary learning for multispectral image denoising,'' in
  \emph{Proc. IEEE Conf. Comput. Vis. Pattern Recognit.}, 2014, pp. 2949--2956.

\bibitem{Zhang2015KTSVD}
Z.~Zhang and S.~Aeron, ``Denoising and completion of 3d data via
  multidimensional dictionary learning,'' in \emph{Proc. 25th Int. Joint Conf.
  Artif. Intell}, 2016, pp. 2371--2377.

\bibitem{he2015total}
W.~He, H.~Zhang, L.~Zhang, and H.~Shen, ``Total-variation-regularized low-rank
  matrix factorization for hyperspectral image restoration,'' \emph{IEEE Trans.
  Geosci. Remote Sens.}, vol.~54, no.~1, pp. 178--188, 2015.

\bibitem{xie2016multispectral}
Q.~Xie, Q.~Zhao, D.~Meng, Z.~Xu, S.~Gu, W.~Zuo, and L.~Zhang, ``Multispectral
  images denoising by intrinsic tensor sparsity regularization,'' in
  \emph{Proc. IEEE Conf. Comput. Vis. Pattern Recognit.}, 2016, pp. 1692--1700.

\bibitem{chang2017hyper}
Y.~Chang, L.~Yan, and S.~Zhong, ``Hyper-laplacian regularized unidirectional
  low-rank tensor recovery for multispectral image denoising,'' in \emph{Proc.
  IEEE Conf. Comput. Vis. Pattern Recognit.}, 2017, pp. 4260--4268.

\bibitem{He2018LLRGTV}
W.~{He}, H.~{Zhang}, H.~{Shen}, and L.~{Zhang}, ``Hyperspectral image denoising
  using local low-rank matrix recovery and global spatial–spectral total
  variation,'' \emph{IEEE J. Sel. Topics Appl. Earth Observ. Remote Sens.},
  vol.~11, no.~3, pp. 713--729, 2018.

\bibitem{wu2018weighted}
Y.~Wu, L.~Fang, and S.~Li, ``Weighted tensor rank-1 decomposition for nonlocal
  image denoising,'' \emph{IEEE Trans. Image Process.}, vol.~28, no.~6, pp.
  2719--2730, 2018.

\bibitem{lv2019denoising}
H.~Lv and R.~Wang, ``Denoising 3d magnetic resonance images based on low-rank
  tensor approximation with adaptive multirank estimation,'' \emph{IEEE
  Access}, vol.~7, pp. 85\,995--86\,003, 2019.

\bibitem{kong2017new}
Z.~Kong, L.~Han, X.~Liu, and X.~Yang, ``A new 4-d nonlocal transform-domain
  filter for 3-d magnetic resonance images denoising,'' \emph{IEEE Trans. Med.
  Imag.}, vol.~37, no.~4, pp. 941--954, 2017.

\bibitem{he2019non}
W.~He, Q.~Yao, C.~Li, N.~Yokoya, and Q.~Zhao, ``Non-local meets global: An
  integrated paradigm for hyperspectral denoising,'' in \emph{Proc. IEEE Conf.
  Comput. Vis. Pattern Recognit.}, 2019, pp. 6861--6870.

\bibitem{gong2020low}
X.~Gong, W.~Chen, and J.~Chen, ``A low-rank tensor dictionary learning method
  for hyperspectral image denoising,'' \emph{IEEE Trans. Signal Process.},
  vol.~68, pp. 1168--1180, 2020.

\bibitem{dong2015low}
W.~Dong, G.~Li, G.~Shi, X.~Li, and Y.~Ma, ``Low-rank tensor approximation with
  laplacian scale mixture modeling for multiframe image denoising,'' in
  \emph{Proc. IEEE Int. Conf. Comput. Vis.}, 2015, pp. 442--449.

\bibitem{rubel2014metric}
A.~S. Rubel, V.~V. Lukin, and K.~O. Egiazarian, ``Metric performance in similar
  blocks search and their use in collaborative 3d filtering of grayscale
  images,'' in \emph{Image Processing: Algorithms and Systems XII}, vol. 9019,
  2014.

\bibitem{foi2016foveated}
A.~Foi and G.~Boracchi, ``Foveated nonlocal self-similarity,'' \emph{Int. J.
  Comput. Vis.}, vol. 120, no.~1, pp. 78--110, 2016.

\bibitem{ehret2017global}
T.~Ehret, P.~Arias, and J.-M. Morel, ``Global patch search boosts video
  denoising,'' in \emph{Proc. Int. Conf. Comput. Vis. Theory Appl.}, vol.~5,
  2017, pp. 124--134.

\bibitem{Foi2020}
Y.~{Mäkinen}, L.~{Azzari}, and A.~{Foi}, ``Collaborative filtering of
  correlated noise: Exact transform-domain variance for improved shrinkage and
  patch matching,'' \emph{IEEE Trans. Image Process.}, 2020.

\bibitem{liu2012tensor}
J.~Liu, P.~Musialski, P.~Wonka, and J.~Ye, ``Tensor completion for estimating
  missing values in visual data,'' \emph{IEEE Trans. Pattern Anal. Mach.
  Intell.}, vol.~35, no.~1, pp. 208--220, 2012.

\bibitem{tucker1966some}
L.~R. Tucker, ``Some mathematical notes on three-mode factor analysis,''
  \emph{Psychometrika}, vol.~31, no.~3, pp. 279--311, 1966.

\bibitem{de2000multilinear}
L.~De~Lathauwer, B.~De~Moor, and J.~Vandewalle, ``A multilinear singular value
  decomposition,'' \emph{SIAM J. Matrix Anal. Appl.}, vol.~21, no.~4, pp.
  1253--1278, 2000.

\bibitem{kilmer2011factorization}
M.~E. Kilmer and C.~D. Martin, ``Factorization strategies for third-order
  tensors,'' \emph{Linear Algebra Appl.}, vol. 435, no.~3, pp. 641--658, 2011.

\bibitem{kilmer2013third}
M.~E. Kilmer, K.~Braman, N.~Hao, and R.~C. Hoover, ``Third-order tensors as
  operators on matrices: A theoretical and computational framework with
  applications in imaging,'' \emph{SIAM J. Matrix Anal. Appl.}, vol.~34, no.~1,
  pp. 148--172, 2013.

\bibitem{donoho1994ideal}
D.~L. Donoho and J.~M. Johnstone, ``Ideal spatial adaptation by wavelet
  shrinkage,'' \emph{biometrika}, vol.~81, no.~3, pp. 425--455, 1994.

\bibitem{muti2008lower}
D.~Muti, S.~Bourennane, and J.~Marot, ``Lower-rank tensor approximation and
  multiway filtering,'' \emph{SIAM J. Matrix Anal. Appl.}, vol.~30, no.~3, pp.
  1172--1204, 2008.

\bibitem{oseledets2011tensor}
I.~V. Oseledets, ``Tensor-train decomposition,'' \emph{SIAM J. Sci. Comput.},
  vol.~33, no.~5, pp. 2295--2317, 2011.

\bibitem{bengua2017efficient}
J.~A. Bengua, H.~N. Phien, H.~D. Tuan, and M.~N. Do, ``Efficient tensor
  completion for color image and video recovery: Low-rank tensor train,''
  \emph{IEEE Trans. Image Process.}, vol.~26, no.~5, pp. 2466--2479, 2017.

\bibitem{burger2012image}
H.~C. Burger, C.~J. Schuler, and S.~Harmeling, ``Image denoising: Can plain
  neural networks compete with bm3d?'' in \emph{Proc. IEEE Conf. Comput. Vis.
  Pattern Recognit.}, 2012, pp. 2392--2399.

\bibitem{chen2016trainable}
Y.~Chen and T.~Pock, ``Trainable nonlinear reaction diffusion: A flexible
  framework for fast and effective image restoration,'' \emph{IEEE Trans.
  Pattern Anal. Mach. Intell.}, vol.~39, no.~6, pp. 1256--1272, 2016.

\bibitem{zhang2017beyond}
K.~Zhang, W.~Zuo, Y.~Chen, D.~Meng, and L.~Zhang, ``Beyond a gaussian denoiser:
  Residual learning of deep cnn for image denoising,'' \emph{IEEE Trans. Image
  Process.}, vol.~26, no.~7, pp. 3142--3155, 2017.

\bibitem{jiang2018denoising}
D.~Jiang, W.~Dou, L.~Vosters, X.~Xu, Y.~Sun, and T.~Tan, ``Denoising of 3d
  magnetic resonance images with multi-channel residual learning of
  convolutional neural network,'' \emph{Jpn. J. Radiol.}, vol.~36, no.~9, pp.
  566--574, 2018.

\bibitem{lefkimmiatis2017non}
S.~Lefkimmiatis, ``Non-local color image denoising with convolutional neural
  networks,'' in \emph{Proc. IEEE Conf. Comput. Vis. Pattern Recognit.}, 2017,
  pp. 3587--3596.

\bibitem{Lefkimmiatis_2018_CVPR}
------, ``Universal denoising networks : A novel cnn architecture for image
  denoising,'' in \emph{Proc. IEEE Conf. Comput. Vis. Pattern Recognit.}, 2018,
  pp. 3204--3213.

\bibitem{zhang2018ffdnet}
K.~Zhang, W.~Zuo, and L.~Zhang, ``Ffdnet: Toward a fast and flexible solution
  for cnn-based image denoising,'' \emph{IEEE Trans. Image Process.}, vol.~27,
  no.~9, pp. 4608--4622, 2018.

\bibitem{yuan2018hyperspectral}
Q.~Yuan, Q.~Zhang, J.~Li, H.~Shen, and L.~Zhang, ``Hyperspectral image
  denoising employing a spatial--spectral deep residual convolutional neural
  network,'' \emph{IEEE Trans. Geosci. Remote Sens.}, vol.~57, no.~2, pp.
  1205--1218, 2018.

\bibitem{maffei2019single}
A.~Maffei, J.~M. Haut, M.~E. Paoletti, J.~Plaza, L.~Bruzzone, and A.~Plaza, ``A
  single model cnn for hyperspectral image denoising,'' \emph{IEEE Trans.
  Geosci. Remote Sens.}, vol.~58, no.~4, pp. 2516--2529, 2019.

\bibitem{claus2019videnn}
M.~Claus and J.~van Gemert, ``Videnn: Deep blind video denoising,'' in
  \emph{Proc. Conf. Comput. Vis. Pattern Recognit. Workshops}, 2019, pp. 1--10.

\bibitem{chang2020spatial}
M.~Chang, Q.~Li, H.~Feng, and Z.~Xu, ``Spatial-adaptive network for single
  image denoising,'' \emph{arXiv preprint arXiv:2001.10291}, 2020.

\bibitem{Vaksman_2020_CVPR_Workshops}
G.~Vaksman, M.~Elad, and P.~Milanfar, ``Lidia: Lightweight learned image
  denoising with instance adaptation,'' in \emph{Proc. IEEE Conf. Comput. Vis.
  Pattern Recognit. Workshops}, June 2020.

\bibitem{zhao2020adrn}
Y.~Zhao, D.~Zhai, J.~Jiang, and X.~Liu, ``Adrn: Attention-based deep residual
  network for hyperspectral image denoising,'' in \emph{Proc. IEEE Int. Conf.
  Acoust., Speech Signal Process.}, 2020, pp. 2668--2672.

\bibitem{chang2018hsi}
Y.~Chang, L.~Yan, H.~Fang, S.~Zhong, and W.~Liao, ``Hsi-denet: Hyperspectral
  image restoration via convolutional neural network,'' \emph{IEEE Trans.
  Geosci. Remote Sens.}, vol.~57, no.~2, pp. 667--682, 2018.

\bibitem{lehtinen2018noise2noise}
J.~Lehtinen, J.~Munkberg, J.~Hasselgren, S.~Laine, T.~Karras, M.~Aittala, and
  T.~Aila, ``Noise2noise: Learning image restoration without clean data,'' in
  \emph{Proc. Int. Conf. Mach. Learn.}, 2018, pp. 2965--2974.

\bibitem{yue2019variational}
Z.~Yue, H.~Yong, Q.~Zhao, D.~Meng, and L.~Zhang, ``Variational denoising
  network: Toward blind noise modeling and removal,'' in \emph{Proc. Advances
  Neural Inf. Process. Syst.}, 2019, pp. 1690--1701.

\bibitem{guo2019toward}
S.~Guo, Z.~Yan, K.~Zhang, W.~Zuo, and L.~Zhang, ``Toward convolutional blind
  denoising of real photographs,'' in \emph{Proc. Conf. Comput. Vis. Pattern
  Recognit.}, 2019, pp. 1712--1722.

\bibitem{manjon2018mri}
J.~V. Manj{\'o}n and P.~Coup{\'e}, ``Mri denoising using deep learning,'' in
  \emph{Patch-Based Techniques in Medical Imaging}, 2018, pp. 12--19.

\bibitem{abbasi2019three}
A.~Abbasi, A.~Monadjemi, L.~Fang, H.~Rabbani, and Y.~Zhang, ``Three-dimensional
  optical coherence tomography image denoising through multi-input
  fully-convolutional networks,'' \emph{Comput. Biol. Med.}, vol. 108, pp.
  1--8, 2019.

\bibitem{yu2019deep}
S.~Yu, B.~Park, and J.~Jeong, ``Deep iterative down-up cnn for image
  denoising,'' in \emph{Proc. Conf. Comput. Vis. Pattern Recognit. Workshops},
  2019.

\bibitem{song2019dynamic}
Y.~Song, Y.~Zhu, and X.~Du, ``Dynamic residual dense network for image
  denoising,'' \emph{Sensors}, vol.~19, no.~17, p. 3809, 2019.

\bibitem{gu2019self}
S.~Gu, Y.~Li, L.~V. Gool, and R.~Timofte, ``Self-guided network for fast image
  denoising,'' in \emph{Proc. IEEE Int. Conf. Comput. Vis.}, 2019, pp.
  2511--2520.

\bibitem{chen2019real}
C.~Chen, Z.~Xiong, X.~Tian, Z.-J. Zha, and F.~Wu, ``Real-world image denoising
  with deep boosting,'' \emph{IEEE Trans. Pattern Anal. Mach. Intell.}, 2019.

\bibitem{anwar2019real}
S.~Anwar and N.~Barnes, ``Real image denoising with feature attention,'' in
  \emph{Proc. IEEE Int. Conf. Comput. Vis.}, 2019, pp. 3155--3164.

\bibitem{Kim_2020_CVPR}
Y.~Kim, J.~W. Soh, G.~Y. Park, and N.~I. Cho, ``Transfer learning from
  synthetic to real-noise denoising with adaptive instance normalization,'' in
  \emph{Proc. IEEE Conf. Comput. Vis. Pattern Recognit.}, 2020, pp. 3482--3492.

\bibitem{quan2020self2self}
Y.~Quan, M.~Chen, T.~Pang, and H.~Ji, ``Self2self with dropout: Learning
  self-supervised denoising from single image,'' in \emph{Proc. IEEE Conf.
  Comput. Vis. Pattern Recognit.}, 2020, pp. 1890--1898.

\bibitem{wei20203}
K.~Wei, Y.~Fu, and H.~Huang, ``3-d quasi-recurrent neural network for
  hyperspectral image denoising,'' \emph{IEEE Trans. Neural Netw. Learn.
  Syst.}, 2020.

\bibitem{zamir2020cycleisp}
S.~W. Zamir, A.~Arora, S.~Khan, M.~Hayat, F.~S. Khan, M.-H. Yang, and L.~Shao,
  ``Cycleisp: Real image restoration via improved data synthesis,'' in
  \emph{Proc. IEEE Conf. Comput. Vis. Pattern Recognit.}, 2020, pp. 2696--2705.

\bibitem{yue2020dual}
Z.~Yue, Q.~Zhao, L.~Zhang, and D.~Meng, ``Dual adversarial network: Toward
  real-world noise removal and noise generation,'' \emph{arXiv preprint
  arXiv:2007.05946}, 2020.

\bibitem{Zamir2020MIRNet}
S.~W. Zamir, A.~Arora, S.~Khan, M.~Hayat, F.~S. Khan, M.-H. Yang, and L.~Shao,
  ``Learning enriched features for real image restoration and enhancement,'' in
  \emph{Proc. Eur. Conf. Comput. Vis.}, 2020.

\bibitem{valsesia2020deep}
D.~Valsesia, G.~Fracastoro, and E.~Magli, ``Deep graph-convolutional image
  denoising,'' \emph{IEEE Trans. Image Process.}, vol.~29, pp. 8226--8237,
  2020.

\bibitem{chen2018image}
J.~Chen, J.~Chen, H.~Chao, and M.~Yang, ``Image blind denoising with generative
  adversarial network based noise modeling,'' in \emph{Proc. IEEE Conf. Comput.
  Vis. Pattern Recognit.}, 2018, pp. 3155--3164.

\bibitem{yan2019unsupervised}
H.~Yan, X.~Chen, V.~Y. Tan, W.~Yang, J.~Wu, and J.~Feng, ``Unsupervised image
  noise modeling with self-consistent gan,'' \emph{arXiv preprint
  arXiv:1906.05762}, 2019.

\bibitem{yeh2018image}
R.~A. Yeh, T.~Y. Lim, C.~Chen, A.~G. Schwing, M.~Hasegawa-Johnson, and M.~Do,
  ``Image restoration with deep generative models,'' in \emph{Proc. IEEE Int.
  Conf. Acoust., Speech Signal Process.}, 2018, pp. 6772--6776.

\bibitem{Lin_2019_CVPR_Workshops}
K.~Lin, T.~H. Li, S.~Liu, and G.~Li, ``Real photographs denoising with noise
  domain adaptation and attentive generative adversarial network,'' in
  \emph{Proc. IEEE Conf. Comput. Vis. Pattern Recognit. Workshops}, June 2019.

\bibitem{abdelhamed2019ntire}
A.~Abdelhamed, R.~Timofte, M.~S. Brown, S.~Yu, B.~Park, J.~Jeong, S.~W. Jung,
  D.~W. Kim, J.~R. Chung, J.~Liu \emph{et~al.}, ``Ntire 2019 challenge on real
  image denoising: Methods and results,'' in \emph{Proc. Conf. Comput. Vis.
  Pattern Recognit. Workshops}, 2019, pp. 2197--2210.

\bibitem{abdelhamed2018high}
A.~Abdelhamed, S.~Lin, and M.~S. Brown, ``A high-quality denoising dataset for
  smartphone cameras,'' in \emph{Proc. IEEE Conf. Comput. Vis. Pattern
  Recognit.}, 2018, pp. 1692--1700.

\bibitem{ulyanov2018deep}
D.~Ulyanov, A.~Vedaldi, and V.~Lempitsky, ``Deep image prior,'' in \emph{Proc.
  IEEE Conf. Comput. Vis. Pattern Recognit.}, 2018, pp. 9446--9454.

\bibitem{zhou1987novel}
Y.~Zhou, R.~Chellappa, and B.~Jenkins, ``A novel approach to image restoration
  based on a neural network,'' in \emph{IEEE International Conference on Neural
  Networks}, vol.~4, 1987, pp. 269--276.

\bibitem{chiang1989multi}
Y.-W. Chiang and B.~Sullivan, ``Multi-frame image restoration using a neural
  network,'' in \emph{Proc. Midwest Symp. Circuits Syst.}\hskip 1em plus 0.5em
  minus 0.4em\relax IEEE, 1989, pp. 744--747.

\bibitem{lecun1998gradient}
Y.~LeCun, L.~Bottou, Y.~Bengio, and P.~Haffner, ``Gradient-based learning
  applied to document recognition,'' \emph{Proc. IEEE}, vol.~86, no.~11, pp.
  2278--2324, 1998.

\bibitem{ji20123d}
S.~Ji, W.~Xu, M.~Yang, and K.~Yu, ``3d convolutional neural networks for human
  action recognition,'' \emph{IEEE Trans. Pattern Anal. Mach. Intell.},
  vol.~35, no.~1, pp. 221--231, 2012.

\bibitem{moeskops2016automatic}
P.~Moeskops, M.~A. Viergever, A.~M. Mendrik, L.~S. De~Vries, M.~J. Benders, and
  I.~I{\v{s}}gum, ``Automatic segmentation of mr brain images with a
  convolutional neural network,'' \emph{IEEE Trans. Med. Imag.}, vol.~35,
  no.~5, pp. 1252--1261, 2016.

\bibitem{shi2016real}
W.~Shi, J.~Caballero, F.~Husz{\'a}r, J.~Totz, A.~P. Aitken, R.~Bishop,
  D.~Rueckert, and Z.~Wang, ``Real-time single image and video super-resolution
  using an efficient sub-pixel convolutional neural network,'' in \emph{Proc.
  IEEE Conf. Comput. Vis. Pattern Recognit.}, 2016, pp. 1874--1883.

\bibitem{windrim2018pretraining}
L.~Windrim, A.~Melkumyan, R.~J. Murphy, A.~Chlingaryan, and R.~Ramakrishnan,
  ``Pretraining for hyperspectral convolutional neural network
  classification,'' \emph{IEEE Trans. Geosci. Remote Sens.}, vol.~56, no.~5,
  pp. 2798--2810, 2018.

\bibitem{jain2009natural}
V.~Jain and S.~Seung, ``Natural image denoising with convolutional networks,''
  in \emph{Proc. Advances Neural Inf. Process. Syst.}, 2009, pp. 769--776.

\bibitem{ioffe2015batch}
S.~Ioffe and C.~Szegedy, ``Batch normalization: Accelerating deep network
  training by reducing internal covariate shift,'' in \emph{Proc. Int. Conf.
  Mach. Learn.}, 2015, pp. 448--456.

\bibitem{nair2010rectified}
V.~Nair and G.~E. Hinton, ``Rectified linear units improve restricted boltzmann
  machines,'' in \emph{Proc. Int. Conf. Mach. Learn.}, 2010, pp. 807--814.

\bibitem{radford2015unsupervised}
A.~Radford, L.~Metz, and S.~Chintala, ``Unsupervised representation learning
  with deep convolutional generative adversarial networks,'' \emph{arXiv
  preprint arXiv:1511.06434}, 2015.

\bibitem{chatterjee2009denoising}
P.~Chatterjee and P.~Milanfar, ``Is denoising dead?'' \emph{IEEE Trans. Image
  Process.}, vol.~19, no.~4, pp. 895--911, 2009.

\bibitem{zhang2016fast}
X.~Zhang and R.~Wu, ``Fast depth image denoising and enhancement using a deep
  convolutional network,'' in \emph{Proc. IEEE Int. Conf. Acoust., Speech
  Signal Process.}, 2016, pp. 2499--2503.

\bibitem{nam2016holistic}
S.~Nam, Y.~Hwang, Y.~Matsushita, and S.~Joo~Kim, ``A holistic approach to
  cross-channel image noise modeling and its application to image denoising,''
  in \emph{Proc. IEEE Conf. Comput. Vis. Pattern Recognit.}, 2016, pp.
  1683--1691.

\bibitem{anaya2018renoir}
J.~Anaya and A.~Barbu, ``Renoir--a dataset for real low-light image noise
  reduction,'' \emph{J. Vis. Commun. Image Represent.}, vol.~51, pp. 144--154,
  2018.

\bibitem{xu2018real}
J.~Xu, H.~Li, Z.~Liang, D.~Zhang, and L.~Zhang, ``Real-world noisy image
  denoising: A new benchmark,'' \emph{arXiv preprint arXiv:1804.02603}, 2018.

\bibitem{Perazzi2016}
F.~Perazzi, J.~Pont-Tuset, B.~McWilliams, L.~Van~Gool, M.~Gross, and
  A.~Sorkine-Hornung, ``A benchmark dataset and evaluation methodology for
  video object segmentation,'' in \emph{Proc. IEEE Conf. Comput. Vis. Pattern
  Recognit.}, 2016, pp. 724--732.

\bibitem{yue2020supervised}
H.~Yue, C.~Cao, L.~Liao, R.~Chu, and J.~Yang, ``Supervised raw video denoising
  with a benchmark dataset on dynamic scenes,'' in \emph{Proc. IEEE Conf.
  Comput. Vis. Pattern Recognit.}, 2020, pp. 2301--2310.

\bibitem{CAVE_0293}
F.~Yasuma, T.~Mitsunaga, D.~Iso, and S.~Nayar, ``{G}eneralized {A}ssorted
  {P}ixel {C}amera: {P}ost-{C}apture {C}ontrol of {R}esolution, {D}ynamic
  {R}ange and {S}pectrum,'' Tech. Rep., Nov 2008.

\bibitem{arad_and_ben_shahar_2016_ECCV}
B.~Arad and O.~Ben-Shahar, ``Sparse recovery of hyperspectral signal from
  natural rgb images,'' in \emph{Proc. Eur. Conf. Comput. Vis..}, 2016, pp.
  19--34.

\bibitem{chakrabarti2011statistics}
A.~Chakrabarti and T.~Zickler, ``Statistics of real-world hyperspectral
  images,'' in \emph{Proc. IEEE Conf. Comput. Vis. Pattern Recognit.}, 2011,
  pp. 193--200.

\bibitem{cocosco1997brainweb}
C.~A. Cocosco, V.~Kollokian, R.~K.-S. Kwan, G.~B. Pike, and A.~C. Evans,
  ``Brainweb: Online interface to a 3d mri simulated brain database,'' in
  \emph{NeuroImage}, 1997.

\bibitem{marcus2007open}
D.~S. Marcus, T.~H. Wang, J.~Parker, J.~G. Csernansky, J.~C. Morris, and R.~L.
  Buckner, ``Open access series of imaging studies (oasis): cross-sectional mri
  data in young, middle aged, nondemented, and demented older adults,''
  \emph{J. Cogn. Neurosci.}, vol.~19, no.~9, pp. 1498--1507, 2007.

\bibitem{Kodak}
Kodak, ``Kodak gallery dataset,'' [Online]. Available:
  \url{http://r0k.us/graphics/kodak}.

\bibitem{Set8}
Derf, ``Derfs test media collection,'' [Online]. Available:
  \url{https://media.xiph.org/video/derf}.

\bibitem{wei2020physics}
K.~Wei, Y.~Fu, J.~Yang, and H.~Huang, ``A physics-based noise formation model
  for extreme low-light raw denoising,'' in \emph{Proc. IEEE Conf. Comput. Vis.
  Pattern Recognit.}, 2020, pp. 2758--2767.

\bibitem{amigo2020configuration}
J.~M. Amigo and S.~Grassi, ``Configuration of hyperspectral and multispectral
  imaging systems,'' in \emph{Data Handling in Science and Technology}, 2020,
  vol.~32, pp. 17--34.

\bibitem{elson2020interventional}
D.~S. Elson, ``Interventional imaging: Biophotonics,'' in \emph{Handbook of
  Medical Image Computing and Computer Assisted Intervention}.\hskip 1em plus
  0.5em minus 0.4em\relax Elsevier, 2020, pp. 747--775.

\bibitem{wang2004image}
Z.~Wang, A.~C. Bovik, H.~R. Sheikh, and E.~P. Simoncelli, ``Image quality
  assessment: from error visibility to structural similarity,'' \emph{IEEE
  Trans. Image Process.}, vol.~13, no.~4, pp. 600--612, 2004.

\bibitem{fang2019perceptual}
Y.~Fang, H.~Zhu, K.~Ma, Z.~Wang, and S.~Li, ``Perceptual evaluation for
  multi-exposure image fusion of dynamic scenes,'' \emph{IEEE Trans. Image
  Process.}, vol.~29, pp. 1127--1138, 2019.

\bibitem{yuhas1990determination}
R.~H. Yuhas, J.~W. Boardman, and A.~F. Goetz, ``Determination of semi-arid
  landscape endmembers and seasonal trends using convex geometry spectral
  unmixing techniques,'' \emph{ratio}, vol.~4, p.~22, 1990.

\bibitem{wald2002data}
L.~Wald, \emph{Data fusion: definitions and architectures: fusion of images of
  different spatial resolutions}.\hskip 1em plus 0.5em minus 0.4em\relax Les
  Presses de l'Ecoledes Mines, 2002.

\bibitem{foi2011noise}
A.~Foi, ``Noise estimation and removal in mr imaging: The
  variance-stabilization approach,'' in \emph{Proc. IEEE Int. Symp. Biomed.
  Imag., Nano Macro}, 2011, pp. 1809--1814.

\bibitem{rabie2005robust}
T.~Rabie, ``Robust estimation approach for blind denoising,'' \emph{IEEE Trans.
  Image Process.}, vol.~14, no.~11, pp. 1755--1765, 2005.

\bibitem{majumdar2018blind}
A.~Majumdar, ``Blind denoising autoencoder,'' \emph{IEEE Trans. Neural Netw.
  Learn. Syst.}, vol.~30, no.~1, pp. 312--317, 2018.

\bibitem{chen2006new}
J.~Chen, J.~Benesty, Y.~Huang, and S.~Doclo, ``New insights into the noise
  reduction wiener filter,'' \emph{IEEE/ACM Trans. Audio, Speech, Language
  Process.}, vol.~14, no.~4, pp. 1218--1234, 2006.

\bibitem{luo2012generalized}
E.~Luo, S.~Pan, and T.~Nguyen, ``Generalized non-local means for iterative
  denoising,'' in \emph{Proc. 20th Eur. Signal Process. Conf.}, 2012, pp.
  260--264.

\bibitem{tirer2018image}
T.~Tirer and R.~Giryes, ``Image restoration by iterative denoising and backward
  projections,'' \emph{IEEE Trans. Image Process.}, vol.~28, no.~3, pp.
  1220--1234, 2018.

\bibitem{zontak2013separating}
M.~Zontak, I.~Mosseri, and M.~Irani, ``Separating signal from noise using patch
  recurrence across scales,'' in \emph{Proc. IEEE Conf. Comput. Vis. Pattern
  Recognit.}, 2013, pp. 1195--1202.

\bibitem{dong2015image}
C.~Dong, C.~C. Loy, K.~He, and X.~Tang, ``Image super-resolution using deep
  convolutional networks,'' \emph{IEEE Trans. Pattern Anal. Mach. Intell.},
  vol.~38, no.~2, pp. 295--307, 2015.

\bibitem{wang2020deep}
Z.~Wang, J.~Chen, and S.~C. Hoi, ``Deep learning for image super-resolution: A
  survey,'' \emph{IEEE Trans. Pattern Anal. Mach. Intell.}, 2020.

\bibitem{abramov2020prediction}
S.~Abramov, V.~Lukin, O.~Rubel, and K.~Egiazarian, ``Prediction of performance
  of 2d dct-based filter and adaptive selection of its parameters,''
  \emph{Electronic Imaging}, vol. 2020, no.~9, 2020.

\bibitem{wu2019learning}
Q.~Wu, W.~Ren, and X.~Cao, ``Learning interleaved cascade of shrinkage fields
  for joint image dehazing and denoising,'' \emph{IEEE Trans. Image Process.},
  vol.~29, pp. 1788--1801, 2019.

\bibitem{liu2020joint}
L.~Liu, X.~Jia, J.~Liu, and Q.~Tian, ``Joint demosaicing and denoising with
  self guidance,'' in \emph{Proc. IEEE Conf. Comput. Vis. Pattern Recognit.},
  2020, pp. 2240--2249.

\bibitem{liu2020connecting}
D.~Liu, B.~Wen, J.~Jiao, X.~Liu, Z.~Wang, and T.~S. Huang, ``Connecting image
  denoising and high-level vision tasks via deep learning,'' \emph{IEEE Trans.
  Image Process.}, vol.~29, pp. 3695--3706, 2020.

\bibitem{kamm1998kronecker}
J.~Kamm and J.~G. Nagy, ``Kronecker product and svd approximations in image
  restoration,'' \emph{Linear Algebra Appl.}, vol. 284, no. 1-3, pp. 177--192,
  1998.

\bibitem{de2000best}
L.~De~Lathauwer, B.~De~Moor, and J.~Vandewalle, ``On the best rank-1 and
  rank-(r1, r2,..., rn) approximation of higher-order tensors,'' \emph{SIAM J.
  Matrix Anal. Appl.}, vol.~21, no.~4, pp. 1324--1342, 2000.

\bibitem{cichocki2015tensor}
A.~Cichocki, D.~Mandic, L.~De~Lathauwer, G.~Zhou, Q.~Zhao, C.~Caiafa, and H.~A.
  Phan, ``Tensor decompositions for signal processing applications: From
  two-way to multiway component analysis,'' \emph{IEEE Sig. Process. Mag.},
  vol.~32, no.~2, pp. 145--163, 2015.

\bibitem{zhang2018tensor}
A.~Zhang and D.~Xia, ``Tensor svd: Statistical and computational limits,''
  \emph{IEEE Trans. Inf. Theory}, vol.~64, no.~11, pp. 7311--7338, 2018.

\end{thebibliography}
